\documentclass[12pt]{article}
\pdfoutput=1
\usepackage{amsmath,epsfig,cite,latexsym,color,amssymb,etoolbox}
\usepackage{mathrsfs}
\usepackage[squaren,mediumqspace]{SIunits}
\usepackage{graphicx}
\usepackage{caption,subcaption}
\usepackage[usenames]{xcolor}
\DeclareGraphicsExtensions{.pdf,.png,.jpg,.eps}
\usepackage{hyperref}
\graphicspath{{./figure/}}

\newcommand{\thdm}{{\text{2HDM}}}
\newcommand{\amu}{a_\mu}
\newcommand{\amub}{a_{\mu}^{\text B}}
\newcommand{\amuf}{a_{\mu}^{\text F}}
\newcommand{\tb}{t_\beta}
\newcommand{\coefftb}{\left(\tb -\frac{1}{\tb}\right)}
\newcommand{\lFA}{\Lambda _5}
\newcommand{\twobo}{2-boson~}
\newcommand{\threebo}{3-boson~}
\newcommand{\GlobalC}{\frac{\ALsq}{576\,\pi ^2\,\CW ^4 \SW ^4}\,\frac{\MMsq}{\MZsq}}

\newcommand{\ZHp}{x_{\MHpm}}
\newcommand{\ZHSM}{x_{H_{\text{SM}}}}
\newcommand{\ZAZ}{x_{\MA}}
\newcommand{\ZHH}{x_{\MH}}
\newcommand{\gw}{g_2}
\newcommand{\vac}{v}
\newcommand{\ALsq}{\alpha ^2}
\newcommand{\CW}{c_{\text{W}}}
\newcommand{\CWsq}{c_{\text{W}} ^{2}}
\newcommand{\SW}{s_{\text{W}}}
\newcommand{\SWsq}{s_{\text{W}} ^{2}}
\newcommand{\MMsq}{m_\mu ^2}
\newcommand{\MZ}{M_{Z}}
\newcommand{\MZsq}{M_{Z} ^2}
\newcommand{\MW}{M_{W}}
\newcommand{\MWsq}{M_{W} ^2}
\newcommand{\HSM}{h_{\text{SM}}}
\newcommand{\Mh}{h}
\newcommand{\MH}{H}
\newcommand{\MA}{A}
\newcommand{\MHpm}{H^{\pm}}
\newcommand{\MHSM}{M_{h_{\text{SM}}}}
\newcommand{\MMh}{M_{\Mh}}
\newcommand{\MMH}{M_{\MH}}
\newcommand{\MMA}{M_{\MA}}
\newcommand{\MMHpm}{M_{\MHpm}}

\newcommand{\yAl}{Y ^{\MA} _l}

\newcommand{\zl}{\zeta_l}
\newcommand{\zu}{\zeta_u}
\newcommand{\zd}{\zeta_d}
\newcommand{\dilog}{\text{Li}_2}
\newcommand{\LL}{\text{L}}
\newcommand{\dmw}{\delta M_W ^2}
\newcommand{\dmz}{\delta M_Z ^2}
\newcommand{\dzaa}{\delta Z_{\text{AA}}}
\newcommand{\dzza}{\delta Z_{\text{ZA}}}
\newcommand{\dmmu}{\delta m_{\mu}}
\newcommand{\dzmul}{\delta Z_{\mu}^{\text{L}}}
\newcommand{\dzmur}{\delta Z_{\mu}^{\text{R}}}
\newcommand{\dze}{\delta Z_e}

\newcommand{\dt}{\delta T_{\text{SM}}}
\newcommand{\dth}{\delta T_{h}}
\newcommand{\dtH}{\delta T_{H}}
\newcommand{\fA}{A_0}
\newcommand{\az}{\amu^{\text{non-Yuk}}}
\newcommand{\aezezeze}{a_{0,0}^{0}}
\newcommand{\aezezez}{a_{0,z}^{0}}
\newcommand{\aezelze}{a_{5,0}^{0}}
\newcommand{\aezelz}{a_{5,z}^{0}}
\newcommand{\aezeze}{a_{0,0}^{1}}
\newcommand{\aezez}{a_{0,z}^{1}}
\newcommand{\aelze}{a_{5,0}^{1}}
\newcommand{\aelz}{a_{5,z}^{1}}
\newcommand{\TF}{{\Phi}}
\newcommand{\Fmaster}{{\mathcal{F}}_m^{0}}
\newcommand{\FmasterHp}{{\mathcal{F}}_m^{\pm}}
\newcommand{\eref}[1]{Eq.\,\eqref{#1}}
\newcommand{\erefs}[2]{Eqs.\,\eqref{#1}\,--\,\eqref{#2}}
\newcommand{\erefand}[2]{Eqs.\,\eqref{#1} and~\eqref{#2}}
\newcommand{\Sref}[1]{Sec.\,\ref{#1}}
\newcommand{\fref}[1]{Fig.\,\ref{#1}}
\newcommand{\frefs}[2]{Figs.\,\ref{#1}\,--\,\ref{#2}}
\newcommand{\frefand}[2]{Figs.\,\ref{#1} and~\ref{#2}}

\begin{document}
\begin{center}
{\Large\bf\boldmath The muon magnetic moment in the $\thdm$:\\complete two-loop result}
\\\vspace{3em}
{Adriano Cherchiglia, Patrick Kneschke, Dominik St\"ockinger,\\ Hyejung St\"ockinger-Kim}\\[2em]
 {\sl Institut f\"ur Kern- und Teilchenphysik, TU Dresden, 01069 Dresden, Germany}
\end{center}
\vspace{2ex}
\begin{abstract}
We study the $\thdm$ contribution to the muon anomalous magnetic moment $\amu$ and present the complete two-loop result, particularly for the bosonic contribution.
We focus on the Aligned $\thdm$, which has general Yukawa couplings
and contains the type I, II, X, Y models as special cases. The result is expressed with physical parameters: three Higgs boson masses, Yukawa couplings, two mixing angles, and one quartic potential parameter. We show that the result can be split into several parts, each of which has a simple parameter dependence, and we document their general behavior. Taking into account constraints on parameters, we find that the full $\thdm$ contribution to $\amu$ can accommodate the current experimental value, and the complete two-loop bosonic contribution can amount to $(2\cdots 4)\times10^{-10}$, more than the future experimental uncertainty.
\end{abstract}
\section{Introduction}
\label{sec:intro}
The measured properties of the Higgs boson discovered at the LHC \cite{Aad:2012tfa,Chatrchyan:2012xdj} are compatible with the Standard Model (SM) \cite{DuhrssenMoriond}. However, there is room for alternative explanations of the Higgs boson and electroweak symmetry breaking in models with extended Higgs sectors. The two-Higgs doublet model ($\thdm$) is a particularly interesting framework to be studied. In a large part of its parameter space it is compatible with experimental data, it can originate from more fundamental theories like the MSSM, and it predicts a multitude of observable effects by which it can be studied and constrained.

Here we focus on the muon anomalous magnetic moment $\amu=(g_\mu-2)/2$ in the $\thdm$. This is one of the most useful precision observables to provide complementary, non-collider constraints of extensions of the SM \cite{CzM,WhitePaper,Stockinger:1900zz}. After significant recent progress on all aspects of the SM prediction, there is a stable $3$--$4\,$$\sigma$ deviation between the SM prediction and the Brookhaven measurement \cite{Bennett:2006fi},
\begin{align}
a_\mu^{\text{Exp$-$SM}}&=
\begin{cases}
(28.7 \pm 8.0 ) \times 10^{-10} \mbox{\cite{Davier}}, \\
(26.1 \pm 8.0 ) \times 10^{-10} \mbox{\cite{HMNT}},
\end{cases}
\label{deviation}
\end{align}
using the indicated references for the hadronic vacuum polarization
contributions.\footnote{The numbers take into account the most recent
  refinements on the QED~\cite{Kinoshita2012} and
  electroweak~\cite{Gnendiger:2013pva} contributions. For further
  recent theoretical progress on QED and hadronic contributions and
  reviews, see Refs.~\cite{Kataev:2012kn,SteinhauserQED}, 
 \cite{JegerlehnerSzafron,Benayoun:2012wc,Kurz:2014wya,Colangelo:2014qya,Colangelo:2014,Pauk:2014rfa,lattice,Ablikim:2015orh,Chakraborty:2015ugp}, 
and~\cite{JegerlehnerNyffeler,Miller:2012opa,Blum:2013xva,Benayoun:2014tra,Melnikov:2016wdt }, 
  respectively.}

Several recent studies \cite{Broggio:2014mna,Wang:2014sda,Ilisie:2015tra,Abe:2015oca,Crivellin:2015hha,Chun:2015hsa,Han:2015yys} have shown that the $\thdm$ has viable parameter regions in which this (or at least most of this) deviation is explained. The existing studies evaluate $\amu$ in the $\thdm$ using one-loop and particular two-loop diagrams, so-called Barr-Zee diagrams. Such Barr-Zee diagrams were first considered in Ref.~\cite{BarrZee} and for $\amu$ in the $\thdm$ in Refs.~\cite{Chang:2000ii,Cheung:2001hz,Wu:2001vq,Krawczyk:2002df}; the most complete calculation is presented in Ref.~\cite{Ilisie:2015tra}. Here we present and document the full two-loop calculation of $\amu$ in the $\thdm$, including Barr-Zee and non-Barr-Zee diagrams.

Our calculation is motivated in two ways. Firstly, the $\thdm$ one-loop contributions are suppressed by two additional powers of the small muon Yukawa coupling. 
Thus the one-loop contributions are parametrically smaller than the two-loop contributions. In this sense our calculation completes the {\em leading-order} prediction of $\amu^\thdm$.

Secondly, new $\amu$ experiments are planned at Fermilab and JPARC \cite{Carey:2009zzb,Iinuma:2011zz}. These promise to reduce the experimental uncertainty significantly, in particular the Fermilab measurement plans to obtain
\begin{align}
\Delta\amu^{\text{Fermilab}} &= 1.6\times10^{-10}.
\end{align}
This highlights the need for reliable and accurate theory predictions also in extensions of the SM. In the electroweak SM, the full two-loop calculation has been done in Refs. \cite{CZKM,CZMV,HSW04,Gnendiger:2013pva}. In other models, such as the MSSM, several classes of two-loop contributions have been evaluated \cite{ArhribBaek,ChenGeng,Cheung:2009fc,HSW03,HSW04,vonWeitershausen:2010zr,Fargnoli:2013zda,Fargnoli:2013zia}. It has been found that each class can give rise to significant corrections, and an analysis of the remaining MSSM theory uncertainty has shown that the future experimental precision can only be matched by a complete two-loop computation \cite{DSreview,Gm2Calc}.

This paper is divided as follows: in \Sref{sec:model} we review the $\thdm$ and introduce the phenomenological constraints adopted in our analysis. In \Sref{sec:aresult} the complete renormalized $\thdm$ two-loop contributions to $\amu$ is presented. Each part of the computation is documented in a series of plots and/or analytic formulas. We perform a numerical analysis of our result in \Sref{sec:numanaly}, showing that the complete two-loop bosonic contribution can amount to
$(2\cdots 4)\times10^{-10}$, i.e. at the level of the precision of the planned Fermilab experiment. We present our conclusion in \Sref{sec:conclusion}. Appendix~\ref{appx:1} contains all analytic formulas of the renormalized bosonic two-loop contributions to $\amu$ while in Appendix~\ref{appx:2} we discuss the cancellation of $\MMA$ dependence in $\yAl$ sector.
\section{Two-Higgs Doublet Model}
\label{sec:model}
\subsection{The model and its parameters}
\label{subsec:model}
The two-Higgs-Doublet Model ($\thdm$) is an extended Standard Model (SM) with two complex scalar doublets 
\begin{align}\label{doublets}
\phi_i =& \binom{a_i ^+}{\frac{1}{\sqrt{2}}(v_i + b_i + i c_i)}, i = 1,2. 
\end{align}
Both scalar doublets are assigned with the same hypercharge as the SM doublet. The vacuum expectation value (VEV) of the SM, $v$, is recovered by the relation $ v^2 = v_1 ^2 + v_2 ^2$. The most general form of the Higgs potential $V(\phi_{1},\phi_{2})$ depends on eleven physical parameters~\cite{Branco:2011iw}. In this work, we consider the CP-conserving case in which all parameters are real. For simplicity, we also impose an approximate $Z_2$ symmetry which restricts the quartic couplings to five (denoted by $\lambda_{1,2,3,4,5}$) while the quadratic couplings are given by three parameters ($m^{2}_{11}$, $m^{2}_{22}$, and $m^{2}_{12}$), the last one breaking the $Z_{2}$ symmetry softly~\cite{Gunion:2002zf,Branco:2011iw}

\begin{align}\label{potential}
V\left(\phi_{1},\phi_{2}\right)=&m^{2}_{11}\phi^{\dagger}_{1}\phi_{1}+m^{2}_{22}\phi^{\dagger}_{2}\phi_{2}
-m^{2}_{12}\left(\phi^{\dagger}_{1}\phi_{2}+\phi^{\dagger}_{2}\phi_{1}\right)+\nonumber\\
&+\frac{\lambda_{1}}{2}\left(\phi^{\dagger}_{1}\phi_{1}\right)^{2}
+\frac{\lambda_{2}}{2}\left(\phi^{\dagger}_{2}\phi_{2}\right)^{2}
+\lambda_{3}\phi^{\dagger}_{1}\phi_{1}\phi^{\dagger}_{2}\phi_{2}\nonumber\\
&+\lambda_{4}\phi^{\dagger}_{1}\phi_{2}\phi^{\dagger}_{2}\phi_{1}
+\frac{\lambda_{5}}{2}\left[\left(\phi^{\dagger}_{1}\phi_{2}\right)^{2}+\left(\phi^{\dagger}_{2}\phi_{1}\right)^{2}\right].
\end{align}
 Through a rotation with angle $\tan\beta \equiv \tb \equiv v_2/v_1$, we can choose new scalar doublets $(\Phi _v, \Phi _\perp)$ as  
\begin{align}\label{newdoubletbasis}
\binom{\Phi _v}{\Phi _\perp} =& \binom{\quad\cos\beta\quad\sin\beta}{-\sin\beta\quad\cos\beta}\,\binom{\phi _1}{\phi _2}.
\end{align}
In the new basis only the doublet $\Phi _v$ contains the VEV and the Goldstone bosons, and the components are explicitly   
\begin{align}\label{newdoubletcomponents}
\Phi_v =& \binom{G ^+}{\frac{1}{\sqrt{2}}(v + S_1 +i\,G^0)}, &\Phi_\perp = \binom{H ^+}{\frac{1}{\sqrt{2}}(S_2 + i\,\MA)}.
\end{align}
$\MHpm$ corresponds to the charged Higgs bosons and $\MA$ to the neutral CP-odd one. $S_1$ and $S_2$ are not mass eigenstates, but they are related to the CP-even neutral mass eigenstates $\Mh, \MH$ through a new mixing angle $\alpha$ as
\begin{align}\label{doubletrelation}
\binom{\MH}{\Mh} = \binom{ \cos(\beta-\alpha)\quad-\sin(\beta - \alpha)}{\sin(\beta - \alpha)\quad\quad\cos(\beta - \alpha)}\,\binom{S_1}{S_2}
\end{align}
If $\beta - \alpha = \frac{\pi}{2}$, the two mass eigenstates are
completely separated in each scalar doublet and the neutral CP-even
Higgs boson $\Mh$ has just the same interactions as the SM Higgs
boson, $\HSM$. We call this the SM-limit, following Ref.~\cite{Gunion:2002zf}. The LHC data allow a small deviation $\eta$~\cite{Celis:2013ixa}, which we define as 
\begin{align}
\beta - \alpha = \frac{\pi}{2} - \eta.
\end{align}
In this work we present the results away from the SM-limit, where $\eta \neq 0$.

Seven of the eight parameters,
$m_{11}^{2},m_{22}^{2},m_{12}^{2},\lambda_{1},\cdots,\lambda_{5}$,
introduced in the $\thdm$ potential~\eref{potential} can be replaced
with physical parameters such as the scalar boson masses, $\MMh$,
$\MMH$, $\MMA$, $\MMHpm$, the mixing angles, $\beta$, $\alpha$, the
VEV, $v$~\cite{Gunion:2002zf,Branco:2011iw}. The tree-level relations
for the $\lambda _i$ can be written as
\newcommand{\cosa}{c _\alpha}
\newcommand{\sina}{s _\alpha}
\newcommand{\cosb}{c _\beta}
\newcommand{\sinb}{s _\beta}
\begin{align}\label{lambdas1}
\lambda _1 =& \frac{\MMH ^2 \cosa ^2 + \MMh ^2 \sina ^2 - m_{12} ^2 \tb}{v ^2 \cosb ^2},\\
\label{lambdas2}
\lambda _2 =& \frac{\MMH ^2 \sina ^2 + \MMh ^2 \cosa ^2 - m_{12} ^2 \tb ^{-1}}{v ^2 \sinb ^2},\\
\label{lambdas3}
\lambda _3 =& \frac{(\MMH ^2 - \MMh ^2) \cosa \sina + 2 \MMHpm ^2 \sinb \cosb - m_{12} ^2}{v ^2 \sinb \cosb},\\
\label{lambdas4}
\lambda _4 =& \frac{(\MMA ^2 - 2 \MMHpm ^2) \sinb \cosb + m_{12} ^2}{v ^2 \sinb \cosb},\\
\label{lambdas5}
\lambda _5 =& \frac{m_{12} ^2 - \MMA ^2 \sinb \cosb}{v ^2 \sinb \cosb}, 
\end{align}
where $\cosa =\cos \alpha$, $\sina = \sin \alpha$, $\cosb = \cos \beta$, and $\sinb = \sin \beta$. 
We are still left with one more free parameter $m_{12}^2$, or equivalently $\lambda _{1}$. 
It is convenient to define the quantity $\lFA$, which absorbs $m_{12} ^2$ or $\lambda _1$, as
\begin{align}
\lFA \equiv \frac{2}{v ^2}\frac{m_{12} ^2}{\sin\beta \cos\beta}. 
\end{align}
The equivalent relation in terms of $\lambda _1$  can be written up to $\eta ^1$-order as 
\footnote{$\lFA$ corresponds to $\lambda_{5}$ in the $\thdm$ model file of FeynArts \cite{FeynArts}.}
\begin{align}
\label{LAMBDA}
\lFA &= \frac{2}{\tb ^2}\Big( \frac{\MMh ^2}{v ^2} - \lambda _1 \Big) + 2\,\frac{\MMH ^2}{v ^2} + 4\,\frac{\eta}{\tb}\Big(\frac{\MMH ^2 - \MMh ^2}{v ^2}\Big). 
\end{align}
All the previous relations hold at tree level and might be modified at
higher orders, depending on the chosen renormalization scheme for the
Higgs sector parameters. Renormalization schemes for the $\thdm$ Higgs
sector parameters have been discussed recently in
Refs.~\cite{Krause:2016oke,Denner:2016etu}. For our purposes it will
turn out that the tree-level relations are sufficient.

We complete the discussion of the $\thdm$ by introducing the fermionic sector. The Yukawa coupling is model-dependent. In the present paper we focus on the Aligned $\thdm$. The Aligned $\thdm$ is very general and contains the usual type I, II, X and Y models as special cases: see Table~\ref{table:yukXYZ}.

In the Aligned $\thdm$ it is only required that the mass matrices and the Yukawa coupling matrices in the most general Yukawa Lagrangian are proportional to each other with proportionality constant, $\zeta _{f}$~\cite{Pich:2009sp}. The aligned Yukawa Lagrangian reads 
\begin{align}\label{yukawaalign}
{\cal{L}} _Y = &\sqrt{2} H ^+ \big({\bar{u}}[V_{\text{CKM}} y _d ^{\MHpm} P _{\text{R}} + y _u ^{\MHpm} V_{\text{CKM}} P _{\text{L}}] d + {\bar{\nu}} y _l ^{\MHpm} P _{\text{R}} l \big)\nonumber\\&- \sum _{f}\,h{\bar{f}} y _f ^{h} P _{\text{R}} f - \sum _{f}\,H{\bar{f}} y _f ^{H} P _{\text{R}} f +i\sum _{f}\,A{\bar{f}} y _f ^{A} P _{\text{R}} f+ h.c.,
\end{align}
where $P_{{\text{R}},{\text{L}}} = \frac{1}{2}(1 \pm \gamma _5)$, and $V_{\text{CKM}}$ is the Cabibbo-Kobayashi-Maskawa matrix. The Yukawa coupling matrices are defined as 
\begin{align}\label{yukawas}
y_{f}^{\cal S}&=\frac{Y_{f}^{\cal S}}{v} M_f,
\end{align}
\noindent
where $M_{f}$ denotes the diagonal $3 \times 3$ fermion mass matrix. We have $f = u, d, l$ and ${\cal S} \in \{\Mh, \MH, \MA, \MHpm \}$. The generation independent coefficients $Y_{f}^{S}$ are specific for each model.

In the Aligned $\thdm$, $Y_{f}^{\cal S}$ are dependent on $(\beta - \alpha)$ and $\zeta _f$, and we have~\cite{Pich:2009sp}
\begin{align}\label{yukawacs}
Y^{\Mh}_{f} =& \sin(\beta-\alpha)+\cos(\beta-\alpha)\zeta_{f}, \nonumber\\
Y^{\MH}_{f} =& \cos(\beta-\alpha)-\sin(\beta-\alpha)\zeta_{f}, \nonumber\\
Y^{\MHpm}_{d,l}=Y^{\MA}_{d,l} =& -\zeta_{d,l}, \quad Y^{\MHpm}_{u}=Y^{\MA}_{u} = \zeta_{u}. 
\end{align}
Since we focus on small deviations from the SM-limit, i.e. small $\eta$, it is useful to expand the coefficients of~\eref{yukawacs} for small $\eta$, 
\begin{align}\label{yukawaaeps}
Y^{\Mh}_{f} =& 1 + \eta \zeta_{f}, \quad Y^{\MH}_{f} = - \zeta_{f} + \eta, \quad Y^{\MHpm}_{f}=Y^{\MA}_{f} =-\Theta^{\MA} _{f}\zeta_{f},\nonumber\\
\Theta^{\MA} _{d,l} =& 1, \qquad \Theta^{\MA} _{u} = -1, \qquad \Theta^{\MH} _{u,d,l} = 1.  
\end{align}
\begin{table}[t]
\begin{center}
\begin{tabular}{lrrrr}
\hline\hline
           &\qquad Type I      &\qquad Type II      &\qquad Type X       &\qquad Type Y        \\ \hline        
$\zeta_{u}$ &\qquad $\cot\beta$ &\qquad $\cot\beta$  &\qquad $\cot\beta$  &\qquad $\cot\beta$  \\ \hline
$\zeta_{d}$ &\qquad $\cot\beta$ &\qquad $-\tan\beta$ &\qquad $\cot\beta$  &\qquad $-\tan\beta$ \\ \hline
$\zeta_{l}$ &\qquad $\cot\beta$ &\qquad  $-\tan\beta$ &\qquad $-\tan\beta$ &\qquad $\cot\beta$  \\ \hline\hline
\end{tabular}
\end{center}
\caption{Relation between the Yukawa aligned parameters $\zeta_{f}$ and the usual type I, II, X, and Y models.}
\label{table:yukXYZ}
\end{table}
The parameters $\zeta _{l,u,d}$ are constrained by experimental results of other physical processes. The detailed explanation on the allowed parameter regions is given in~\Sref{subsec:const}. Types I, II, X, Y are recovered by assigning specific values of the aligned parameters $\zeta_{f}$ as listed in Table~\ref{table:yukXYZ}.

\subsection{Constraints}
\label{subsec:const}
Following the presentation of Ref.~\cite{Broggio:2014mna}, we introduce some constraints to restrict the allowed parameter region. They are mainly theoretical and electroweak (EW) constraints. As theoretical constraints, we consider the requirements of stability, and perturbativity that the scalar potential must retain. Regarding EW constraints, we assure that the allowed range for masses of the new scalars does not violate the experimental measured values of EW precision observables such as $\MWsq$ or $\sin\theta_{W}$. 

\subsubsection{Theoretical constraints}
\label{sec:theoconst}

The theoretical constraints faced by the $\thdm$ are of two different natures. The first is related to the stability of the potential, requiring that a vacuum minimum exists and that this minimum is the global minimum of the system. The second is related to perturbativity, requiring that none of the couplings exceeds a given maximal value. For the CP-conserving potential~\eref{potential}, all these requirements are translated into relations between the different $\lambda_{i}$ introduced on the potential as below~\cite{Branco:2011iw,Barroso:2013awa,Ferreira:2009jb}:
\begin{itemize}
\item Stability
\end{itemize}
\begin{align}\label{const1}
&\lambda_{1,2} > 0,\quad \lambda_{3} \geq - \sqrt{\lambda_{1}\lambda_{2}},\quad
\lambda_{3}+\lambda_{4}-|\lambda_{5}|\geq -  \sqrt{\lambda_{1}\lambda_{2}}.
\end{align}

\begin{itemize}
\item Global minimum
\end{itemize}
\begin{align}\label{const2}
m_{12}^{2}\left(m^{2}_{11}-m^{2}_{22}\sqrt{\lambda_{1}/\lambda_{2}}\right)\left(\tb-(\lambda_{1}/\lambda_{2})^{1/4}\right)>0.
\end{align}

\begin{itemize}
\item Perturbativity
\end{itemize}
\begin{align}\label{const3}
|\lambda_{i}|<\lambda_{\text{max}}.
\end{align}

As Ref.~\cite{Ferreira:2009jb,Broggio:2014mna}, we adopt
$\lambda_{\text{max}}=4\pi$. In the phenomenological analysis we
employ~\erefs{lambdas1}{lambdas5} to translate the constraints
of~\erefs{const1}{const3} into those on the physical mass
parameters. Since we do not assume the $\thdm$ to be necessarily a
fundamental theory valid up to very high energy scales, we require the
validity of the above conditions only for the tree-level parameters.
For constraints from requiring conditions on running, high-scale
parameters see particularly Ref.~\cite{Ferreira:2009jb}. 

\subsubsection{Electroweak and experimental constraints}
\label{sec:ewconst}
Regarding electroweak precision data, we will include the constraints on the Peskin-Takeuchi parameters S, T and U~\cite{Peskin:1990zt,Agashe:2014kda}
\begin{align}
S=-0.03\pm0.10, \quad\quad T=0.01\pm0.12, \quad\quad U = 0.05\pm0.10.
\end{align}
To implement them in our phenomenological analysis, we use 2HDMC-1.7.0~\cite{Eriksson:2009ws,Eriksson:2010zzb} to restrict the allowed parameter space on the masses of the scalars. We also include the model-independent constraint obtained by LEP on the mass of the charged scalar~\cite{Agashe:2014kda}
\begin{align}
\MMHpm\geq 80\;\text{GeV}.
\end{align}

Finally we introduce the constraints on the aligned parameters $\zeta_{f}$. As discussed in~\cite{Celis:2013ixa}, in order to avoid conflict with current LHC data they should satisfy
\begin{align}
 0 < |\zu| < 1.2,  \quad \quad \quad  0 < |\zd| < 50,  \quad \quad \quad 0 < |\zl| < 100.
\end{align}
\section{2HDM Two-loop Contributions}
\label{sec:aresult}
The purpose of our study is to present the complete two-loop $\thdm$ contribution to $\amu$. 
The renormalized two-loop result $\amu^{\thdm, 2}$ is the sum of the one-loop contribution $\amu ^{\thdm, 1}$, two-loop bosonic and fermionic contributions and a shift from using the Fermi constant   
\begin{align}\label{amuren2}
\amu ^{\thdm , 2} = \amu ^{\thdm, 1} + \amub + \amuf + \amu ^{\Delta  r  {\text{-shift}}}. 
\end{align}
The actual renormalized two-loop contributions, $\amub$ and $\amuf$,
are obtained from the sum of the appropriate two-loop  and  one-loop
counterterm diagrams. The one-loop contribution and $\amu ^{\Delta r
  {\text{-shift}}}$ are discussed in~\Sref{subsec:oneloop},  the
counterterm parts  in~\Sref{subsec:cts}. The bosonic and fermionic
results are presented in~\Sref{subsec:bloop} and~\Sref{subsec:floop},
respectively. 

In the EW SM, it is sufficient to evaluate the full result only up to
order $m_\mu ^2 / \MWsq$ and neglect higher order terms of ${\cal O}(m
_\mu ^4)$. In the $\thdm$, however, there are potentially
non-negligible terms of this order. Hence we evaluate $\amu ^{\thdm,
  1}$ up to ${\cal O}(m _\mu ^4)$, but at the two-loop level terms up
to ${\cal O}(m _\mu ^2)$ are sufficient. We furthermore expand the results in the
small parameter $\eta=\alpha-\beta+\pi/2$ up to the order $\eta$, and
we set the mass of the Higgs boson $\Mh$ to the mass of the observed
Higgs boson, $\MMh=\MHSM$.

Our calculational procedure is based on the one described in  Refs.\cite{HSW03,HSW04} using TwoCalc \cite{Weiglein:1993} for evaluating two-loop integrals and in-house routines for reduction to master integrals, large mass expansion, and analytical simplification.

\subsection{One-loop contribution}
\label{subsec:oneloop}
The $\thdm$ one-loop result is expressed as \cite{Lautrup:1971jf,Leveille:1977rc,Dedes:2001nx}
\newcommand{\foneloop}{F}
\begin{align}\label{oneloopresult}
\amu ^{\thdm, 1} =& \frac{G_F \, m _\mu ^2}{4\, \sqrt{2}\, \pi ^2}\sum _{\cal S} (Y _l ^{\cal S})^2  \frac{m _\mu ^2}{M _{\cal S} ^2} \foneloop _{\cal S} \Big(\frac{m _\mu ^2}{M _{\cal S} ^2} \Big),
\end{align}
where ${\cal S} \in \{\Mh, \MH, \MA, \MHpm\}$. $G _F$ is the muon decay constant. The $Y _l ^{\cal S}$ are given in~\eref{yukawaaeps}. 

For each Higgs boson the loop-function $\foneloop _{\cal S}$ is defined as
\begin{align}
\label{oneloopa}
\foneloop _{\Mh / \MH}(x) =& \int _0 ^1 d u \frac{u ^2 ( 2 - u )}{1 - u + x u ^2} &\simeq&\quad -\ln (x) - \frac{7}{6} + {\cal O}(x),\\
\label{oneloopb}
\foneloop _{\MA} (x) =& \int _0 ^1 d u \frac{- u ^3}{1 - u + x u ^2} &\simeq&\quad \ln (x) + \frac{11}{6} + {\cal O}(x),\\
\label{oneloopc}
\foneloop _{\MHpm} (x) =& \int _0 ^1 d u \frac{- u (1 - u)}{1 - (1 - u) x} &\simeq&\quad -\frac{1}{6} + {\cal O}(x). 
\end{align}
The right column shows the approximations in the small $x$ limit~\cite{Broggio:2014mna}. 

The numerical values and signs of the different contributions can be easily read off from the following approximation, using $\hat{x} _{\cal S} \equiv M_{\cal S}/100\;\text{GeV}$, and neglecting terms of order $\eta$,  
\begin{align}
\amu ^{\thdm, 1} \simeq& \Big(\frac{\zl}{100}\Big) ^2 \, 10 ^{-10} \, \Big\{ 
\frac{3.3 + 0.5 \ln( \hat{x} _\MH )}{\hat{x} _\MH ^2}
- \frac{3.1 + 0.5 \ln( \hat{x} _{\MA} )}{\hat{x} _{\MA} ^2}-\frac{0.04}{\hat{x} _{\MHpm} ^2} 
\Big\}
\end{align}

At this point we also remark that the EW SM one-loop result is evaluated in terms of the muon decay constant $G _F$,
\begin{align}
\label{oneloopGf}
\amu ^{\text{EW} (1)} = \frac{G_F \, m _\mu ^2}{8 \,\sqrt{2}\, \pi ^2} \Big( \frac{5}{3} + \frac{1}{3}(1 - 4 \SWsq) ^2 \Big).
\end{align} 
$G _F$ is related to the input parameter of the on-shell renormalization scheme as 
\begin{align}
\label{Gf}
\frac{G_F}{1 + \Delta r} = \frac{\pi\,\alpha}{\sqrt{2}\; \SWsq \MWsq}.
\end{align} 
As a result of this, if the on-shell scheme is used to define the counterterms for the two-loop calculation, there is an additional contribution $\amu ^{\text{EW} (1)} \times (- \Delta r)$; see also\cite{CZKM,CZMV,HSW04,Gnendiger:2013pva}.

The extra contribution for the $\thdm$ is then given by
\begin{align}
\label{amushift}
\amu ^{\Delta r {\text{-shift}}} =& \amu ^{\text{EW} (1)} \times (-\Delta r ^{\thdm}), 
\end{align}
where $\Delta r ^{\thdm}$ is the extra $\thdm$ contribution to $\Delta r$. It is discussed in detail in \cite{Bertolini:1985ia,LopezVal:2012zb}. In accordance with Ref.\cite{LopezVal:2012zb} we verified that in all the parameter space relevant for our analysis $\Delta r ^{\thdm}$ is at most of the order of $10 ^{-3}$, and thus $|\amu ^{\Delta r {\text{-shift}}}| \leq 2 \times 10^{-12}$. 
\subsection{Counterterm contribution}
\label{subsec:cts}
The $\thdm$ counterterm diagrams involve the renormalization constants in Table~\ref{table:rc}. 
\begin{table}[t]
\begin{center}
\begin{tabular}{l l}
\hline\hline
Mass renormalization constants:\qquad\qquad & $\dmw$, $\dmz$, $\dmmu$\\
Field renormalization constants:\qquad\qquad & $\dzaa$, $\dzza$, $\dzmul$, $\dzmur$\\
Tadpole renormalization constants:\qquad\qquad & $\dth$, $\dtH$\\ 
\hline\hline
\end{tabular}
\end{center}
\caption{Renormalization constants.}
\label{table:rc}
\end{table}
These renormalization constants are defined in the on-shell renormalization scheme~\cite{Denner:1991kt,Santos:1996vt}.
In terms of these the electric charge renormalization constant $\dze$ is derived as 
\begin{align}
\dze =& -\frac{1}{2}\,\left(\dzaa + \frac{\SW}{\CW}\,\dzza\right),
\end{align}
where $\dzaa$ is the photon field renormalization constant and $\dzza$ the photon-$Z$ mixing renormalization constant. For the mass and field renormalization constants several useful statements can be made. They are obtained from self-energy diagrams with external SM particles. In the expansion in $\eta$ up to ${\cal O}(\eta ^1)$ each mass and field renormalization constant can be decomposed into the SM and additional $\thdm$ contributions. These additional $\thdm$ contributions to the renormalization constants are obtained by computing the loop diagrams containing the new scalar bosons in the $\thdm$. For these renormalization constants the fermionic contributions are the same in both the SM and $\thdm$. Therefore the additional $\thdm$ contributions from these diagrams arise entirely from the bosonic parts. 

The tadpole renormalization constants should be treated separately. The tadpole renormalization constants are determined in such a way that the one-point Green functions of the CP-even Higgs fields vanish. In the CP-conserving $\thdm$ there are two tadpole renormalization constants, $\dth$ and $\dtH$, whereas in the SM we have one tadpole renormalization constant $\dt$. The contributions from gauge($g$) and Goldstone($G$) bosons and fermions($f$) are related to the SM counterpart by simple rescaling of couplings as 
\begin{align}\label{tadpolegGf}
\dth ^{(g/G/f)} = \sin(\beta - \alpha) \dt ^{(g/G/f)},\,\,\,\dtH ^{(g/G/f)} = \cos(\beta - \alpha) \dt ^{(g/G/f)}.
\end{align}
However, the Higgs loops of the tadpole diagrams are proportional to the triple Higgs couplings, are $\tb$-dependent, and do not satisfy such a simple relation. 
\begin{figure}[]
\centering
\begin{subfigure}[]{.3\textwidth}
\includegraphics[scale=.5]{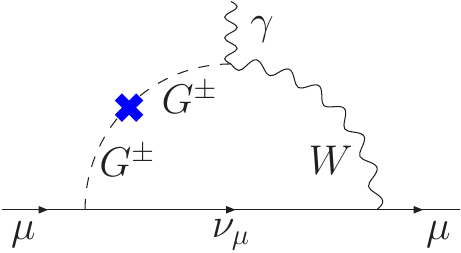}
\subcaption{}
\label{fig:ctGG}
\end{subfigure}
\begin{subfigure}[]{.3\textwidth}
\includegraphics[scale=.5]{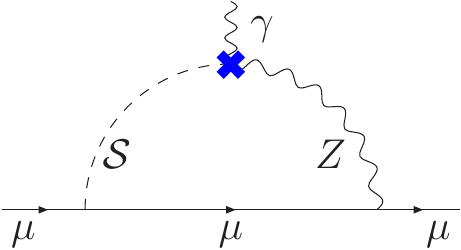}
\subcaption{}
\label{fig:ctH}
\end{subfigure}
\begin{subfigure}[]{.3\textwidth}
\includegraphics[scale=.5]{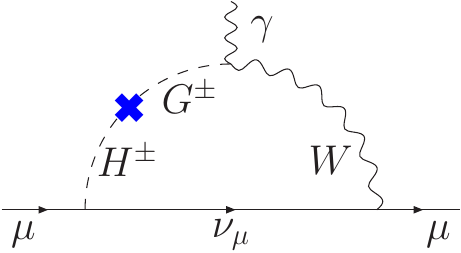}
\subcaption{}
\label{fig:ctHG}
\end{subfigure}
\caption{Counterterm Feynman diagrams.}
\label{fig:thdmct}
\end{figure}

We now turn to the counterterm diagrams.~\fref{fig:thdmct} shows sample diagrams. It is convenient to classify the $\thdm$ counterterm diagrams into three groups. The first group encompasses the SM-like counterterm diagrams without Higgs boson inside. The second group contains the counterterm diagrams with neutral physical Higgs bosons. The third group consists of the counterterm diagram with $G ^\pm$-$H ^\pm$ mixing counterterm vertex. In the following we explain them one after the other and provide the explicit results. 
\begin{itemize}
\item SM-like counterterms without physical Higgs bosons:\\
The first group encompasses the SM counterterm diagrams which do not contain physical Higgs bosons. The results of SM counterterm diagrams are found in Ref.~\cite{CZKM}. The additional $\thdm$ contributions from these counterterm diagrams are obtained by applying the corresponding additional $\thdm$ renormalization constants. This is straightforward for all diagrams except~\fref{fig:ctGG}. This diagram is the only one which contains the tadpole renormalization constants. In the following we explain the cancellation of the gauge and Goldstone boson contributions as well as the fermion contributions.  

The additional $\thdm$ contribution from this diagram is the difference between the $\thdm$ and SM results,  
\begin{align}\label{tadpoleGGresult}
\amu ^{\text{CT }(GG)} =& \frac{(\delta t _{GG} ^{\thdm} - \delta t _{GG} ^{\text{SM}})}{v}\frac{\alpha}{96 \pi \SWsq}\frac{m_\mu ^2}{\MW ^4}\,\left\{2 + \varepsilon\left(3 - 2 \LL(\MWsq)\right)\right\}, 
\end{align}
where 
\begin{align}
\LL(M ^2) \equiv \gamma_{E} - \ln(4\pi) + \ln(M ^2/\mu^2). 
\end{align}
The counterterm vertices of the $G^\pm$-$G ^\pm$ propagator for the SM and $\thdm$ are 
\begin{align}\label{tadpoleGG}
\delta t _{GG} ^{\text{SM}} =& \dt,\quad
\delta t _{GG} ^{\thdm} = \cos(\beta-\alpha)\,\dtH + \sin(\beta-\alpha)\,\dth.
\end{align}
Hence, from~\erefand{tadpolegGf}{tadpoleGG} we find that the tadpole renormalization constants with gauge/Goldstone bosons and fermion loops drop out from the result,~\eref{tadpoleGGresult}. Consequently no new $\thdm$ contributions are obtained from the fermion, gauge boson, and Goldstone boson loops. The additional $\thdm$ contribution of~\fref{fig:ctGG} and all other diagrams of this group arises from the physical Higgs boson loop contributions to the tadpole, field and mass renormalization constants.  
\item Counterterm with neutral Higgs bosons:\\
The second group of counterterm diagrams is shown in~\fref{fig:ctH}. These diagrams are dependent on Yukawa coupling and proportional to $\dzza$, which is the same in both the SM and the $\thdm$. The difference arises from the Yukawa coupling and the second neutral CP-even $\thdm$ Higgs boson. The fermionic loop contribution to $\dzza$ are zero, therefore the diagrams of this group do not contribute to the fermion contributions. 

The additional $\thdm$ contribution is obtained when the SM Higgs boson contribution is subtracted from the $\thdm$ results, $\amu^{{\text{CT}}(\MH)} + \amu^{{\text{CT}}(\Mh)} - \amu^{{\text{CT}}(\HSM)}$, where the explicit result of~\fref{fig:ctH} for an arbitrary scalar field ${\cal S}$ is 
\begin{align}
\amu^{\text{CT}({\cal S})} =& \,C_{\cal S}\,Y_l ^{\cal S}\, \frac{\alpha}{32 \pi} \frac{(1-4\,\SWsq)}{\CW ^3\,\SW ^3}\frac{\MMsq}{(M_{\cal S} ^2 - \MZsq)^2}\, \dzza \nonumber\\
&\times\left\{\MZsq - M_{\cal S}^2 + M_{\cal S}^2\,\ln(M_{\cal S}^2/\MZsq) \right.\nonumber\\
&\quad\,\,+ \frac{\varepsilon}{2}\,(3(\MZsq - M_{\cal S} ^2) + 3\,M_{\cal S}^{2}\,\LL(M_{\cal S} ^2) - M_{\cal S} ^2\,\LL (M_{\cal S} ^2)^2 \nonumber\\
&\left.\qquad\quad\,\,\,- (M_{\cal S}^{2} + 2\, \MZsq)\,\LL(\MZsq) +  M_{\cal S}^2\,\LL(\MZsq)^2)\right\}.   
\end{align}
${\cal S}$ can be any of the neutral Higgs bosons in the SM and the $\thdm$: $\HSM$, $\Mh$, $\MH$. The contribution of the CP-odd neutral Higgs $\MA$ is zero. 
The coefficient, $C_{\cal S}$, is derived from the gauge coupling. It
is $1$ for the SM Higgs $\HSM$, $\sin(\beta-\alpha)$ for $\Mh$ and
$\cos(\beta - \alpha)$ for $\MH$. $Y _l ^{\cal S}$ is derived from the
Yukawa coupling constant and listed in~\eref{yukawaaeps}. For the SM Higgs $\HSM$, $Y _l ^{\HSM} =1$.
\item Counterterm diagram with $G ^\pm$-$H ^\pm$ mixing:\\
The third group consists of the diagram of~\fref{fig:ctHG} which is proportional to the $G ^\pm$-$H ^\pm$ mixing counterterm vertex. This counterterm diagram does not appear in the SM. The explicit analytic result reads 
\begin{align}
\amu^{\text{CT }(GH)} =& \frac{\delta t_{HG}}{v}\frac{\alpha}{16 \pi \SWsq}\frac{m_\mu ^2}{\MW\,(\MWsq - \MMHpm ^2)^3}\,\zl
\nonumber\\
\times&\bigg[(\MMHpm ^4  - \MW ^4)\big(1 + \varepsilon\,\LL(\MWsq)\big)\nonumber\\
&-\MMHpm ^2 \MWsq\,\ln(\MMHpm ^2/\MWsq)\big(2 + \varepsilon\{3 -\LL(\MMHpm ^2) - \LL(\MW ^2)\}\big)\nonumber\\
&+ \frac{\varepsilon}{2}(\MMHpm ^2 - \MW ^2)(\MMHpm ^2 + 5\,\MW ^2) \bigg] 
\end{align}
where $\delta t_{HG} = \cos(\beta-\alpha)\,\dth - \sin(\beta-\alpha)\dtH$. $\zl$-dependency arises from the charged Higgs boson coupling to the muon in the Aligned $\thdm$.~\fref{fig:ctHG} is the only counterterm diagram which contributes to the fermionic two-loop result.  
\end{itemize}

\subsection{Bosonic loop contribution}
\label{subsec:bloop}
\subsubsection{\twobo and \threebo diagrams}
\label{sec:b1}

\begin{figure}[]
\centering
\begin{subfigure}[]{.23\textwidth}
\centering
\includegraphics[scale=.35]{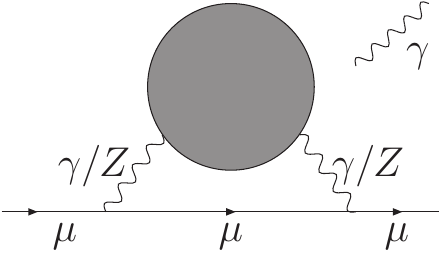}
\subcaption{}
\label{fig:THDM2gZ}
\end{subfigure}
\begin{subfigure}[]{.23\textwidth}
\centering
\includegraphics[scale=.35]{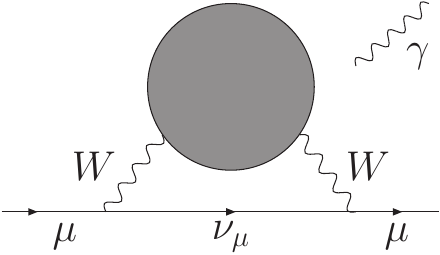}
\subcaption{}
\label{fig:THDM2WW}
\end{subfigure}
\begin{subfigure}[]{.23\textwidth}
\centering
\includegraphics[scale=.35]{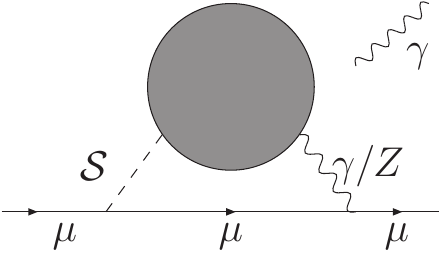}
\subcaption{}
\label{fig:THDMSZg}
\end{subfigure}
\begin{subfigure}[]{.23\textwidth}
\centering
\includegraphics[scale=.35]{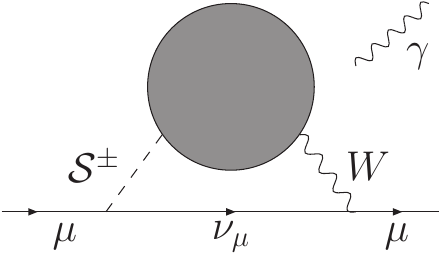}
\subcaption{}
\label{fig:THDMSpmW}
\end{subfigure}
\caption{Generic \twobo Feynman diagrams. The gray loops denote any bosonic loop.}
\label{fig:2boson}
\end{figure}

\begin{figure}[]
\centering
\begin{subfigure}[]{.23\textwidth}
\centering
\includegraphics[scale=.35]{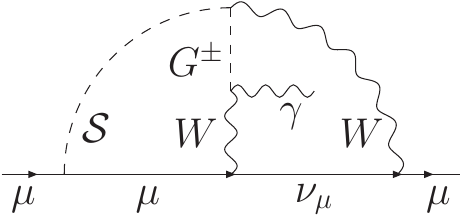}
\subcaption{}
\label{fig:THDMWG}
\end{subfigure}
\begin{subfigure}[]{.23\textwidth}
\centering
\includegraphics[scale=.35]{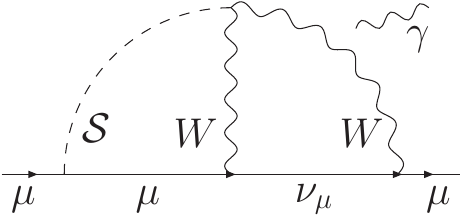}
\subcaption{}
\label{fig:THDMWW}
\end{subfigure}
\begin{subfigure}[]{.23\textwidth}
\centering
\includegraphics[scale=.35]{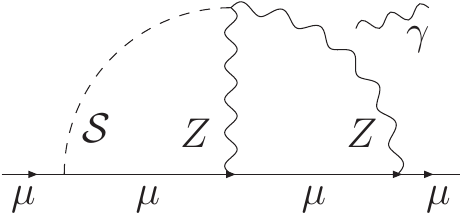}
\subcaption{}
\label{fig:THDMZZ1}
\end{subfigure}
\begin{subfigure}[]{.23\textwidth}
\centering
\includegraphics[scale=.35]{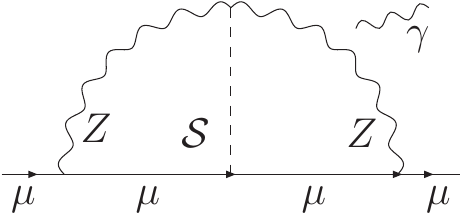}
\subcaption{}
\label{fig:THDMZZ2}
\end{subfigure}
\caption{\threebo Feynman diagrams are mediated either with $W$ or $Z$ bosons. They contain only the neutral physical CP-even Higgs bosons.}
\label{fig:3boson}
\end{figure}
We classify the bosonic two-loop diagrams according to the number of bosons coupling to the muon line. With this criterion it is possible to group the diagrams into \twobo and \threebo types. The \twobo type denotes all diagrams in which two internal bosons couple to the muon line. The generic diagrams of the \twobo diagrams are shown in~\fref{fig:2boson}. Gray circles in~\fref{fig:2boson} denote all possible bosonic loops. These \twobo diagrams contain the so-called Barr-Zee diagrams, which have already been computed and intensively discussed in the literature \cite{Krawczyk:2002df,ArhribBaek,ChenGeng,HSW03,Cheung:2009fc,Ilisie:2015tra}. The \twobo diagrams also contain self-energy type diagrams in which the external photon couples to the muon line. 

The \threebo diagrams have a more complicated structure and involve three internal bosons which couple to the muon line.~\fref{fig:3boson} shows all \threebo diagrams which contribute to the difference between the \thdm\ and the SM. In addition, diagrams with four bosons coupling to the muon line exist but do not contribute to the difference between the SM and $\thdm$ at the order of ${\cal O} (m _\mu ^2)$.

We especially computed the \threebo diagrams shown
in~\fref{fig:3boson} for the first
time.~\frefand{fig:THDMWG}{fig:THDMWW} are dependent on the $W$ boson
and~\frefand{fig:THDMZZ1}{fig:THDMZZ2} on the $Z$ boson. While
computing the diagrams in~\fref{fig:3boson}, we should pay attention
to two interactions. One is the muon Yukawa coupling to the neutral
scalar bosons, $\Mh$ or $\MH$, and the other is the Higgs-gauge
interaction of the two neutral Higgs bosons. The gauge interaction to
$\MH$ is suppressed by $\eta$. In the SM-limit the contribution from
this interaction becomes zero, whereas the gauge interaction to $\Mh$
recovers the SM value. The explicit result of~\frefand{fig:THDMWG}{fig:THDMWW} reads
\vspace{-0.08cm}
\newcommand{\ys}{y_{\cal S}} 
\begin{align}
\label{analyticW}
\amu ^{W, {\cal S}} =& C _{\cal S} Y _l ^{\cal S} \GlobalC \Big\{ \frac{3}{\varepsilon} - 6 \LL (\MWsq) - \frac{55}{2} + \frac{32}{\ys} - \frac{4 \pi ^2}{3}\Big(\frac{4 + 3 \ys}{\ys ^2}\Big) \nonumber\\
& - \Big( 35 + \frac{32}{\ys}\Big) \ln (\ys) + \Big( 6 + \frac{32}{\ys ^2} + \frac{24}{\ys} - 32 \ys \Big) \dilog (1-\ys) \nonumber\\
& + \Big(\frac{10 + 70 \ys - 32 \ys ^2 }{(\ys - 4)\ys}\Big) \TF(\sqrt{\ys},1,1),
\end{align}
where $S \in {\Mh, \MH}$, and $\ys \equiv \frac{M _{\cal S} ^2}{\MWsq}$. $\TF(x,y,z)$ is defined in Appendix~\ref{appx:1}. We have $C _\Mh = 1$, $Y _l ^{\Mh} = 1 + \eta \zl$ for $\Mh$ and $C _\MH = \eta$, $Y _l ^{\MH} = \eta - \zl$ for $\MH$ up to ${\cal O} (\eta)$. For the SM Higgs boson, $Y _l ^{\HSM} = 1$, and $C _{\HSM} = 1$. The divergent part of~\eref{analyticW} drops out in the final result of the difference of the SM and $\thdm$. Note that the result of~\fref{fig:THDMWW} alone is finite. In the off-SM scenario, $\eta \neq 0$, the result of~\fref{fig:3boson} for ${\cal S} = \Mh$ results in additional EW contributions. 

The additional $\thdm$ contribution from the diagrams of~\frefand{fig:THDMWG}{fig:THDMWW} is obtained when the SM Higgs boson result of~\eref{analyticW} is subtracted from the sum of the $\Mh$ and $\MH$ contributions, $\amu ^{W,\Mh} + \amu ^{W,\MH} - \amu ^{W, \HSM}$. After employing the known SM parameters we obtain the numerical result 
\begin{align}
\label{Wnum}
\amu ^{W,\MH} + \amu ^{W,\Mh} - \amu ^{W, \HSM} = (3 \cdots -4.6)\times 10^{-12}\, \eta \zl,  
\end{align}
for $50 < \MMH < 500\;{\text{GeV}}$ and $\MMh = \MHSM = 125\;{\text{GeV}}$. The maximum value of~\eref{Wnum} for a fixed $\eta \zl$ is $|-5.1 \eta \zl| \times 10 ^{-12}$ for $\MMH \sim 950\;{\text{GeV}}$.~\eref{Wnum} vanishes when $\MMH = \MMh$. 

For the case of $Z$ boson dependent non Barr-Zee diagrams,~\frefand{fig:THDMZZ1}{fig:THDMZZ2}, the analytic result for an arbitrary Higgs boson ${\cal S}$ reads 
\newcommand{\xs}{x _{\cal S}}
\begin{align}\label{analyticZ}
\amu ^{Z, {\cal S}} =& C _{\cal S} Y _l ^{\cal S} \GlobalC \Big( f_{a}(\xs) + \SWsq (1 - 2\SWsq) f_{b}(\xs)\Big),\\
f_a (\xs) =& \frac{3(4 - \xs)}{\xs} - \frac{\pi ^2}{2}\Big(\frac{4 + 3 \xs}{\xs ^2}\Big) -\frac{3(4 + \xs)}{\xs}\ln (\xs)\nonumber\\
& + \frac{12 + 9 \xs - 3 \xs ^3}{\xs ^2} \dilog (1 - \xs) + \frac{3(2 +\xs)}{\xs}\TF(\sqrt{\xs},1,1),\\
f_b (\xs) =& \frac{\pi ^2(8 + 6 \xs - 12 \xs ^4 + 3 \xs ^5)}{\xs ^2} + \frac{6(-8 + 2 \xs + 3 \xs ^2)}{\xs}\nonumber\\
& + \frac{12(4 + \xs + 3 \xs ^2)}{\xs} \ln (\xs) + 9(-4 + \xs) \xs ^2 \ln (\xs) ^2 \nonumber\\
& + \frac{12(-4 - 3 \xs + 4 \xs ^3 - 12 \xs ^4 + 3 \xs ^5)}{\xs ^2} \dilog ( 1 - \xs ) \nonumber\\
& + \frac{6( 4 + 2 \xs - 6 \xs ^2 + 3 \xs ^3)}{\xs} \TF(\sqrt{\xs},1,1),
\end{align}
and $\xs \equiv \frac{M_{\cal S} ^2}{\MZsq}$. $C _{\cal S}$ and $Y _l ^{\cal S}$ for $\Mh$ and $\MH$ are the same as in~\eref{analyticW}. 

Like~\eref{Wnum} the additional $\thdm$ contribution from the diagrams of~\frefand{fig:THDMZZ1}{fig:THDMZZ2} is obtained as
\begin{align}
\label{numZ}
\amu ^{Z, \MH} + \amu ^{Z, \Mh} - \amu ^{Z, \HSM} = (5.6 \cdots -5.6) \times 10 ^{-13} \eta \zl, 
\end{align}
for $50 < \MMH < 500\;{\text{GeV}}$ and $\MMh = \MHSM =
125\;\text{GeV}$. When $\MMH > 125\;{\text{GeV}}$,~\eref{numZ} becomes
negative. The W boson result is approximately a factor 10 larger than
the Z boson result; it can become significant for large values of $\zl$ and $\eta$.

\subsubsection{Analytic results}
\label{sec:b2} 

In this section we present the complete renormalized bosonic $\thdm$ contribution. The bosonic result is expanded with respect to the parameter $\eta$ introduced in~\Sref{sec:model}, and terms up to $\eta ^1$ are taken. In the SM-limit, $\eta\to 0$, the interactions of $\Mh$ to the gauge bosons or fermions become just those of the SM. 

For the discussion of the complete result we do not use the \twobo and
\threebo separation. Instead, we divide the renormalized bosonic
contribution into two parts. One part,  $\amu ^{\text{EW add.}}$, is
defined by the Feynman diagrams
containing only gauge/Goldstone/$\Mh$ bosons, i.e.~purely SM-like
diagrams.
The other part is defined by those diagrams which include at least one of the new $\thdm$ Higgs bosons, $\MH, \MA, \MHpm$. This part can be again divided into Yukawa-dependent and Yukawa-independent parts. Considering this classification we can write the bosonic contribution as 
\begin{align}
\label{amub}
\amub =& \amu ^{\text{EW add.}} + \az + \amu ^{\text{Yuk}},
\end{align}
where $\az$ denotes Yukawa-independent $\thdm$ Higgs contributing part and $\amu ^{\text{Yuk}}$ the Yukawa-dependent part. In the following we explain each of the contributions explicitly. 
\begin{itemize}
\item  $\amu ^{\text{EW add.}}$\\
We start with the computation of $\amu ^{\text{EW add.}}$. The additional
$\thdm$ EW contribution, $\amu ^{\text{EW add.}}$ is obtained by
subtracting the Feynman diagram result with SM physical Higgs boson
$\HSM$  from the $\thdm$ diagrams which include only $\Mh$. The
diagrams of~\fref{fig:THDMSZg} and~\fref{fig:3boson} with ${\cal
  S}=\Mh$ contribute to this difference at the order $\eta\zl$ due
to the different Yukawa couplings in the two models. 
The diagrams of \frefand{fig:THDM2gZ}{fig:THDM2WW} as well as \fref{fig:THDMSpmW} with charged Goldstone boson, 
${\cal S} ^{\pm} = G^\pm$ and with $\Mh$ in the gray loop also contribute to $\amu ^{\text{EW add.}}$, however only starting at the order $\eta^2$; hence we
neglect them.
The only counterterm diagram contributing to $\amu ^{\text{EW add.}}$ is the diagram of~\fref{fig:ctH} with $\Mh$.

After summing up the two-loop and the counterterm results and
employing the SM parameters we obtain finally the complete result
\begin{align}
\label{smlikenum}
\amu ^{\text{EW add.}} =& 2.3 \times 10^{-11} \,\eta\,\zl.
\end{align}
The sign of $\amu ^{\text{EW add.}}$ is dependent on $\eta\,\zl$. Even
though $\eta$ must be small, the appearance of $\zl$ can  enhance the
contribution of $\amu ^{\text{EW add.}}$. 
\item $\az$\\
Now we turn to  $\az$ in~\eref{amub}. 
\begin{figure}[]
\centering
\begin{subfigure}{0.45\textwidth}
       \centering
        \includegraphics[scale=0.45]{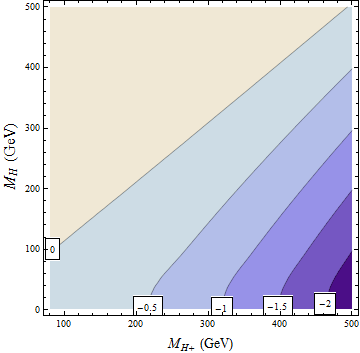}
\subcaption{$\MMA = 10\;\text{GeV}$}
\end{subfigure}
\begin{subfigure}{0.45\textwidth}
       \centering
        \includegraphics[scale=0.45]{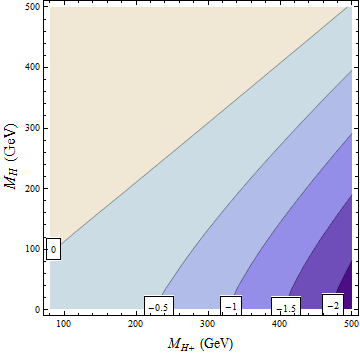}
\subcaption{$\MMA = 50\;\text{GeV}$}
\end{subfigure}
\\\vspace{.5cm}
\begin{subfigure}{0.45\textwidth}
       \centering
        \includegraphics[scale=0.45]{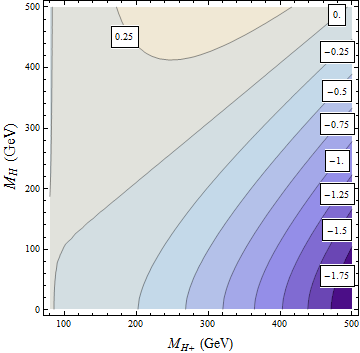}
\subcaption{$\MMA=100\;\text{GeV}$}
\end{subfigure}
\begin{subfigure}{0.45\textwidth}
       \centering
        \includegraphics[scale=0.45]{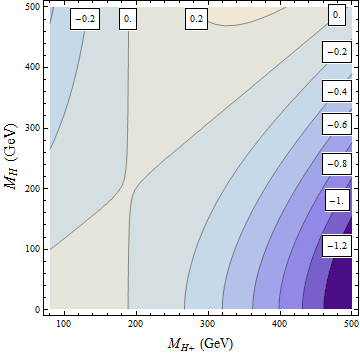}
\subcaption{$\MMA=200\;\text{GeV}$}
\end{subfigure}
 \caption{Plots of $\az$ for different values of  $\MMA=10,50,100,200\;\text{GeV}$. The results should be multiplied by a factor $10^{-10}$. The contour line value for fixed $\MMH$ and $\MMHpm$ decreases as $\MMA$ increases. As $\MMA$ becomes larger, $\az$ becomes more sensitive to the difference of the neutral and charged Higgs boson masses: compare the right-bottom areas of the plots. For a given $\MMA$ value $|\az|$ increases as $\MMHpm - \MMH$ becomes larger.}
\label{fig:a0}
\end{figure}
It comes from the Feynman diagrams without Yukawa couplings containing at least one of the new $\thdm$ physical Higgs bosons, $\MH, \MA, \MHpm$. 
The Feynman diagrams of~\frefand{fig:THDM2gZ}{fig:THDM2WW} with $\MH / \MA / \MHpm$ in the gray loops contribute to $\az$. 

$\az$ is dependent on parameters, $\MMH$, $\MMA$, and $\MMHpm$, but
not on $\tb$ and $\lFA$.  It also does not gain terms linearly dependent on the parameter $\eta$. We should stress that $\az$ is the only part dependent on $\MMA$ in the bosonic contributions. 
The explicit analytic result is found in Appendix~\ref{appx:1}. 

\fref{fig:a0} shows the change of $\az$ for different $\MMA$
values. For $\MMA < 100\;{\text{GeV}}$ and $\MMH,\MMHpm >
100\;{\text{GeV}}$, $\az$ has the same sign of the difference between
$\MMH$ and $\MMHpm$. $\az$ depends mainly on the difference between
the masses of the three Higgs bosons. In the largest part of the
parameter space in the Figure,  $\az$ is negative and amounts up to   $ -2\,\times\,10^{-10}$.  
\item $\amu ^{\text{Yuk}}$\\
The terms contained in $\amu ^{\text{Yuk}}$ in~\eref{amub} are from those diagrams with Yukawa contributions and the corresponding counterterms. 
\begin{figure}[]
\centering
\begin{subfigure}{0.4\textwidth}
       \centering
        \includegraphics[scale=0.45]{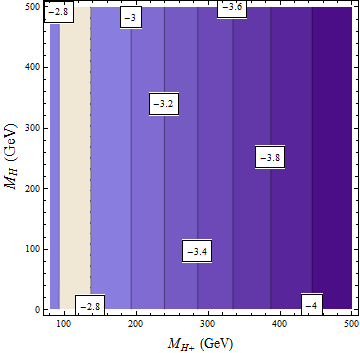}
\subcaption{$\aezezeze$}
\end{subfigure}\qquad
\begin{subfigure}{0.4\textwidth}
       \centering
        \includegraphics[scale=0.45]{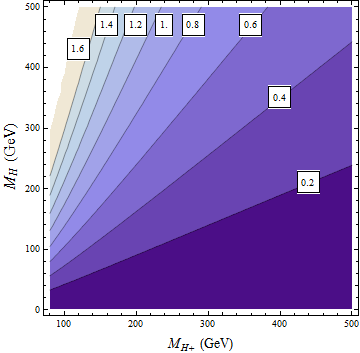}
\subcaption{$\aezezez$}
\end{subfigure}\\\vspace{.5cm}
\begin{subfigure}{0.4\textwidth}
       \centering
        \includegraphics[scale=0.45]{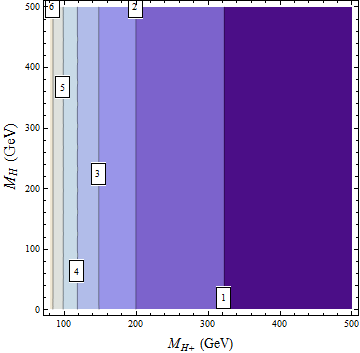}
\subcaption{$\aezelze$}
\end{subfigure}\qquad
\begin{subfigure}{0.4\textwidth}
       \centering
        \includegraphics[scale=0.45]{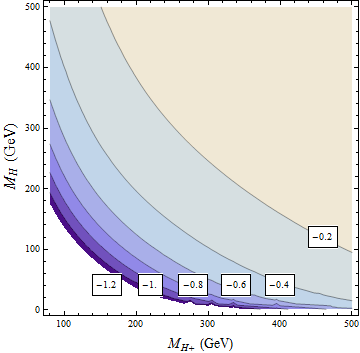}
\subcaption{$\aezelz$}
\end{subfigure}
\caption{Plots of the $\eta ^0$-order coefficients, $a _{i,j} ^{0}$ in~\eref{ayuk2}. The values of contour lines should be multiplied by $10 ^{-12}$. The values of these plots are not suppressed by $\eta$. $\aezezeze$ and $\aezelze$ are only dependent on $\MMHpm$. On the given parameter space $\aezezeze$ is negative whereas $\aezelze$ positive. As $\MMHpm$ increases, $|\aezezeze|$ increases, but $|\aezelze|$ decreases. Although the magnitudes of $|\aezezez|$ and $|\aezelz|$ are smaller than those of $\aezezeze$ and $\aezelze$, they are enhanced by large $\tb$ and $\zl$.}
\label{fig:aii0}
\end{figure}
Among the \twobo diagrams the Feynman diagrams
of~\frefand{fig:THDMSZg}{fig:THDMSpmW} with ${\cal S} = \MH$ and
${\cal S} ^{\pm} = \MHpm$ contribute to $\amu ^{\text{Yuk}}$. The diagrams
of~\fref{fig:THDMSZg} also contribute if ${\cal S} = \Mh$ and the
gray loop contains at least one of the new physical $\thdm$ Higgs
bosons. These diagrams include triple or quartic scalar boson
couplings. The \threebo diagrams of~\fref{fig:3boson} with $\MH$
contribute to $\amu ^{\text{Yuk}}$, too. 

Clearly, all diagrams with $\MH$ or $\MHpm$ and gauge bosons are
suppressed by $\eta$ but enhanced by $\zl$. The diagrams without gauge
bosons involve triple Higgs couplings and are of particular interest. 
A closer look at the triple Higgs coupling constants helps to analyze the $\tb$-dependency. The triple Higgs couplings constants in the $\thdm$ are  
\begin{align}
\label{triph}
g_{\Mh,H^\pm,H^\mp} \propto& \left\{\vac\,\left( \lFA - \frac{\MMh^2}{\vac^2} - 2\frac{\MMHpm^2}{\vac^2}\right)\right.\nonumber\\
& \left. \,\,+\, \eta\,\left(\tb - \frac{1}{\tb}\right)\frac{\vac}{2}\left(2\,\frac{\MMh^2}{\vac^2} - \lFA\right)\right\}\\
\label{tripH}
g_{\MH,H^\pm,H^\mp} \propto& \left\{
\left(\tb - \frac{1}{\tb}\right)\frac{\vac}{2}\left( \lFA -  2\,\frac{\MMH^2}{\vac^2} \right)\right.\nonumber\\
&\left.\,\, + \eta\,v\left( \lFA - \frac{\MMH^2}{\vac^2} - 2\frac{\MMHpm^2}{\vac^2}\right)\right\}
\end{align}
The triple Higgs coupling constants show that the $\tb$-dependency
comes only in the form of $(\tb - \frac{1}{\tb})$, which leads to a
large $\tb$-enhancement.
In the actual Feynman diagrams with triple Higgs couplings, the
coupling \eref{triph} appears multiplied 
with $Y_l^{\Mh}$, and the coupling \eref{tripH} is multiplied with $Y _l ^{\MH}$ and $Y _l ^{\MA}$. 
This allows to read off which
combinations of the parameters $\zl$, $\eta$, $\lFA$ appear in these
diagrams. With these considerations, we can rewrite $\amu ^{\text{Yuk}}$ as 
\begin{align}
\label{ayuk2}
\amu ^{\text{Yuk}} =& \aezezeze + \aezezez\,\coefftb\,\zl + \aezelze\,\lFA + \aezelz\,\coefftb\,\lFA\,\zl + \nonumber\\
+& \left(\aezeze\,\coefftb + \aezez\,\zl + \aelze\,\coefftb\,\lFA + \aelz\,\lFA\,\zl\right)\,\eta. 
\end{align}
The notation is such that the terms with
superscript $^0$ are independent of $\eta$, the terms with superscript
$^1$ are linear in $\eta$. The subscript $_z$ denotes terms enhanced
by $\zl$, the subscript $_5$ denotes terms $\propto\lFA$.
All terms here arise from diagrams with triple Higgs couplings except
the $\aezez$ term. 
The results of the \threebo diagrams~\erefand{analyticW}{analyticZ} for $\MH$ are included
 in $\aezez$. 
The parameter dependence of each coefficient $a _{i,j} ^{k}$ is rather
simple. $\aezezeze$ and $\aezelze$ are dependent only on $\MMHpm$ and
the rest dependent only on $\MMH$ and $\MMHpm$. In Appendix~\ref{appx:1} we
present the explicit expression of the coefficients $a_{i,j}^{k}$ as
well as $\az$, and in Appendix~\ref{appx:2} we show that there is no
dependence on $\MMA$. 

The plots in~\fref{fig:aii0} show the complete mass dependence of the
coefficients $a ^{0} _{i,j}$. $\aezezeze$ and $\aezelze$ arise from
the Feynman diagrams containing the muon Yukawa interaction to $\Mh$
and the $\eta$-independent part of \eref{triph}, therefore are
dependent only on $\MMHpm$ and neither enhanced by $\tb$ nor by
$\zl$. In contrast, $\aezezez$ and $\aezelz$ arise from diagrams
involving the triple Higgs coupling \eref{tripH} and appear enhanced
by large $\tb$ and $\zl$ in \eref{ayuk2}. 

The plots in~\fref{fig:aii1} show the change of $\aezeze$, $\aezez$,
$\aelze$ and $\aelz$, the $\eta$-suppressed terms. 
The coefficient $\aezez$, which gets contributions from a larger
class of diagrams, can be numerically larger than the other coefficients.
\end{itemize}
\begin{figure}[]
\centering
\begin{subfigure}{0.4\textwidth}
       \centering
        \includegraphics[scale=0.45]{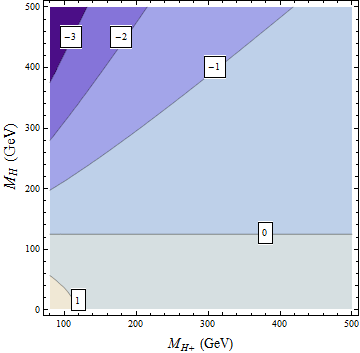}
\subcaption{$\aezeze$}
\label{fig:aezeze}
\end{subfigure}\qquad
\begin{subfigure}{0.4\textwidth}
       \centering
        \includegraphics[scale=0.45]{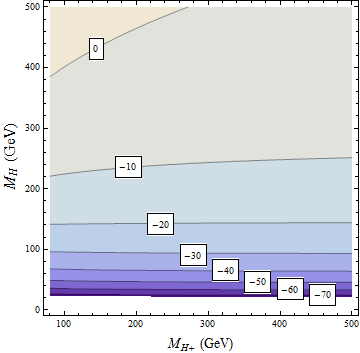}
\subcaption{$\aezez$}
\label{fig:aezez}
\end{subfigure}\\\vspace{.5cm}
\begin{subfigure}{0.4\textwidth}
       \centering
        \includegraphics[scale=0.45]{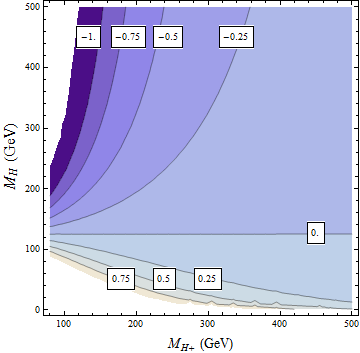}
\subcaption{$\aelze$}
\label{fig:zelze}
\end{subfigure}\qquad
\begin{subfigure}{0.4\textwidth}
       \centering
        \includegraphics[scale=0.45]{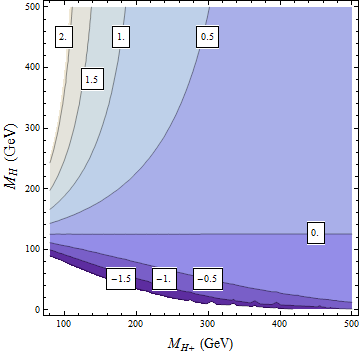}
\subcaption{$\aelz$}
\label{fig:aelz}
\end{subfigure}
\caption{Plots of $\aezeze$, $\aezez$, $\aelze$ and $\aelz$. The results must be multiplied by a factor $10^{-12}$. Terms with these coefficients in~\eref{ayuk2} are suppressed by $\eta$. $\aezeze$ and $\aelze$ are enhanced by large $\tb$ whereas $\aezez$ and $\aelz$ by $\zl$.}
\label{fig:aii1}
\end{figure}

\subsection{Fermionic loop contribution}
\label{subsec:floop}
In this section we present the fermionic loop contribution to $\amu$.
Due to the higher order muon mass suppression (considering terms up to $m_{\mu}^2$ order), all diagrams contain only one scalar boson, which interacts with the incoming/outgoing muon and the fermion in the inner loop. 
Thus, the result is always proportional to the product of two Yukawa couplings $Y^{\cal S}_{l}Y^{\cal S}_{f}$. 

\begin{figure}[t]
\centering
\begin{subfigure}[]{.3\textwidth}
\centering
\includegraphics[scale=.5]{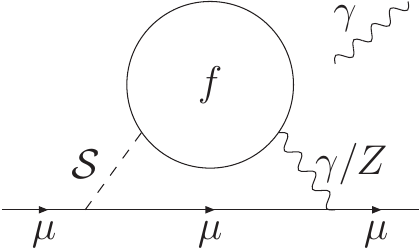}
\subcaption{}
\label{fig:fn1}
\end{subfigure}
\begin{subfigure}[]{.3\textwidth}
\centering
\includegraphics[scale=.5]{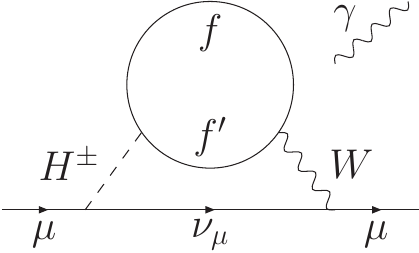}
\subcaption{}
\label{fig:fc1}
\end{subfigure}
\begin{subfigure}[]{.3\textwidth}
\centering
\includegraphics[scale=.5]{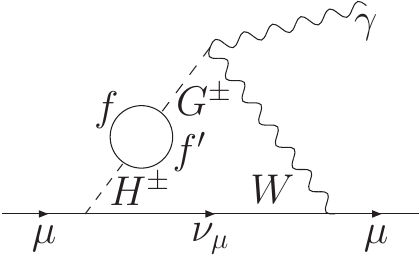}
\subcaption{}
\label{fig:fc2}
\end{subfigure}
 \caption{(a)Generic two-loop diagram with fermion loops and neutral Higgs bosons. The photon can couple with any charged particle inside. When the photon couples with the fermion loop, we obtain Barr-Zee diagrams. (b) Generic fermionic two-loop diagram with charged Higgs bosons. Barr-Zee diagrams are diagrams where the photon generates an effective photon-vector-scalar interaction. (c) $\gamma\,W\,G^\pm$ vertex diagram with charged Higgs bosons.}
\label{fig:floop}
\end{figure}

The fermionic two-loop Feynman diagrams contain either neutral or charged Higgs bosons. \fref{fig:fn1} shows the generic diagrams for neutral Higgs bosons while \fref{fig:fc1} is related to charged bosons.
When the external photon couples with the muon line we obtain self-energy type diagrams, and the sum of these vanishes. The remaining diagrams are Barr-Zee diagrams.

Our result for neutral Higgs bosons is coincident with previous analysis\footnote{We report a minus sign difference to the result presented in \cite{Chang:2000ii} regarding the Z boson contribution.}~\cite{Chang:2000ii,Cheung:2001hz,Ilisie:2015tra}, and the explicit form is
\begin{align}
\amu^{\text{F, N}}&= \sum_{{\cal S}=\{\Mh,\MH,\MA\}}\sum_{f=\{u,d,l\}}\left[f^{\cal S}_{f}(M_{\cal S},m_{f})\right]Y^{\cal S}_{f}Y^{\cal S}_{l}\nonumber\\
&\equiv \sum_{{\cal S}=\{\Mh,\MH,\MA\}}\sum_{f=\{u,d,l\}}\left[f^{\gamma}_{\cal S}(M_{\cal S},m_{f})+f^{Z}_{\cal S}(M_{\cal S},m_{f})\right]Y^{\cal S}_{f}Y^{\cal S}_{l},
\label{eq:floopn}
\end{align}
\noindent
where
\begin{align}
f^{\gamma}_{\cal S}(M_{\cal S},m_{f})=&\frac{\alpha^{2}m^{2}_{\mu}}{4\pi^{2}\MW^{2}\SWsq}\left(Q_{f}^{2}N_{c}^{f}\right)\left(\frac{m_{f}^{2}}{M_{\cal S}^{2}}\right)\mathcal{F}_{\cal S}(M_{\cal S},m_{f}),\\
f^{Z}_{\cal S}(M_{\cal S},m_{f})=&\frac{\alpha^{2}m^{2}_{\mu}}{4\pi^{2}\MW^{2}\SWsq}
\left(-\frac{N_{c}^{f}Q_{f}g_{v}^{l}g_{v}^{f}}{\SWsq\CWsq}\right)\nonumber\\
&\times\frac{m^{2}_{f}}{(M_{\cal S}^{2}-\MZ^{2})}\left[\mathcal{F}_{\cal S}(M_{\cal S},m_{f})-\mathcal{F}_{\cal S}(\MZ,m_{f})\right].
\end{align}
\noindent
For ${\cal S}=\{\Mh,\MH\}$ we have
\begin{align}
\mathcal{F}_{\cal S}(M_{\cal S},m_{f})&=-2+\ln\left(\frac{M_{\cal S}^{2}}{m_{f}^{2}}\right)-\left(\frac{M_{\cal S}^{2}-2m_{f}^{2}}{M_{\cal S}^{2}}\right)\,\frac{\TF(M_{\cal S},m_{f},m_{f})}{M_{\cal S}^{2}-4m_{f}^{2}},
\end{align}
and for ${\cal S}=\MA$
\begin{align}
\mathcal{F}_{\cal S}(M_{\cal S},m_{f})&=\frac{\TF(M_{\cal S},m_{f},m_{f})}{M_{\cal S}^{2}-4m_{f}^{2}}.
\end{align}
\noindent
A sum over all types of fermions is implicit. 
$Q_{f}$ denotes the charge of the respective fermion $f$, and $N_{c}^{f}$ the color factor. We also define $g_{v}^{f} \equiv \frac{T_{3}}{2}-Q_{f}\SWsq$, and $\TF(M_{\cal S},m_{f},m_{f})$ is defined in Appendix~\ref{appx:1}. 
Both $\gamma$ and $Z$ bosons contribute to the fermionic loop result with neutral Higgs bosons. However, the result from the $Z$ boson is suppressed by factor $g_v^{f}$, which is $-1/4 + \SWsq \sim -0.02$ for leptons, compared to the result from the diagrams with photon. Hence the $Z$ contributions are always smaller than those of the photon. 

\begin{figure}[]
\centering
\begin{subfigure}{0.4\textwidth}
  \centering
        \includegraphics[scale=0.45]{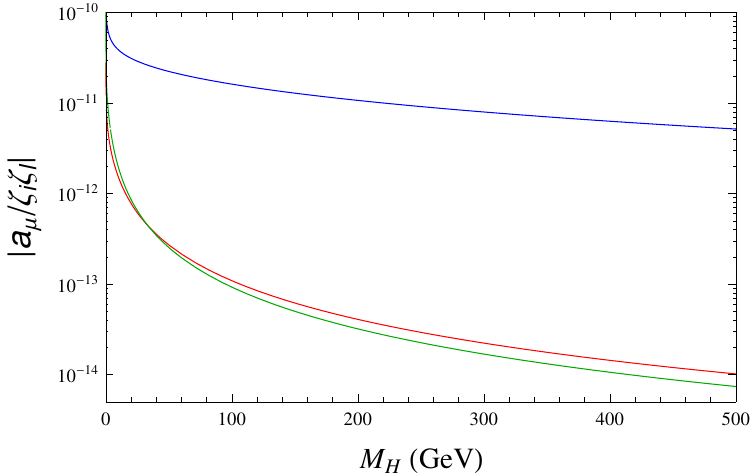}
\subcaption{}
\label{fig:ferma}
\end{subfigure}\qquad
\begin{subfigure}{0.4\textwidth}
  \centering
        \includegraphics[scale=0.45]{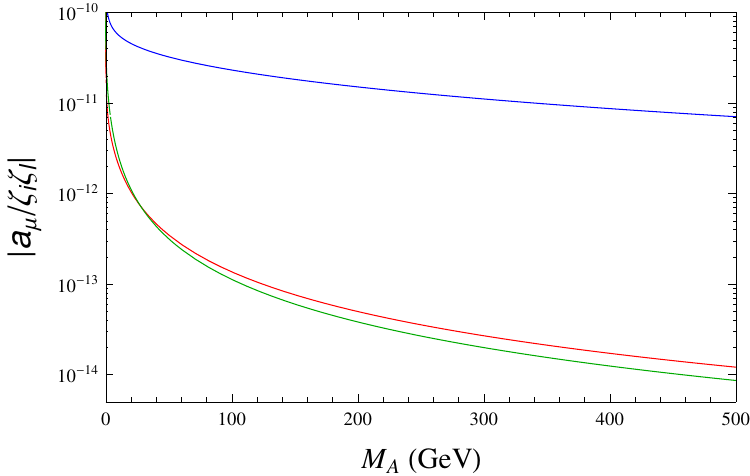}
\subcaption{}
\label{fig:fermb}
\end{subfigure}\\\vspace{.5cm}
\begin{subfigure}{0.4\textwidth}
  \centering
        \includegraphics[scale=0.45]{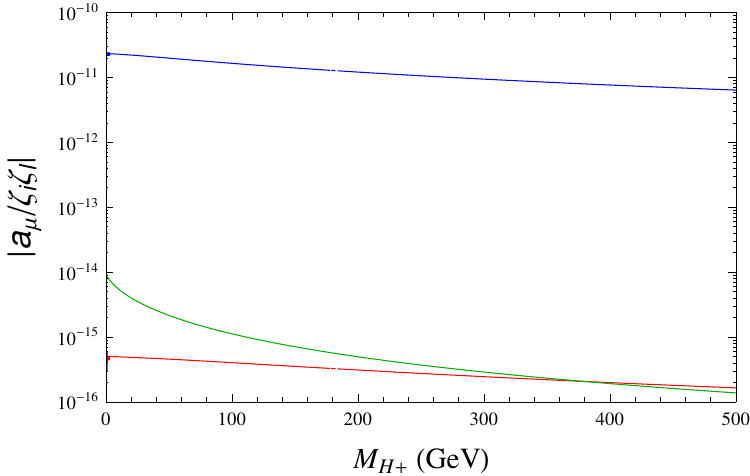}
\subcaption{}
\label{fig:fermc}
\end{subfigure}\qquad
\begin{subfigure}{0.4\textwidth}
  \centering
        \includegraphics[scale=0.45]{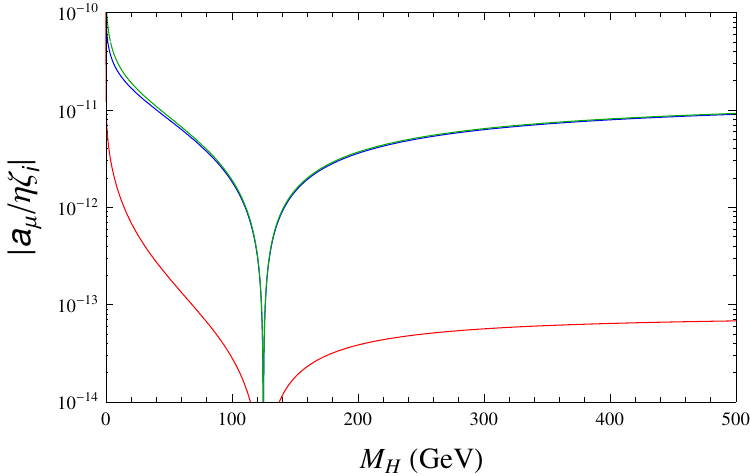}
\subcaption{}
\label{fig:fermd}
\end{subfigure}
 \caption{Fermionic contributions to $\amu$. The blue/red/green line refers to diagrams proportional to $\zu$/$\zd$/$\zl$, respectively. The first three graphs (a)--(c) show the contributions with $\MH$, $\MA$, $\MHpm$ at order $\eta^{0}$. The plot (d) shows the corrections at $\eta$ order from diagrams with CP-even bosons.}   
\label{fig:floopplots}
\end{figure}

Now we turn to the fermionic two-loop contributions with charged Higgs bosons. \frefand{fig:fc1}{fig:fc2} show the corresponding Feynman diagrams. Especially the result of \fref{fig:fc2} is divergent, the corresponding counterterm diagram is shown in \fref{fig:ctHG}. The renormalized two-loop result is obtained by summing up the two-loop and the counterterm diagrams. 
These diagrams were computed in the context of SUSY models long ago \cite {ArhribBaek,ChenGeng} in which case a type II structure for the Yukawas needed to be assumed. In the case of a general model (Aligned Model, for instance) the analysis was only recently performed \cite{Ilisie:2015tra}. 
We also recover the analytic result presented in \cite{ArhribBaek,ChenGeng,Ilisie:2015tra}, explicitly
\begin{align}\label{eq:floopc}
\amu^{\text{F,C}}=\sum_{f=\{u,d,l\}}f^{\MHpm}_{f}(\MMHpm,M_{f})Y^{\MA}_{f}Y^{\MA}_{l},
\end{align}
\noindent
where $M_{f}$ corresponds to pairs of fermions masses as $M_{u}=\{(m_{u},m_{d}),(m_{c},m_{s}),(m_{t},m_{b})\}$, $M_{d} = M_{u}$, $M_{l}=\{(m_{e},0),(m_{\mu},0),(m_{\tau},0)\}$, and \eref{eq:floopc} contains an implicit sum over pairs. We neglect neutrino masses and generation mixing. We also introduced the definitions below in \eref{eq:floopc}
\begin{align}
&f^{\MHpm}_{f}(M_{\MHpm},M_{f})=\frac{\alpha^{2}m^{2}_{\mu}}{32\pi^{2}\MW^{2}\SW^{4}}\frac{N^{f}_{c}m_{f}^{2}}{(M_{\MHpm}^{2}-\MW^{2})}\left[\mathcal{F}^{\MHpm}_{f}(M_{f})-(M_{\MHpm} \rightarrow \MW)\right],\\
&x_{f}\equiv\frac{m^{2}_{f}}{M_{\MHpm}^{2}}, \quad  y\equiv\left(x_{u}-x_{d}\right)^2-2\left(x_{u}+x_{d}\right)+1, \quad s\equiv \frac{(Q_{u}+Q_{d})}{4} \nonumber\\
&c\equiv\left[\left(x_{u}-x_{d}\right)^2-Q_{u}x_{u}+Q_{d}x_{d}\right], \quad \bar{c}\equiv\left[\left(x_{u}-Q_{u}\right)x_{u}-\left(x_{d}+Q_{d}\right)x_{d}\right], \nonumber\\ 
&\mathcal{F}^{\MHpm}_{l}(M_{l})=x_{l}+x_{l}\left(x_{l}-1\right)\left[\mbox{Li}_{2}(1-1/x_{l})-\frac{\pi^{2}}{6}\right]+\left(x_{l}-\frac{1}{2}\right)\ln(x_{l}),\\
&\mathcal{F}^{\MHpm}_{d}(M_{d})=-(x_{u}-x_{d})+\left[\frac{\bar{c}}{y}-c\left(\frac{x_{u}-x_{d}}{y}\right)\right]\TF(x_{d}^{1/2},x_{u}^{1/2},1)\nonumber\\
&\quad\quad\quad\quad\quad\quad+c\left[\mbox{Li}_{2}\left(1-\frac{x_{d}}{x_{u}}\right)-\frac{1}{2}\ln(x_{u})\ln\left(\frac{x_{d}}{x_{u}}\right)\right]\nonumber\\
&\quad\quad\quad\quad\quad\quad+\left(s+x_{d}\right)\ln(x_{d})+\left(s-x_{u}\right)\ln(x_{u}),\\
&\mathcal{F}^{\MHpm}_{u}(M_{u})=\mathcal{F}^{\MHpm}_{d}(x_{u},x_{d})\left(Q_{u}\rightarrow 2 +  Q_{u}, Q_{d} \rightarrow 2 + Q_{d}\right)\nonumber\\
&\quad\quad\quad\quad\quad\quad-\frac{4}{3}\left(\frac{x_{u}-x_{d}-1}{y}\right)\TF(x_{d}^{1/2},x_{u}^{1/2},1)\nonumber\\
&\quad\quad\quad\quad\quad\quad-\frac{1}{3}\left[\ln^2(x_{d})-\ln^2(x_{u})\right].
\end{align}
Summing the results of \erefand{eq:floopn}{eq:floopc} and subtracting the corresponding SM-Higgs contribution gives the full renormalized two-loop $\thdm$ fermionic contribution
\begin{align}
\amuf= \sum_{f=\{u,d,l\}} \Bigg[ \sum_{{\cal S}=\{\Mh,\MH,\MA\}} f_{f}^{\cal S}(M_{\cal S},m_{f})Y^{\cal S}_{f}Y^{\cal S}_{l} &+ f^{\MHpm}_{f}(M_{\MHpm},M_{f})Y^{\MA}_{f}Y^{\MA}_{l}\nonumber\\&-f_{f}^{\HSM}(\MHSM,m_{f})\Bigg]. 
\label{eq:floopsum1}
\end{align}
After applying the Aligned $\thdm$ Yukawa coupling constants in \eref{yukawaaeps} we can rewrite \eref{eq:floopsum1} with $\zeta_{f}$, and the result reads 
\begin{align}
\amuf=&\sum_{f=\{u,d,l\}}\left[\sum_{{\cal S}=\{\MH,A\}}\Theta^{\cal S}_{f}f_{f}^{\cal S}(M_{\cal S},m_{f})\zeta_{f}\zeta_{l} + \Theta^{\MA}_{f} f^{\MHpm}_{f}(M_{\MHpm},M_{f})\zeta_{f}\zeta_{l}\right]\nonumber\\
 &+\sum_{f=\{u,d,l\}} \left[\eta \left(f_{f}^{\Mh}(M_{\Mh},m_{f})-f_{f}^{\MH}(M_{\MH},m_{f})\right)\left(\zeta_{f}+\zeta_{l}\right)\right],
\label{eq:floopsum2}
\end{align}
where $\Theta^{\MA} _{u} = -1$, otherwise $\Theta^{\cal S} _{f} = 1$.  
Each function $f_f ^i(M_{i})$ in \eref{eq:floopsum2} is dependent on only one mass parameter, $M_{\cal S}$, and this enables us to analyze the individual Higgs boson contributions to the fermionic loop contribution in \fref{fig:floopplots}.
The first line of \eref{eq:floopsum2} contains terms bilinear in the $\zeta_{f}$, and they are shown in the first three plots of \fref{fig:floopplots}. The terms in the second line are proportional to $\zeta_{f}\eta$ and are illustrated in the fourth plot, \fref{fig:fermd}.

In all cases the contribution from the top loop (blue line) is significantly larger, as expected by the factor $m_{t}^{2}/(M_{\cal S}^{2}-M_{B}^{2})$ in the analytic formulas ($M_{B}$ is the mass of the internal gauge boson involved). However, as discussed in \Sref{sec:ewconst}, $\zu$ is constrained to be at most $\zu\simeq 1$, meaning that the tau loop, enhanced by $\zl^2$, plays the decisive role. Another characteristic shared by~\frefs{fig:ferma}{fig:fermc} is that they all decrease with the mass of the scalars.~\fref{fig:fermd} shows the contribution proportional to $\eta$ which comes from diagrams involving CP-even scalar bosons. As presented in \eref{eq:floopsum2} there is a difference between the $\Mh$ and $\MH$ results, explaining why the $\eta^{1}$ contribution vanishes as $\MMh$ approaches $\MMH$. For all plots, we have rescaled $\amu$ to the aligned parameters. Finally, in all graphs the contributions can be both positive or negative. The signs depend on the alignment parameters and can be read off from Table \ref{tab:signs}. 
\begin{table}[h]
\parbox{0.45\textwidth}{
\centering
\begin{tabular}{|c|c|c|c|}
\hline
                                  & $ f^{\MH}$              & $f^{\MA}$               & $f^{\MHpm}$           \\ \hline        
$\zeta_{u}\zeta_{l}$ & $-$ & $-$    & $-$  \\ \hline
$\zeta_{d}\zeta_{l}$ & $-$ & $+$   & $-$ \\ \hline
$\zeta_{l}^2$ & $-$ &  $+$ & $+$  \\ \hline
\end{tabular}
}
\hfill
\parbox{0.45\textwidth}{
\centering
\begin{tabular}{|c|c|c|}
\hline
                                  & $ m_{\MH} < m_{\Mh} $             & $m_{\Mh} < m_{\MH} $            \\ \hline        
$\eta\zeta_{u}$ & $+$ & $-$   \\ \hline
$\eta\zeta_{d}$ & $+$ & $-$  \\ \hline
$\eta\zeta_{l}$ & $+$ &  $-$   \\ \hline
\end{tabular}
}
\caption{Relation between signs of the aligned parameters and the functions depicted in \fref{fig:floopplots}.}
\label{tab:signs}
\end{table}

\section{Numerical Analysis}
\label{sec:numanaly}
In this section we present the numerical analysis of our result. Our aim is to study how large the bosonic contribution, fully computed for the first time, can be. We will show that there are regions of the parameter space in which $\amub$ amounts to $(2 \cdots 4)\times 10^{-10}$. Although always smaller than the fermionic contribution, it proves to be relevant for a precise determination of the $\thdm$ contribution to the muon anomalous magnetic moment. We also analyze the impact of deviations from the SM-limit by studying different values for the expansion parameter $\eta$.

For the analysis we choose physical free input parameters, the masses of the different scalars ($\MMH$, $\MMA$, $\MMHpm$), the alignment parameters ($\zeta_{l,u,d}$), $\tb$, the expansion parameter $\eta$, and $\lFA$. As presented in~\Sref{sec:model}, the last parameter can be expressed in terms of $\lambda_{1}$, which is directly constrained by stability and perturbativity. Therefore, for the numerical analysis, it will be useful to replace the parameter $\lFA$ with $\lambda_{1}$. As discussed in~\Sref{subsec:const} we adopt the following allowed range for the parameters, 
\begin{align}
& 125<\MMH<500\;\text{GeV},\quad \MMA<500\;\text{GeV},\quad 80<\MMHpm<500\;\text{GeV}, \nonumber\\
& 0 < |\zeta_{u}|<1.2,\qquad\quad 0 < |\zeta_{d}|<50,\quad\, 0 < |\zeta_{l}|<100, \nonumber\\
& 1 < \tb < 100,\qquad\quad\,\, 0 < |\eta| <0.1,\quad\, 0 < \lambda_{1} < 4 \pi.
\label{values}
\end{align}
\begin{figure}[]
\centering
\begin{subfigure}{0.45\textwidth}
       \centering
        \includegraphics[scale=0.12]{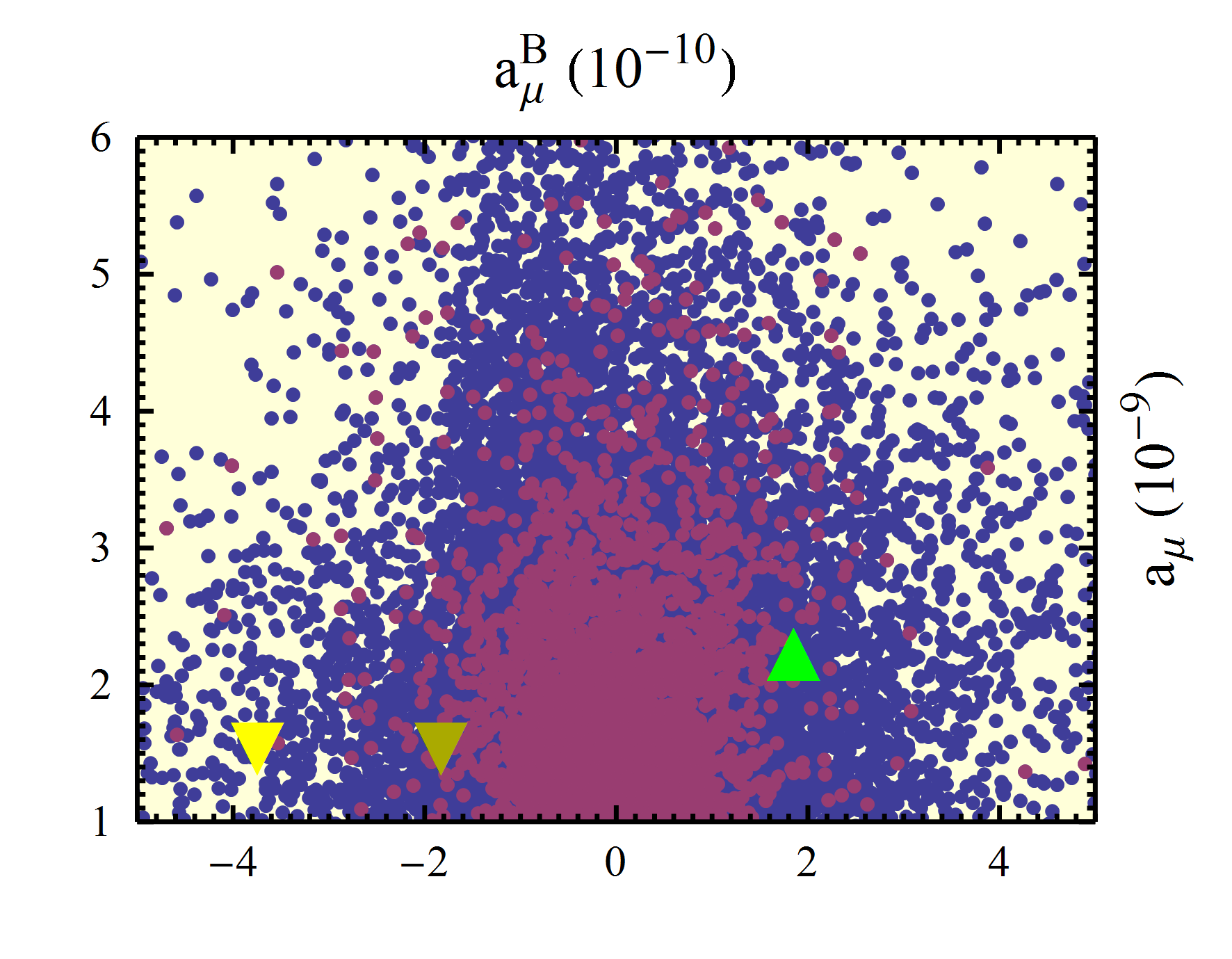}
\subcaption{$\eta = 0$}
\label{scat:0}
\end{subfigure}\qquad\quad
\begin{subfigure}{0.45\textwidth}
       \centering
        \includegraphics[scale=0.12]{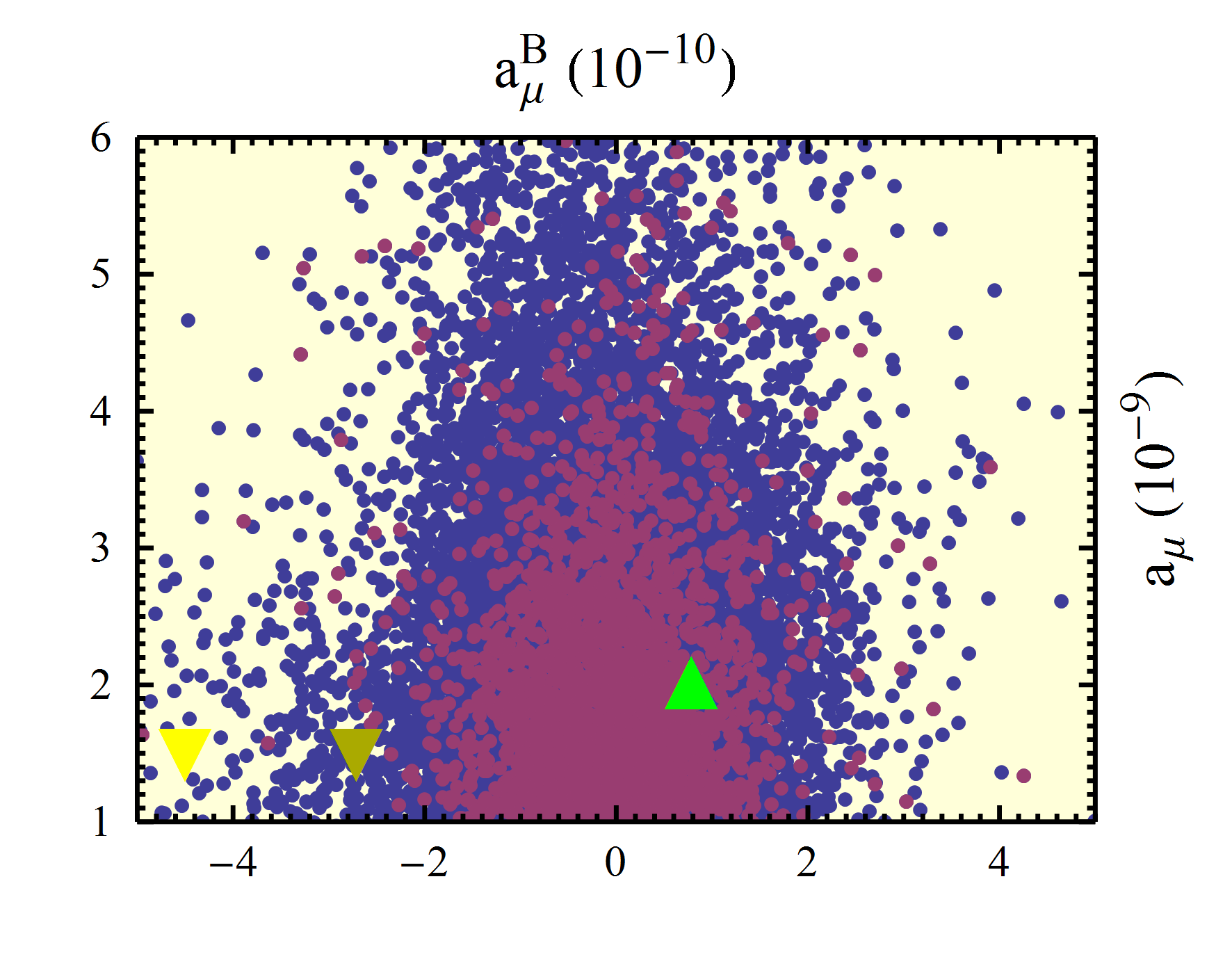}
\subcaption{$\eta = 0.1$}
\label{scat:1p}
\end{subfigure}\\\vspace{.5cm}
\centering
\begin{subfigure}{0.45\textwidth}
       \centering
        \includegraphics[scale=0.12]{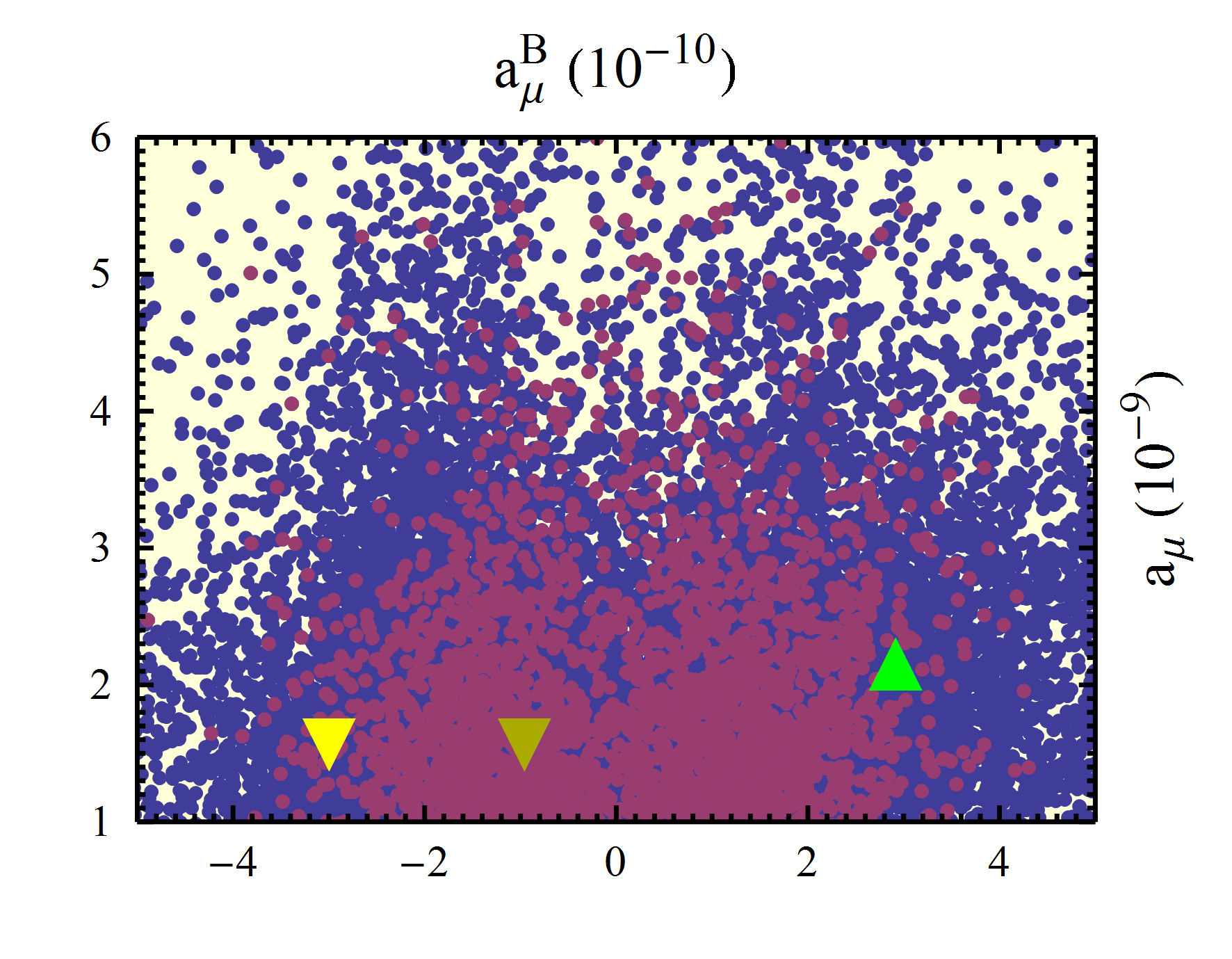}
\subcaption{$\eta=-0.1$}
\label{scat:1m}
\end{subfigure}
 \caption{Scatter plots showing possible values for $\amu$ and $\amub$ evaluated at different $\eta$ values. Blue dots represent points in the general allowed parameter space~\eref{values} while red dots represent the remaining ones after the constraints are applied. Green and yellow triangles are representative points discussed in the text.}
\label{scat}
\end{figure}
Since we want to study the impact of the SM-limit deviation to $\amu$,
hereafter we will choose specific values for $\eta$ 
and compare how the results differ.

We perform a scan over the above region, computing  for each point the
value of the full $\amu$ as well the contribution only due to
two-loop bosonic Feynman diagrams. Our results of the full scan are
depicted as blue points in the plots of~\fref{scat} (for the three values $\eta=0$, $\eta = 0.1$, $\eta=-0.1$). We then apply the further experimental/theoretical constraints discussed in~\Sref{subsec:const}. The survival sample is depicted as red points in~\fref{scat}. 

As can be readily seen, although the values for the full $\amu$ can be
large, the contribution from $\amub$ can amount to $(2 \cdots 4)\times 10^{-10}$. One can also notice a difference in behavior
between the SM-limit case and the one in which $\eta$ is negative. In
the latter case, one observes that the range of values for $\amub$ is significantly larger, spreading over the x-axis, while in the former it is constrained inside the region with absolute value $2\times 10^{-10}$.  

In order to obtain a better insight into the bosonic contribution, we
choose a sample point for which the muon anomaly can be explained and
vary the parameters affecting mainly $\amub$. For comparison with the previous analysis~\cite{Broggio:2014mna}, we consider as starting value a parameter space point allowed by the type X model. 

In the type X model, the explanation to the $\amu$ deviation comes
mainly from fermionic contributions containing a tau loop. The reason
is that, in this model, only the Yukawa coupling of leptons is
enhanced, see Table~\ref{table:yukXYZ}. In the Aligned Model, a type X
scenario is recovered if $\zu = \zd = 1/\tb$, and $\zl$ is identified
as $-\tb$. In Ref.~\cite{Broggio:2014mna}, it was found that the
anomaly could be explained for low values of the CP-odd scalar mass
($\MMA<100\;\text{GeV}$), large $\tb$, and values of the masses of the
CP-even and charged scalar of the order of $200\;\text{GeV}$. In that
reference, the type X model was considered and only fermionic
contributions were included. Therefore, it is particularly simple to
translate parameter points to the Aligned $\thdm$, by identifying $\tb$ as $-\zl$.  
After these considerations, we choose as representative point the one defined by\footnote{It should be noticed that any other point considered in~\cite{Broggio:2014mna} for which $\amu$ is explained at $1\,\sigma$ level could be chosen as well. The behavior of all plots as well as all further discussions remain essentially the same. Furthermore the recent references~\cite{Abe:2015oca,Han:2015yys} also considered $\tau$-decay as a parameter constraint in the $\thdm$, which disfavors a significant part of the preferred parameter space. Nevertheless, reference~\cite{Han:2015yys} found the general parameter region represented by~\eref{repr:point} to be viable.}

\begin{align}
\MMA=50\;\text{GeV}, \quad \MMH=\MMHpm=200\;\text{GeV}, \quad \zl = -100, \quad \zu = \zd = 0.01.
\label{repr:point}
\end{align}

\begin{figure}[]
\centering
\begin{subfigure}{0.45\textwidth}
       \centering
        \includegraphics[scale=.11]{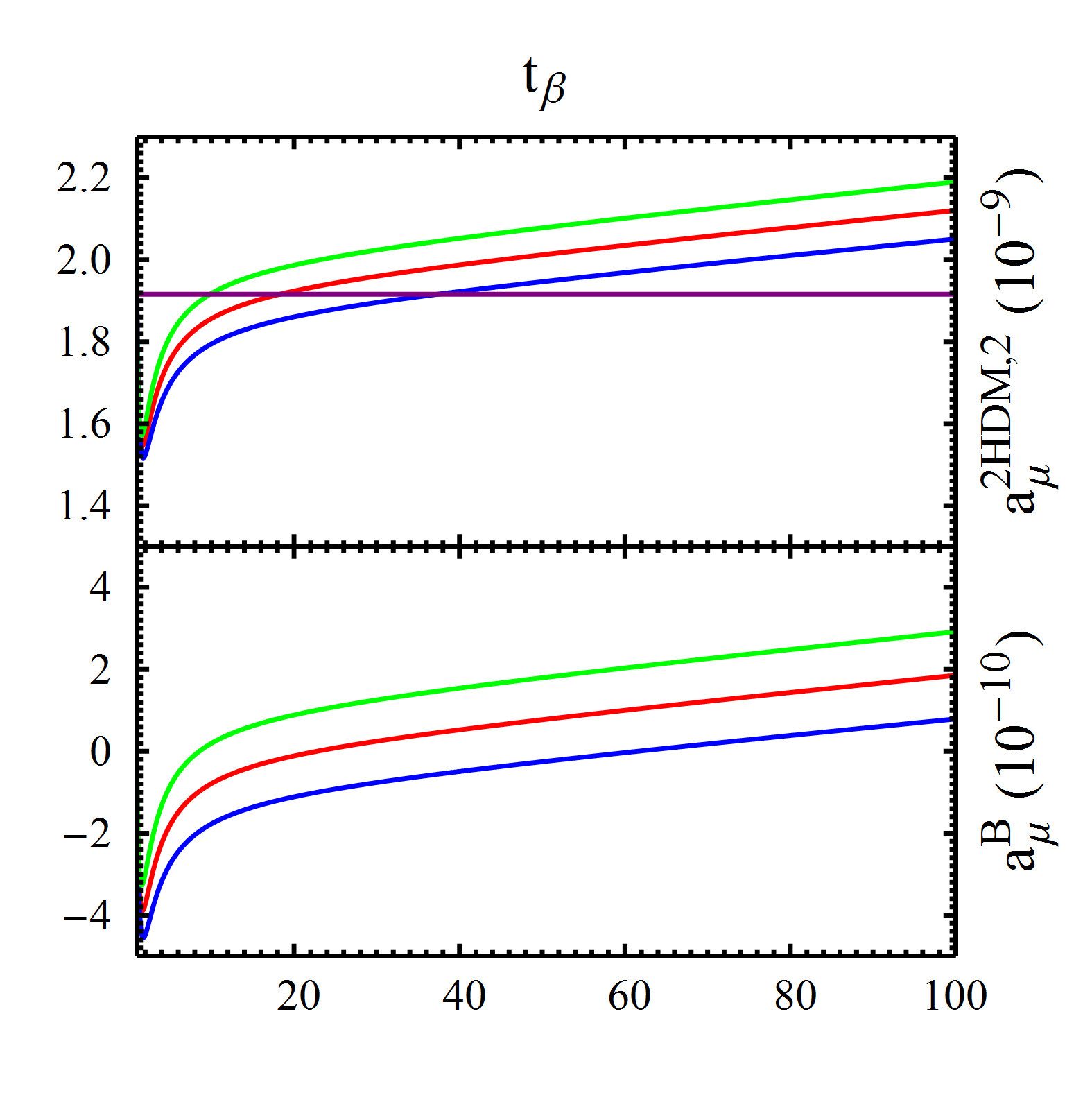}
\subcaption{}
\label{MA50:HighTB}
\end{subfigure}
\begin{subfigure}{0.45\textwidth}\qquad\quad
       \centering
        \includegraphics[scale=.11]{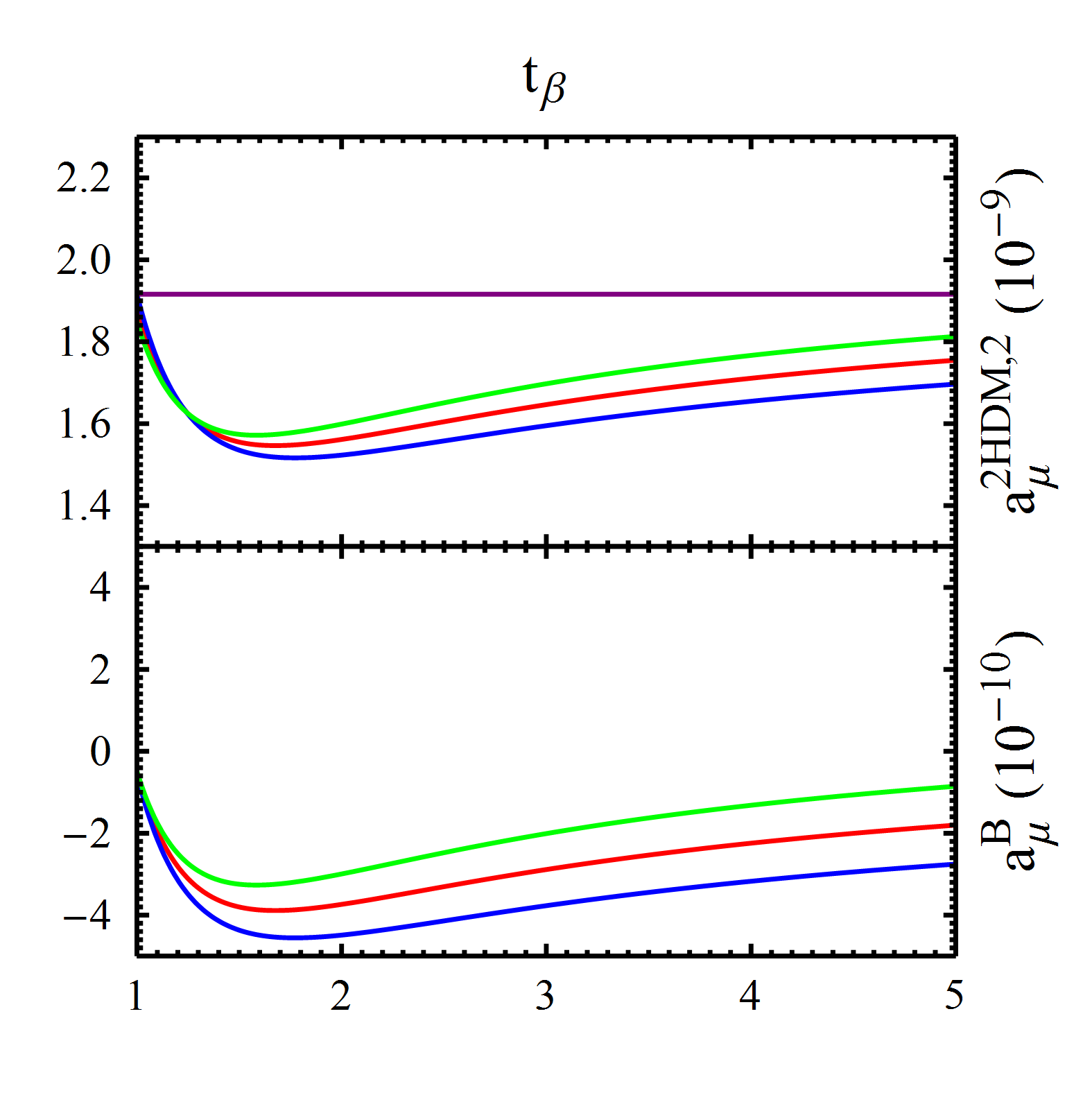}
\subcaption{}
\label{MA50:LowTB}
\end{subfigure}\\\vspace{.5cm}
\centering
\begin{subfigure}{0.45\textwidth}
       \centering
        \includegraphics[scale=.11]{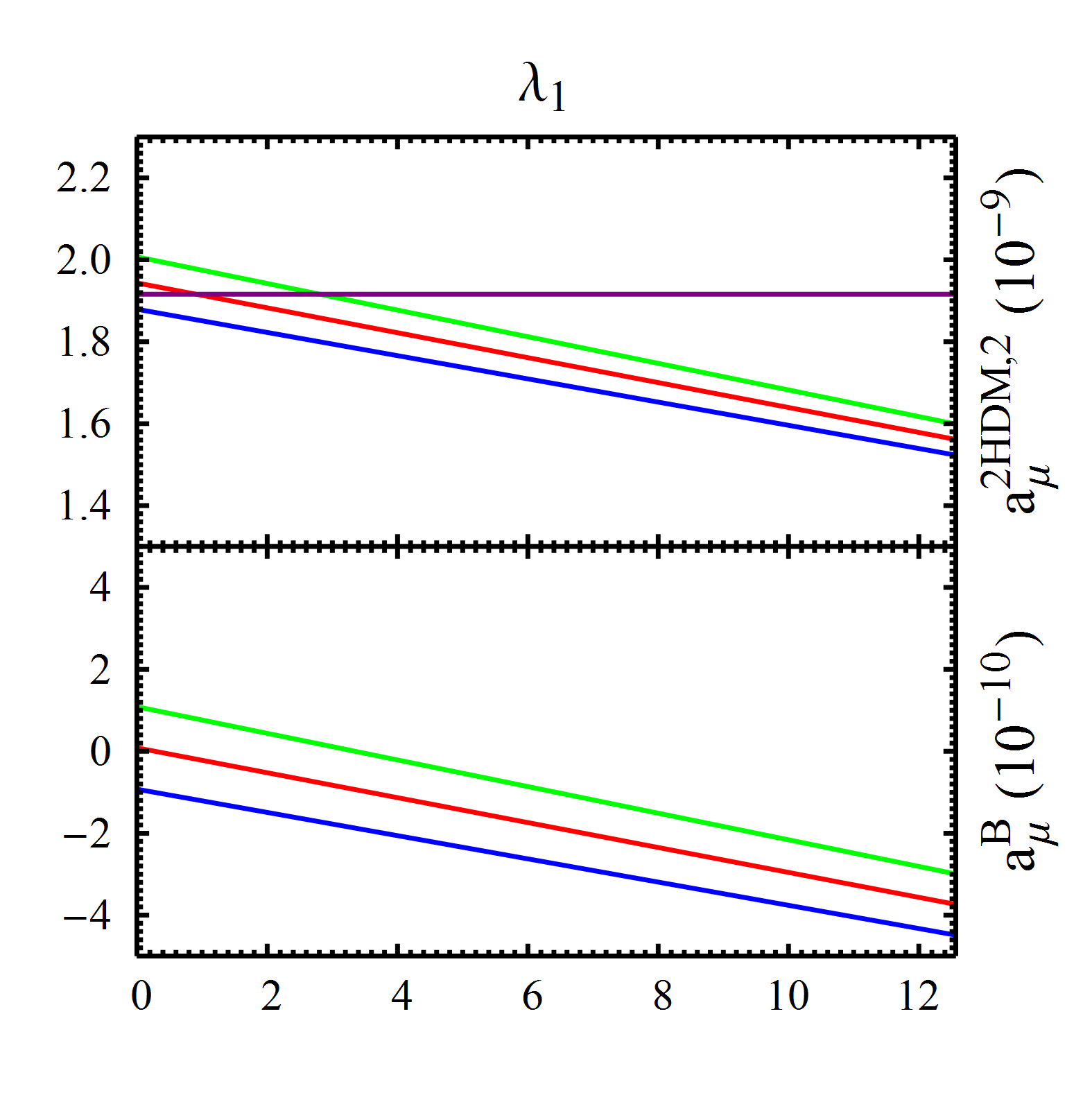}
\subcaption{}
\label{MA50:L1TB2}
\end{subfigure}
\begin{subfigure}{0.45\textwidth}\qquad\quad
       \centering
        \includegraphics[scale=.11]{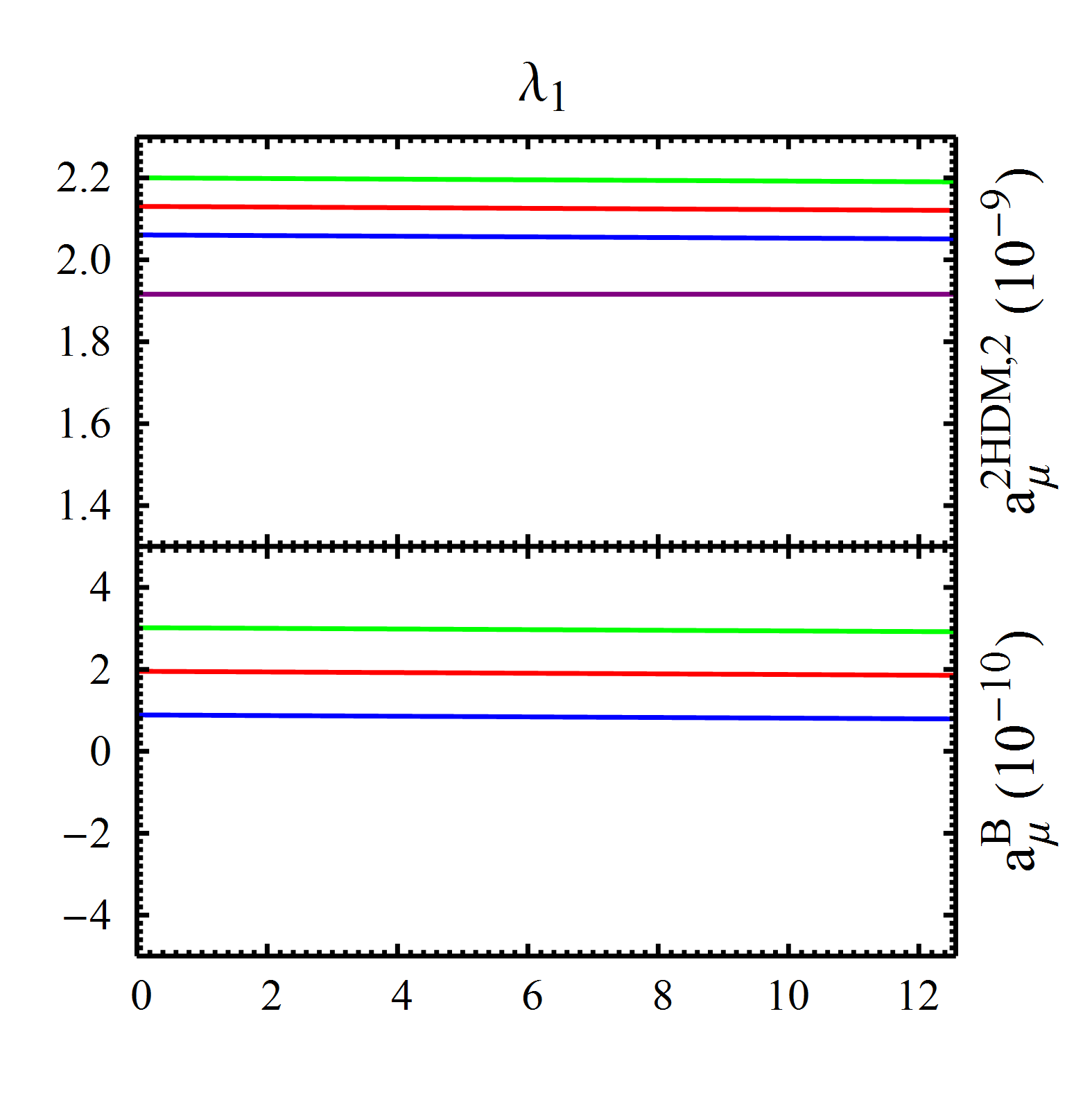}
\subcaption{}
\label{MA50:L1TB100}
\end{subfigure}
\caption{Plots showing the behavior of $\amu^{\thdm, 2}$, and $\amub$. Each red/blue/green line is for $\eta = 0/0.1/-0.1$. $\tb$ varies for (a) and (b), and $\lambda_{1}$ for (c) and (d). We consider the representative mass parameter point in~\eref{repr:point}. $\lambda_{1}=4\pi$ for (a) and (b). We employ $\tb=2$ and $\tb=100$ for (c) and (d) respectively.}
\label{MA50}
\end{figure}

\begin{figure}[]
\begin{subfigure}{0.45\textwidth}
       \centering
        \includegraphics[scale=.11]{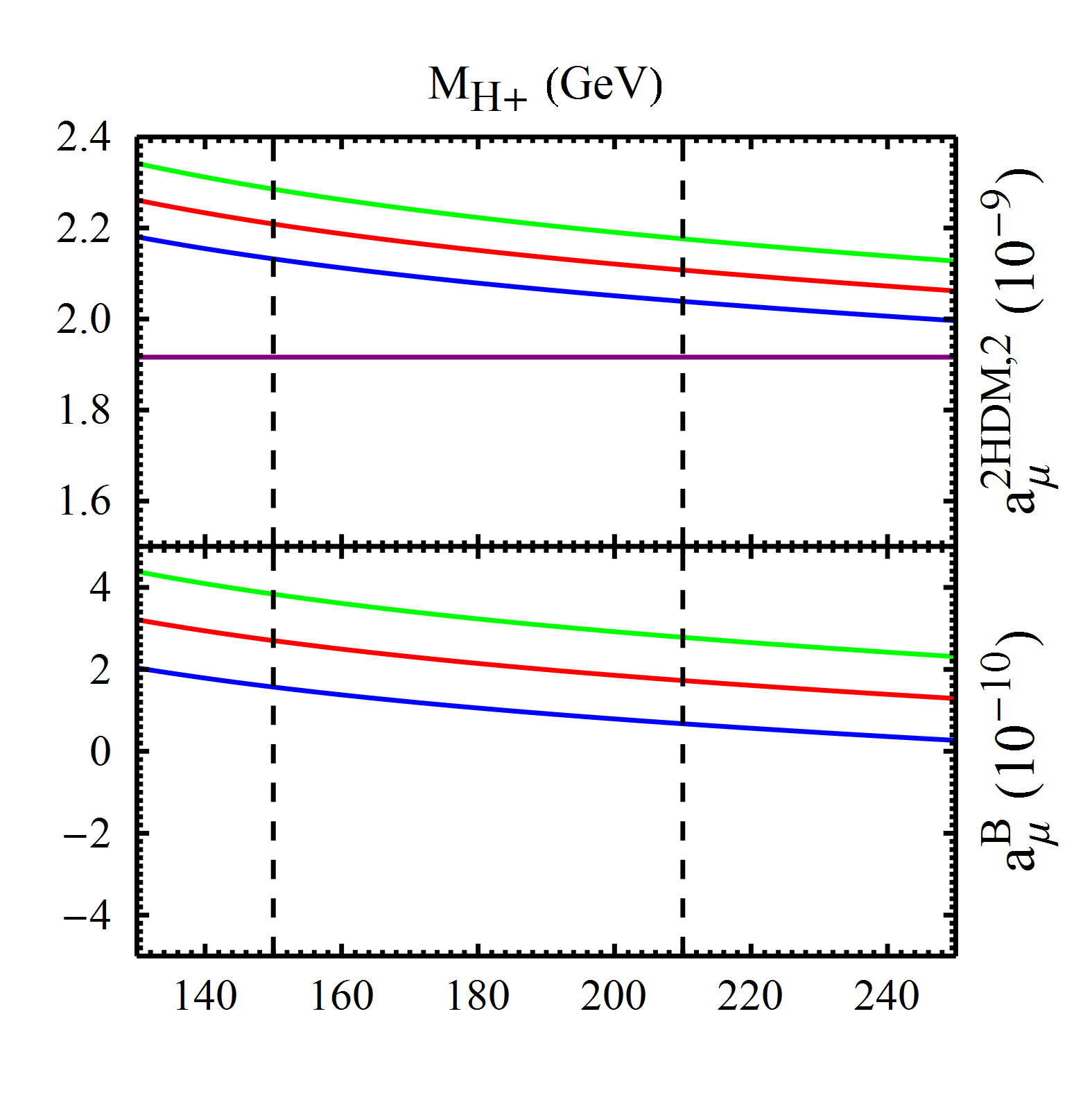}
\subcaption{}
\label{MA50:MHp}
\end{subfigure}
\begin{subfigure}{0.45\textwidth}
       \centering
        \includegraphics[scale=.11]{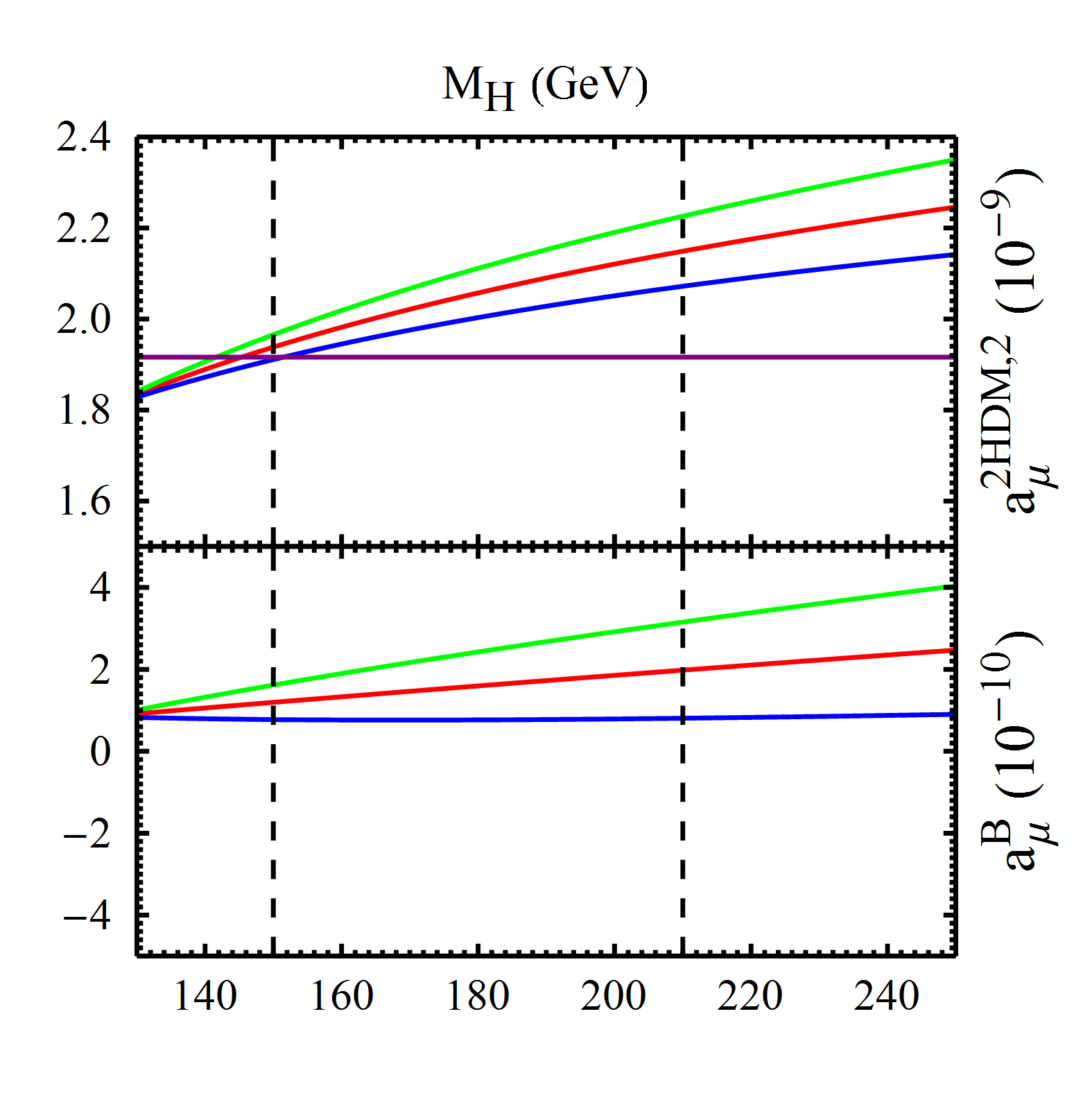}
\subcaption{}
\label{MA50:MHH}
\end{subfigure}
 \caption{Plots showing the behavior of $\amu^{\thdm, 2}$, and $\amub$. Each red/blue/green line is for $\eta = 0/0.1/-0.1$. $\MMHpm$ and $\MMH$ vary in (a) and (b) respectively. We set $\lambda_{1}=4\pi$, and $\tb=100$. The inside regions between the dashed lines are allowed by constraints. The purple line is a reference value as explained in the text.
}
\label{MA50B}
\end{figure}
In the Aligned model, the values for $\tb$, $\lambda_{1}$, and $\eta$
remain free. The first two are only related to the bosonic
contribution via triple Higgs couplings, $\eta$ affects the bosonic
and fermionic contributions.~\fref{MA50} shows
the results from varying  these three parameters, and thus
particularly the impact of the bosonic contribution to $\amu$. In all plots we depict $\amu ^{\thdm, 2}$ on the upper graph, and $\amub$ on the lower one. The upper plots contain as a reference line the value for $\amu$ used in~\cite{Broggio:2014mna}, which takes into account only fermionic contributions for $\eta=0$. The $\eta$-dependence is depicted by red lines ($\eta=0$), blue lines ($\eta=0.1$), and green lines ($\eta=-0.1$). We proceed to explain each of the graphs individually.

\frefs{MA50:HighTB}{MA50:LowTB} show the behavior as a function of $\tb$ for $\lambda_{1}=4\pi$. \footnote{The analysis is unaltered for other choices to $\lambda_{1}$, only the absolute value of the bosonic contribution is modified.} As expected from the scatter plots, the variation of $\amub$ is in the range $(2\cdots4)\times 10^{-10}$, and the contribution can either be negative or positive. The behavior can be understood by analyzing the formula for $\lFA$,~\eref{LAMBDA}, the general formula for $\amub$,~\eref{amub}, and the values of the different coefficients,~\frefs{fig:a0}{fig:aii1}. 

There are two regions: small $\tb$ and large $\tb$. For small $\tb$, $\lFA$ is dominated by the negative term proportional to $\lambda _1$ and several bosonic contributions are suppressed by $(\tb - 1/\tb)$ which vanishes as $\tb \rightarrow  1$. This explains the linear behavior in~\fref{MA50:L1TB2} and the peak in~\fref{MA50:LowTB}. 

For large $\tb$, $\lFA \simeq 2\MMH^2/v^2 \simeq 1.32$, and the prefactor $(\tb - 1/\tb) \simeq \tb$. This explains the linear behavior of the contributions for $\tb > 20$ in~\fref{MA50:HighTB} and the independence of $\lambda_{1}$ in~\fref{MA50:L1TB100}. 

Regarding the $\eta$-dependence, the dominant terms depending on
$\eta$ are $\amu ^{\text{EW add.}}$,~\eref{smlikenum}, and
$\aezez$,~\fref{fig:aezez}. For the present parameter region, the
coefficients of $\eta \, \zl$ are approximately $(2.3 - 1)\times 10^{-11}$. This explains that shifting $\eta$ by $0.1$ decreases $\amub$ by $10 ^{-10}$ in all plots. 

In order to compare our analysis of \fref{MA50} with the scatter plots
of \fref{scat}, we show three representative points in~\fref{scat}. The first, in green, is the representative
point just discussed for the large $\tb$ regime ($\tb=100$, $\lambda_{1}$ arbitrary). The other two, yellow and dark yellow, are
related to the low $\tb$ regime and have $\tb=2$ and two different values
of $\lambda_{1}$, $\lambda_{1}=4\pi$ (yellow) and $\lambda_{1}=2\pi$
(dark yellow). As can be
seen, the green triangle for $\eta=-0.1$ and $\eta=0$ is close to the
border of the constrained sample depicted in red while, for
$\eta=0.1$, the green triangle is well inside the allowed area. It is
instructive to notice that, for negative values of $\eta$, there is a
considerable sample of allowed points with similar values for
$\amub\simeq (2\cdots 3)\times 10^{-10}$. This behavior is explained
by observing~\frefs{MA50:HighTB}{MA50:L1TB100} which show that for any
value of $\lambda_{1}$ there is a large interval for $\tb$, $40 < \tb <100$, allowing $\amub>2\times10^{-10}$. This situation should be
contrasted with the low $\tb$ regime, represented by the yellow
triangles. While the $\eta=-0.1$ case still has a considerable amount
of points with similar values, the ($\eta=0,\; 0.1$) cases represent
rare points in the constrained sample for $\lambda_{1}=4\pi$, and
points close to the border of the allowed area for
$\lambda_{1}=2\pi$. The explanation can be found
in~\frefs{MA50:LowTB}{MA50:L1TB2} which show that values for $\amub$
similar to the ones of the light yellow triangle can only be obtained for a small range of $\tb$, $1 < \tb < 5$, and large values of $\lambda_{1}$, $\lambda_{1}\simeq 4\pi$. These observations explain why the scatter plot for negative $\eta$,~\fref{scat:1m}, has more allowed points with values for $|\amub|$ of order $ (2\cdots 4)\times 10^{-10}$ if compared with the other cases.

Finally we discuss the plots of~\fref{MA50B}. In both cases we study
the behavior of $\amu^{\thdm, 2}$ and $\amub$ as  functions of one of
the masses of the scalars ($\MMHpm$, and $\MMH$ respectively) where
the region delimited by the dashed lines is allowed by theoretical and
EW constraints. The other mass and aligned parameters are kept fixed
as in the representative point~\eref{repr:point}. Regarding $\tb$, we
choose $\tb = 100$, corresponding to a type X parameter point. Since we are in the large $\tb$ limit, $\lambda_{1}$ has no significant influence. We adopt $\lambda_{1}=4\pi$. As can be seen in~\fref{MA50:MHp}, there is a slight mass dependence in $\amu^{\thdm, 2}$. To illustrate the mass dependence we first remark that, for the parameter region we are considering, only the coefficients enhanced by $\zl$ and $\tb$ are important, namely $\aezezez$ and $\aezelz$. Using the definitions in Appendix~\ref{appx:1} and considering the large $\tb$ region, one has
\begin{align}
\amub|_{\eta = 0} \simeq& (\aezezez+ \lFA  \aezelz)\zl\tb\nonumber\\
=& \left[- b(\ZHH,0) - \frac{\lFA}{2}\right]\left[\Fmaster(\ZHH, \ZHp)+\FmasterHp(\ZHH, \ZHp)\right]\zl\tb\nonumber\\
\simeq& -6.3\times 10^{-7} \MMH^2 \left[\Fmaster(\ZHH, \ZHp)+\FmasterHp(\ZHH, \ZHp)\right]\zl\tb,
\label{amu:apr}
\end{align}
where $x_{\cal S}\equiv M_{\cal S}^{2}/M_{Z}^{2}$, and we used $\lFA \simeq 2 \MMH^2 / v^2$. The term containing the functions $\Fmaster$, $\FmasterHp$ is always positive and depends on the inverse of the scalar masses. 

Therefore, if $\MMH$ is kept fixed and $\MMHpm$ increases,  $|\amub|$
will decrease, explaining the behavior observed in~\fref{MA50:MHp}. In
contrast, if $\MMHpm$ is kept fixed and $\MMH$ increases, the explicit
dependency on $\MMH ^2$ coming from $\lFA$ and the coefficient
$b(\ZHH,0)$ leads to an increase of $\amub$ with $\MMH$ in~\fref{MA50:MHH}. 

Regarding the full $\amu$, we verified that the fermionic
contributions essentially do not depend on $\MMHpm$ due to the small
$\zu$, but they depend on $\MMH$. As can be noticed analyzing the
plots of~\fref{fig:floopplots} and  Table~\ref{tab:signs}, the
fermionic contributions from $\MH$ diagrams are negative and decrease in modulus with $\MMH$. Therefore, the net result will be an increase in $\amu$  as observed in~\fref{MA50:MHH}. 

Finally, it can also be noticed that the plots for non-zero values of $\eta$ tend to the case $\eta=0$ as $\MMH$ approaches $\MMh$. This behavior can be understood by observing that in this case the two mass-degenerate CP-even scalars together behave exactly SM-like. 
\newpage
\section{Conclusion}
\label{sec:conclusion}
We presented the full two-loop $\thdm$ contributions to $\amu$,
providing the complete analytic result and a numerical analysis. We confirmed the previous results of the fermion-loop and the bosonic Barr-Zee type contributions. We calculated the remaining diagrams including all \threebo diagrams, which involve three internal boson couplings to the muon line. 

The analytic results are expressed in terms of physical parameters. 
The full bosonic result depends on the three additional Higgs boson
masses, $\tb$, $\sin\alpha$, the alignment parameter $\zeta _l$ and
the quartic scalar coupling $\lambda _1$. We always expand in the
small parameter $\eta=\alpha-\beta+\pi/2$, the deviation from the
SM-limit. 
The bosonic contributions are especially dependent on $\tb$ and
$\lambda _1$, whereas fermionic ones are not. This dependency arises
from the triple Higgs couplings in the bosonic Feynman diagrams. 

We split the bosonic result into several parts, see \eref{amub}
and \eref{ayuk2}. Each term has a straightforward dependence on $\tb$
and $\zl$ and depends only on a subset of masses. The compact analytic
expression of each term is provided in Appendix~\ref{appx:1}. We
documented the parameter dependence in a series of Figures
in~\Sref{subsec:bloop}.  

We also confirmed the previous result of the fermionic contribution. 
Particularly, we presented its analytic form without one-dimensional integral relations in~\Sref{subsec:floop} and 
gave an overview of the numerical behavior.  
The fermionic result involves all three alignment parameters
$\zeta _{l,u,d}$, but the leading contributions are the $\zl$ dependent terms. 

We also investigated the impact of the scenario with a deviation from
the SM-limit of the Higgs couplings, $\eta=\alpha-\beta+\pi/2\ne0$.
For this case, we obtain additional  contributions from the SM-like
Higgs boson, $\amu ^{\text{EW add.}}$. This term is proportional to
$\eta\,\zl$ and gives the dominant $\eta$-dependent bosonic
contributions. Its coefficient is dependent only on the SM parameters
and can be found in~\eref{smlikenum}. 

In the numerical evaluation we confirmed that the fermionic $\thdm$
contribution can be of the order of the deviation~\eref{deviation}. A
series of  plots shows that in parameter regions with large fermionic
contributions, the complete bosonic result can yield 
additional contributions in the range $(2 \cdots 4) \times 10 ^{-10}$,
i.e. at the level of the precision of the planned Fermilab
experiment.  Allowing the SM-like Higgs couplings to deviate from the SM-limit, i.e. $\eta\ne0$, and non-zero values of $\lambda _1$, can slightly increase the bosonic contributions. 

\section*{Acknowledgments}
The figures of \Sref{sec:numanaly} have been created using the SciDraw scientific figure
preparation system \cite{Caprio}. The authors acknowledge financial support from DFG Grant
STO/876/6-1, and CNPq, Conselho Nacional de Desenvolvimento Cient\'{\i}fico e Tecnol\'{o}gico - Brazil.
\appendix
\section{Analytic results}
\label{appx:1}
\newcommand{\vx}{u}
\newcommand{\vy}{\omega}
\newcommand{\AYone}{a^1}
\newcommand{\AYtwo}{a^2}
\newcommand{\AYthree}{a^3}
\newcommand{\fZone}{\frac{7}{2} - \frac{25}{2 \CWsq} + 4\CWsq - 4\CW ^4}
\newcommand{\fZtwo}{2 (17 - 24 \CWsq + 56 \CW ^4 - 128 \CW ^6 + 64 \CW^8)}
\newcommand{\fZthree}{\frac{25 - 32 \CWsq + 4 \CW ^4}{\CWsq \SWsq}}
\newcommand{\fZfour}{\frac{13}{2} - 15\CWsq + 10\CW ^4}
\newcommand{\fZfive}{\frac{\CWsq(5 - 16\CWsq + 8\CW ^4)}{\SWsq}}
\newcommand{\fZsix}{\frac{7 - 14\CWsq + 4\CW ^4}{4 \CWsq \SWsq}}
\newcommand{\fZseven}{1 - 6\CWsq + 4 \CW ^4}
\newcommand{\fZeight}{\frac{13 - 20 \CWsq + 4\CW ^4}{\CWsq\SWsq}}
\newcommand{\fZnine}{7 - 12\CWsq + 8\CW ^4}
\newcommand{\fZones}{f_1}
\newcommand{\fZtwos}{f_2}
\newcommand{\fZthrees}{f_3}
\newcommand{\fZfours}{f_4}
\newcommand{\fZfives}{f_5}
\newcommand{\fZsixs}{f_6}
\newcommand{\fZsevens}{f_7}
\newcommand{\fZeights}{f_8}
\newcommand{\fZnines}{f_9}
\newcommand{\FFTF}{{\mathcal{T}}_0}
\newcommand{\FFPL}{{\mathcal{T}}_1}
\newcommand{\FFApm}{{\mathcal{T}}_2^\pm}
\newcommand{\FFAp}{{\mathcal{T}}_2^+}
\newcommand{\FFAm}{{\mathcal{T}}_2^-}
\newcommand{\FFB}{{\mathcal{T}}_4}
\newcommand{\FFC}{{\mathcal{T}}_5}
\newcommand{\FFD}{{\mathcal{T}}_6}
\newcommand{\FFE}{{\mathcal{T}}_7}
\newcommand{\FFF}{{\mathcal{T}}_8}
\newcommand{\FFG}{{\mathcal{T}}_9}
\newcommand{\FFH}{{\mathcal{T}}_{10}}
\newcommand{\fSone}{{\mathcal{S}}_1}
\newcommand{\fStwo}{{\mathcal{S}}_2}
%
%
\newcommand{\fYone}{-\frac{\CWsq(\CWsq-1)}{2\AL\pi(\ZHH+2\ZHp)}}
\newcommand{\fYYone}{\frac{9}{2}\frac{(\CWsq-1)}{\AL\pi}}
\newcommand{\fYtwo}{\frac{\CWsq(\CWsq-1)}{\AL\pi(\ZHSM+2\ZHp)}}
\newcommand{\fYYtwo}{0}
\newcommand{\fYthree}{-\frac{\ZHH}{(\ZHH+2\ZHp)}}
\newcommand{\fYYthree}{9\frac{\ZHH}{\CWsq}}
\newcommand{\fYfour}{-\frac{\CWsq(\CWsq-1)}{\AL\pi(\ZHH+2\ZHp)}}
\newcommand{\fYYfour}{9\frac{(\CWsq-1)}{\AL\pi}}
\newcommand{\fYfive}{\frac{\CWsq(\CWsq-1)}{2\AL\pi(\ZHH+2\ZHp)}}
\newcommand{\fYYfive}{0}
\newcommand{\fYsix}{1}
\newcommand{\fYYsix}{\frac{9}{\CW^2}(\ZHH+2\ZHp)}
\newcommand{\fYseven}{\frac{\ZHH}{(\ZHH+2\ZHp)}}
\newcommand{\fYYseven}{0}
\newcommand{\FYA}{{\mathcal{F}}_1}
\newcommand{\FN}{{\mathcal{F}}_2}
\newcommand{\FNch}{{\mathcal{F}}_3}
\newcommand{\Aones}{A_1}
\newcommand{\Atwos}{A_2}
\newcommand{\Athrees}{A_3}
\newcommand{\Afours}{A_4}
\newcommand{\Afives}{A_5}
\newcommand{\Asixs}{A_6}
\newcommand{\Asevens}{A_7}
\newcommand{\Aone}{-9 \CW^2 \vx^3+9 \CW^2 \vx^2 \left(3 \CW^2+\vy\right)+27 \CW^4 \vx \left(\vy-\CW^2\right)\nonumber\\ &+9 \left(\CW^8-4 \CW^6 \vy+3 \CW^4 \vy^2\right)}
\newcommand{\Atwo}{\frac{9 \CW^4 \vy}{2}-9 \vx^2 \left(5 \CW^2+\vy\right)+\vx \left(36 \CW^4+\frac{153 \CW^2 \vy}{4}\right)+9 \vx^3}
\newcommand{\Athree}{9 \CW^2 \vx^2-\frac{9}{2} \CW^2 \vx \left(4 \CW^2+\vy\right)}
\newcommand{\Afour}{-\frac{9}{2} \vx^2 \vy \left(2 \CW^4+9 \CW^2 \vy+2 \vy^2\right)\nonumber\\ &+\frac{9}{8} \vx \vy \left(32 \CW^6+13 \CW^4 \vy+35 \CW^2 \vy^2\right)+9 \vx^3 \vy^2}
\newcommand{\Afive}{-9 \vx^3 \left(\CW^2+\vy\right)-9 \vx \left(3 \CW^6+2 \CW^2 \vy^2\right)\nonumber\\ &+9 \vx^2 \left(3 \CW^4+4 \CW^2 \vy+\vy^2\right)+\frac{9}{2} \CW^4 \left(2 \CW^4-6 \CW^2 \vy+\vy^2\right)}
\newcommand{\Asix}{-9 \vx^4 \left(9 \CW^2+\vy\right)+\vx \left(81 \CW^6 \vy-225 \CW^8\right)+9 \CW^8 \left(\vy-\CW^2\right)\nonumber\\ &-\frac{9}{2} \vx^2 \left(3 \CW^6+37 \CW^4 \vy\right)+\vx^3 \left(198 \CW^4+72 \CW^2 \vy\right)+9 \vx^5}
\newcommand{\Aseven}{-9 \CW^2 \vx^4+18 \CW^2 \vx^3 \left(2 \CW^2+\vy\right)+36 \vx \left(\CW^8-2 \CW^6 \vy\right)\nonumber\\ &-9 \CW^2 \vx^2 \left(6 \CW^4-\CW^2 \vy+\vy^2\right)-9 \CW^2 \left(\CW^2-3 \vy\right) \left(\CW^3-\CW \vy\right)^2}
\newcommand{\Fones}{F_1}
\newcommand{\Ftwos}{F_2}
\newcommand{\Fthrees}{F_3}
\newcommand{\Ffours}{F_4}
\newcommand{\Ffives}{F_5}
\newcommand{\Fsixs}{F_6}
\newcommand{\Fsevens}{F_7}
\newcommand{\Feights}{F_8}
\newcommand{\Fnines}{F_9}
\newcommand{\Ftens}{F_{10}}
\newcommand{\Felevens}{F_{11}}
\newcommand{\Ftwelves}{F_{12}}
\newcommand{\Fthirteens}{F_{13}}
\newcommand{\Ffourteens}{F_{14}}
\newcommand{\Ffifteens}{F_{0}}
\newcommand{\Ffifteen}{\frac{3\CW^4\left(-640+576\CW^2+7\pi^2\right)}{4}}
\newcommand{\Fone}{96 \CW^6 \left(11-53 \CW^2+36 \CW^4\right)}
\newcommand{\Ftwo}{-\frac{3}{4} \CW^2 \left(-66 \CW^2-48 \CW^4+672 \CW^6\right)}
\newcommand{\Fthree}{-\frac{3}{4} \CW^2 \left(109-430 \CW^2+120 \CW^4\right)}
\newcommand{\Ffour}{96 \CW^6 \left(-11+9 \CW^2\right)}
\newcommand{\Ffive}{\frac{45 \CW^4}{2}+192 \CW^6}
\newcommand{\Fsix}{\frac{3}{4} \CW^2 \left(157+90 \CW^2\right)}
\newcommand{\Fseven}{-\frac{3}{4} \left(18+61 \CW^2\right)}
\newcommand{\Feight}{\left(-7+61 \CW^2-162 \CW^4+96 \CW^6\right)}
\newcommand{\Fnine}{(1 - 5\CW^2 + 10\CW^4)}
\newcommand{\Ften}{-1728 \CW^8 \left(-1+\CW^2\right)}
\newcommand{\Feleven}{3 \CW^6 \left(-899+768 \CW^2\right)}
\newcommand{\Ftwelve}{\left(387 \CW^4-363 \CW^6\right)}
\newcommand{\Fthirteen}{\frac{9}{2} \CW^2 \left(57+106 \CW^2\right)}
\newcommand{\Ffourteen}{-\frac{15}{2} \left(7+45 \CW^2\right)}
\allowdisplaybreaks

Here we provide the full analytic result of the complete renormalized bosonic two-loop contributions $\amub$, in the decomposition of~\eref{amub}. We begin with required loop function (defined first in Ref.~\cite{Davydychev:1992mt}):
\begin{align}
\label{eq:TFex}
\TF(m_1,m_2,m_3) &=
\frac{\lambda}{2}
\Bigg[
2\ln(\alpha_+)\ln( \alpha_-)
-\ln\left(\frac{m_1^2}{m_3^2}\right)\ln\left(\frac{m_2^2}{m_3^2}\right)
\nonumber\\
&-2\dilog(\alpha_+)-2\dilog(\alpha_-)+\frac{\pi^2}{3}
\Bigg]
,
\\
\lambda
&=\sqrt{m_1^4+m_2^4+m_3^4-2m_1^2m_2^2-2m_2^2m_3^2-2m_3^2m_1^2}\;,
\\
\alpha_\pm &=\frac{m_3^2 \pm m_1^2 \mp m_2^2 - \lambda}{2m_3^2}.
\end{align}
The coefficient $\az$ of the contribution without Yukawa couplings is given by
\begin{align}\label{eq:analyA00}
\az =& \GlobalC\,\left\{
\left(\frac{\ZAZ-\ZHH}{\ZAZ-\ZHp}\right)\FFAp(\ZAZ,\ZHH) + \FFAm(\ZHH,\ZHp) \right.\nonumber\\
&+\left(\frac{\ZAZ-\ZHH}{\ZAZ-\ZHp}\right)\FFB(\ZAZ,\ZHp) + \FFB(\ZHH,\ZAZ) + \FFC(\ZHp,\ZHH) + \FFC
(\ZHp,\ZAZ) \nonumber\\
&+ \FFAp(\ZHp,\ZHH) + \FFAp(\ZHp,\ZAZ) + \FFD(\ZAZ,\ZHp) + \FFD(\ZHH,\ZHp) \nonumber\\
&+ \FFE(\ZAZ,\ZHH) + \FFE(\ZHp,\ZHp)(1-2\CWsq)^2 + \FFF(\ZAZ,\ZHp) + \FFF(\ZHH,\ZHp) \nonumber\\
&-\frac{16}{3}\CWsq\SWsq(1 + 8\CWsq - 8 \CW ^4) + \frac{8 \CW ^4 \SW ^4}{5 \ZHp}  +  \fZtwos \ZHp - \fZthrees \ZHp ^2 \nonumber\\
&+ \fZones(\ZAZ ^2 + \ZHH ^2) + \fZthrees \ZHp (\ZAZ + \ZHH) + \fZfours(\ZAZ + \ZHH) - \fZfives\ZAZ\ZHH\nonumber\\
&\left.+\FFPL(\ZAZ,\ZHp) + \FFPL(\ZHH,\ZHp) + \FFTF(\ZAZ, \ZHp) + \FFTF(\ZHH, \ZHp)
\right\}.
\end{align}
The abbreviations appearing in $\az$ are
\begin{align}
\FFTF(\vx,\vy) =& \frac{9}{\CW ^4}\frac{(\vx - \vy)(\CWsq (\vx - \vy)(\vx + 2\vy) 
- (\vx - \vy)^3 + \CW ^4 \vy)}{\CW ^4 + (\vx - \vy)^2 - 2\CWsq (\vx + \vy)}\nonumber\\
&\times\TF(\sqrt{\vx},\sqrt{\vy},\CW), \\
\FFPL(\vx,\vy) =& \frac{9}{\CW ^4}(\vx-\vy)(\CWsq \vy - (\vx - \vy)^2)\dilog(1-\vx/\vy),\\
\FFApm(\vx,\vy) =&\ln(\vx)\left(\frac{6 \vx^2 + \CWsq(\vx - \ZHp) + 2\CW ^4(\vx - \ZHp)}{2 (\vx - \vy)}\right. \nonumber\\ 
&+ \fZsixs\frac{(\vx - \ZHp)^2 (3 \CW ^4 + 3 \CWsq (\vx - \ZHp) + (\vx - \ZHp)^2}{\CWsq (\vx - \vy)}\nonumber\\ 
&\pm \fZsevens\frac{3 \vx ^2 (\vx - \ZHp)}{(\ZAZ - \ZHH)(\vx - \vy)} - \fZeights \frac{3 \vx (\vx - \ZHp)^2}{2 (\vx - \vy)}\nonumber\\ 
&-\left. \fZnines \frac{3 \vx (\vx - \ZHp)}{2 (\vx - \vy)} \right),\\
\FFB(\vx,\vy) =& \frac{(\vx - \vy)\ln(\vx)}{4} \fZfives (\ZAZ ( 3 + 2\ZHH) - \ZAZ ^2 + 3 \ZHH - \ZHH ^2 - 3), \\
\FFC(\vx,\vy) =& \ln(\vx)\left( \frac{3}{2}\vx + \frac{\fZsixs}{\CWsq}((\vx - \vy)^3 + 3 \CWsq (\vx - \vy)^2 + 3\CW ^4(\vx - \vy))\right.\nonumber\\
&\left. - \frac{3}{2}\fZeights\vx(\vx-\vy) - \frac{\CWsq}{2} - \CW^4 \right), \\
\FFD(\vx,\vy) =& \frac{9}{2}\left(
\frac{(\vx - \vy)(\vx^2 - 2\vx\vy + \vy(\vy - \CWsq))}{\CW ^4}\ln\left(\frac{\vx}{\vy}\right)\ln\left(\frac{\vy}{\CWsq}\right)\right.\nonumber\\
&\left.+ \frac{\ln(\CWsq)}{\CWsq}(2\vx^2 + \vx(\CWsq - 4 \vy) - \vy(\CWsq - 2\vy)) \right),\\
\FFE(\vx,\vy) =& -\frac{\fZfives}{2}( 2(\vx + \vy) - (\vx - \vy)^2 -1 )\ln\left(\frac{\fSone(\vx,\vy)}{2\sqrt{\vx\vy}}\right)\nonumber\\
&\times\left(\vx + \vy - 1 - \frac{4\vx\vy}{\fSone(\vx,\vy)}\right),\\
\fSone(\vx,\vy) =& \vx + \vy -1 + \sqrt{1 + (\vx - \vy)^2 - 2(\vx + \vy)},\\
\FFF(\vx,\vy) =& 2 \fZsixs ( 4\vx\vy - (\vx + \vy - \CWsq)^2)\ln\left(\frac{\fStwo(\vx,\vy)}{2\sqrt{\vx\vy}}\right)\nonumber\\
&\times\left(\frac{(\vx + \vy)}{\CWsq} - \frac{4 \vx \vy}{\CWsq\fStwo(\vx,\vy)} - 1\right),\\
\fStwo(\vx,\vy) =& \vx + \vy - \CWsq + \sqrt{(\vx + \vy - \CWsq)^2 - 4\vx\vy}. 
\end{align}
The following coefficients depend only on known SM parameters; we provide therefore also approximate numerical values:
\begin{align}
\fZones =& \fZone &=&-12  ,\\
\fZtwos =& \fZtwo &=&-9.1 ,\\
\fZthrees =& \fZthree &=&15 ,\\
\fZfours =& \fZfour &=&-0.9,\\
\fZfives =& \fZfive &=&-9,\\
\fZsixs =& \fZsix &=&-2,\\
\fZsevens =& \fZseven &=& -1.2,\\
\fZeights =& \fZeight &=& -0.7,\\
\fZnines =& \fZnine &=& 2.5.
\end{align}

The coefficients of the Yukawa-dependent terms in~\eref{ayuk2} are given by
\begin{align}
\aezezeze & =  b(\ZHSM,\ZHp)\Fmaster(\ZHSM, \ZHp);\\
\aezezez & =  - b(\ZHH,0)\left[\Fmaster(\ZHH, \ZHp)+\FmasterHp(\ZHH, \ZHp)\right];\\
\aezelze & = \Fmaster(\ZHSM, \ZHp);\\
\aezelz & =  -\frac{1}{2}\left[\Fmaster(\ZHH, \ZHp)+\FmasterHp(\ZHH, \ZHp)\right];\\
\aezeze & =    b(\ZHH,0)\Fmaster(\ZHH, \ZHp) - (\ZHH \rightarrow \ZHSM);\\
\aezez & =   -\bigg[b(\ZHH,\ZHp)\left(\Fmaster(\ZHH, \ZHp)+\FmasterHp(\ZHH, \ZHp)\right)\nonumber\\ & \quad- \FNch(\ZHH, \ZHp) - (\ZHH \rightarrow \ZHSM)\bigg] + \FN(\ZHH);\\
\aelze & =   \frac{\Fmaster(\ZHH, \ZHp)}{2} - (\ZHH \rightarrow \ZHSM);\\
\aelz & =  -\Fmaster(\ZHH, \ZHp)-\FmasterHp(\ZHH, \ZHp) - (\ZHH \rightarrow \ZHSM).
\end{align}
The appearing abbreviations are given by
\begin{align}
b(\vx,\vy) =&\frac{\alpha\pi}{\CW^{2}(-1+\CW^2)}(\vx+2\vy),\\
\Fmaster(\vx, \vy)  =&  \GlobalC\, \left(\frac{1}{\alpha\pi}\frac{ \CW^2 (-1+\CW^2)}{(\vx + 2\vy)}\right)\FYA(\vx,\vy),\\
\FmasterHp(\vx, \vy) =&  \GlobalC\, \left(-\frac{9(-1+\CW^2)}{\alpha\pi}\right)\left(\frac{\FFG(\vx,\vy)}{2}+\FFH(\vx,\vy)\right),\\
\FYA(\vx, \vy) =& -72\CWsq (-1 + \CWsq)\frac{\vx + 2\vy}{\vx} - 
   36\CWsq(-1 + \CWsq)\frac{\vx + 2\vy}{\vx}\ln(\vy) \nonumber\\
& + 9 (-8\CW ^4 - 3 \vx + 2\CWsq(4 + \vx))\frac{(\vx + 2\vy)}{2(\vx - 1)\vx}\ln(\vx) \nonumber\\
& - 9 (3 - 10\CWsq + 8\CW ^4)\frac{\vy(\vx + 2\vy)}{(4 \vy - 1)(\vx - 1)} \TF(\sqrt{\vy}, \sqrt{\vy}, 1) \nonumber\\
& + 9 (8\CW ^4 + 3\vx - 2\CWsq(4 + \vx))\frac{\vy(\vx + 2\vy)}{(4\vy - \vx)(\vx - 1)\vx^2} \TF(\sqrt{\vx}, \sqrt{\vy}, \sqrt{\vy}),\\
\FFG(\vx,\vy) =&- \frac{2 \left(\CW^4 \vy+\CWsq \left(\vx^2+\vx \vy-2 \vy^2\right)-(\vx-\vy)^3\right)
   \TF\left(\sqrt{\vx},\sqrt{\vy},\CW\right)}{(\CWsq-\vy) \left(\CW^4-2 \CWsq (\vx+\vy)+(\vx-\vy)^2\right)}\nonumber\\
& + \frac{2 \CW^4 \left(\vx^2-4 \vx \vy+2 \vy^2\right) \TF\left(\sqrt{\vx},\sqrt{\vy},\sqrt{\vy}\right)}{\vy^2
   \left(\vy-\CW^2\right) (\vx-4 \vy)}\nonumber\\
& - \frac{2 \left(\CW^2 \vx (\vx-2 \vy)+\vy (\vx-\vy)^2\right) \dilog\left(1-\frac{\vx}{\vy}\right)}{\vy^2},\\
\FFH(\vx,\vy) =&\frac{\vx^2-\CWsq\vy-2\vx\vy+\vy^2}{2(\CWsq-\vy)}\ln\left(\frac{\vy}{\vx}\right)\ln\left(\frac{\vy}{\CWsq}\right)\nonumber\\
& +\frac{\CWsq(\CWsq+2\vx-2\vy)}{2(\CWsq-\vy)}\ln\left(\frac{\vy}{\CWsq}\right)+\frac{\CWsq}{\vy}\vx\ln\left(\frac{\vy}{\vx}\right)+\frac{\CWsq}{\vy}(\vy-\vx),\\
\FN(\vx) = &F^{\text W}(\vx) + F^{\text Z}(\vx) +\nonumber\\
& + \GlobalC\, \Bigg\{\frac{8\CW^6\pi^2}{\vx^{2}}+\frac{\Ffifteens}{\vx}+\frac{393\CW^2}{8}\nonumber\\
& + \left(\frac{\Fones}{\vx}+\Ftwos+\Fthrees\vx\right)\frac{\ln(\CWsq)}{(4\CWsq-1)(4\CWsq-\vx)} \nonumber\\
& + \left(\frac{\Ffours}{\vx}+\Ffives+\Fsixs\vx+\Fsevens\vx^2\right)\frac{\ln(\vx)}{(\vx-1)(4\CWsq-\vx)} \nonumber\\ 
& - \frac{3}{2}\left(\frac{32\CW^6}{\vx^2}+\frac{21\CW^4}{\vx}+15\CWsq-35\vx\right)\dilog\left(1-\frac{\vx}{\CWsq}\right)\nonumber\\
& + (\Feights+\Fnines\vx)\frac{9\CWsq(-3+4\CW^2)}{2}\frac{\TF\left(\CW,\CW,1\right)}{(4\CWsq-1)^2(\vx-1)}\nonumber\\
& +\left[\frac{\Ftens}{\vx^2}+\frac{\Felevens}{\vx} +\Ftwelves + \Fthirteens\vx + \Ffourteens\vx^{2} + \frac{105\vx^{3}}{2}\right]\frac{\TF\left(\sqrt{\vx},\CW,\CW\right)}{(4\CWsq-\vx)^2(\vx-1)}\Bigg\},
\end{align}
\begin{align}
\Ffifteens =& \Ffifteen &=& -55.9 ,\\
\Fones =& \Fone &=& -380  ,\\
\Ftwos =& \Ftwo &=& -137 ,\\
\Fthrees =& \Fthree &=& +88.8 ,\\
\Ffours =& \Ffour &=& -180,\\
\Ffives =& \Ffive &=& +103,\\
\Fsixs =& \Fsix &=& +132,\\
\Fsevens =& \Fseven &=& -49.0,\\
\Feights =& \Feight &=& -12.3,\\
\Fnines =& \Fnine &=& +3.15,\\
\Ftens =& \Ften &=& +140,\\
\Felevens =& \Feleven &=& -425,\\
\Ftwelves =& \Ftwelve &=& +63.4,\\
\Fthirteens =& \Fthirteen &=& +486,\\
\Ffourteens =& \Ffourteen &=& -314,
\end{align}
\begin{align}
F^{\text Z}(\vx) =& \GlobalC\, \Bigg\{Z_{1} \vx \dilog\left(1-\vx\right)\nonumber\\
&+\frac{Z_{2}}{2 \vx^2}\Big[6 (-4+\vx) \vx+\pi ^2 (4+3 \vx)+6 \vx (4+\vx) \ln(\vx)\nonumber\\
&-6 (4+3 \vx) \dilog\left(1-\vx\right)+6 \vx (2+\vx) \TF\left(\sqrt{\vx},1,1\right)\Big]\nonumber\\
&+ Z_{3} \vx \Big[6+\pi ^2 (-4+\vx) \vx+3 \ln(\vx) (4+(-4+\vx) \vx \ln(\vx))\nonumber\\
&+12 (-4+\vx) \vx \dilog\left(1-\vx\right)+6 (-2+\vx) \TF\left(\sqrt{\vx},1,1\right)\Big],
\end{align}
\begin{align}
Z_{1} =& 3 (17 - 48 \CW^2 + 32 \CW^4) &=& -2.9, \\
Z_{2} =& \left(5-12 \CW^2+8 \CW^4\right) &=& 0.50, \\
Z_{3} =& 3 \left(1-3 \CW^2+2 \CW^4\right) &=& -0.37,
\end{align}
\begin{align}
F^{\text W}(\vx)=&\GlobalC\, \Bigg\{
-\frac{57 \CW^2}{2}-\frac{4 \CW^6 \pi ^2}{\vx^2}+\frac{3 \CW^4 \left(32-3 \pi ^2\right)}{4 \vx}\nonumber\\
&+\frac{3 \left(16 \CW^6+9 \CW^4 \vx+12 \CW^2 \vx^2-19 \vx^3\right) \dilog\left(1-\frac{\vx}{\CW^2}\right)}{2 \vx^2}\nonumber\\
&+\frac{3 \CW^2 \left(16 \CW^2+19 \vx\right) \left(\ln\left(\CW^2\right)-\ln(\vx)\right)}{2 \vx}\nonumber\\
&+\frac{3 \left(4 \CW^4-50 \CW^2 \vx+19 \vx^2\right) \TF\left(\sqrt{\vx},\CW,\CW\right)}{2 \left(4 \CW^2-\vx\right) \vx},\\
\FNch(\vx,\vy) = & \GlobalC\,\Bigg[ \frac{9 \vx \left(2 \CW^2-\vx+\vy\right)}{\vy} \nonumber\\
& + \left[\Aones(\vx,\vy)\ln\left(\frac{\vx}{\CWsq}\right)+ 9 \CW^4\left(\CW^4-4\CW^2\vy+3\vy^2\right)\ln(\CWsq)\right]\frac{\ln\left(\vy/\CWsq\right)}{2\vy^2(\CWsq-\vy)}\nonumber\\
& + \Atwos(\vx,\vy)\frac{\ln\left(\vx\right)}{\vy(4\CWsq-\vx)} + \Athrees(\vx,\vy)\frac{\ln\left(\vy\right)}{\vy(\CWsq-\vy)}\nonumber\\
&  + \Afours(\vx,\vy)\frac{\ln\left(\CWsq\right)}{\vy^2(4\CWsq-\vx)(\CWsq-\vy)} + \frac{\Afives(\vx,\vy)}{\CWsq\vy^2} \dilog\left(1-\frac{\vx}{\CWsq}\right)\nonumber\\
&  +\frac{\Asixs(\vx,\vy)}{\vx\CWsq(4\CWsq-\vx)^2(\CWsq-\vy)}\TF\left(\sqrt{\vx},\CW,\CW\right)\nonumber\\
& +\frac{\Asevens(\vx,\vy)}{\vy^2 \left(\CW^2-\vy\right) \left(\CW^4-2 \CW^2 (\vx+\vy)+(\vx-\vy)^2\right)}\TF\left(\sqrt{\vx},\sqrt{\vy},\CW\right)\Bigg],\\
\Aones(\vx,\vy) =& \Aone ,\\
\Atwos(\vx,\vy) =& \Atwo ,\\
\Athrees(\vx,\vy) =& \Athree ,\\
\Afours(\vx,\vy) =& \Afour, \\
\Afives(\vx,\vy) =& \Afive,\\
\Asixs(\vx,\vy) =& \Asix ,\\
\Asevens(\vx,\vy) =& \Aseven.
\end{align}
\section{Cancellation of \texorpdfstring{$\MMA$}{MA} dependence in \texorpdfstring{$\yAl$}{YAl} sector}
\label{appx:2}
In the bosonic contributions in~\eref{amub}, only the coefficient $\az$ depends on $\MMA$, whereas the Yukawa-coupling dependent parts are independent of $\MMA$. Here we provide details on this cancellation.
\begin{figure}[t]
\centering
\begin{subfigure}[]{.3\textwidth}
\includegraphics[scale=.5]{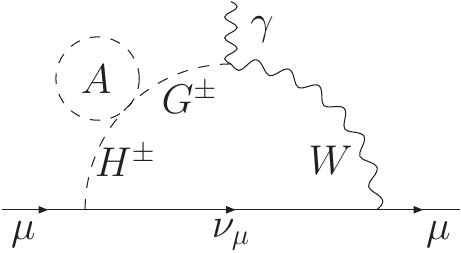}
\subcaption{}
\label{fig:MA01}
\end{subfigure}
\begin{subfigure}[]{.3\textwidth}
\includegraphics[scale=.5]{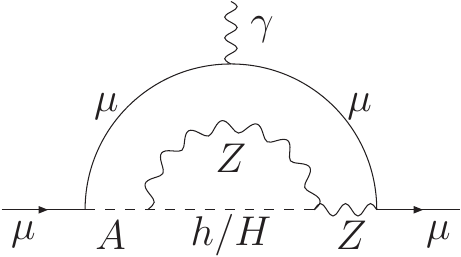}
\subcaption{}
\label{fig:MA02}
\end{subfigure}
\begin{subfigure}[]{.3\textwidth}
\includegraphics[scale=.5]{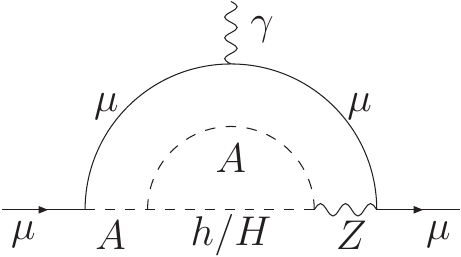}
\subcaption{}
\label{fig:MA03}
\end{subfigure}
\caption{The Feynman diagrams containing $\MA$ with Yukawa couplings.}
\label{fig:THDMA0}
\end{figure}

~\fref{fig:THDMA0} shows the only remaining two-loop Feynman diagram with $\MMA$ and Yukawa coupling dependence. The sum of the diagrams of~\frefand{fig:MA02}{fig:MA03} is zero, therefore these give no contributions. The remaining possible $\MMA$ dependence can arise from the diagram of~\fref{fig:MA01}. However, as we show in the following this contribution cancels out with the tadpole counterterm contribution.

The sum of the two-loop Feynman diagram~\fref{fig:MA01} and the counterterm diagram~\fref{fig:ctHG} with tadpole counterterm containing only $\MA$-loop can be illustrated as
\vspace{.2cm}
\begin{align}
\setlength{\unitlength}{1cm}
\begin{picture}(50,0)
\put(0.2,-0.7){\includegraphics[scale=.3]{{Fig.THDMHp.3.A0}.pdf}}
\put(2.6,-0.4){$+$}
\put(3,-0.7){\includegraphics[scale=.3]{{Fig.THDMCT13HG}.pdf}}
\put(5.4,-0.4){$=$}
\put(5.8,-0.7){\includegraphics[scale=.3]{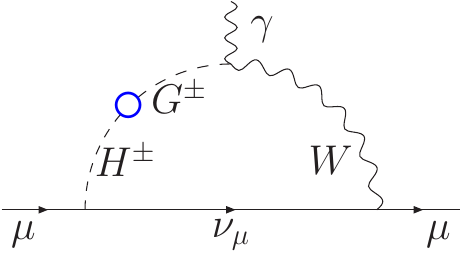}}
\put(8.2,-0.4){$\times \, \Big\{$}
\put(8.9,-0.6){\includegraphics[scale=.3]{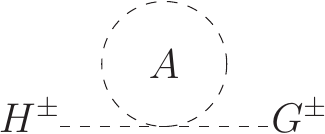}}
\put(10.6,-0.4){$+$}
\put(11.1,-0.4){\includegraphics[scale=.3]{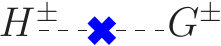}}
\put(12.2,-0.4){$\Big\}.$}
\end{picture}
\end{align}
The explicit form of the $H ^\pm$-$G ^\pm$ mixing propagator counterterm is 
\begin{align}\label{eq:deltatHG1}
\setlength{\unitlength}{1cm}
\begin{picture}(50,0)
\put(2.2,-.3){$\Big($}
\put(2.5,-.4){\includegraphics[scale=.6]{{Fig.THDMCTVHG}.pdf}}
\put(4.8,-.3){$\Big)$}
\put(5.3,-.3){$=$}
\put(6,-.3){$\left(\frac{i \gw}{2\,M_{\rm W}}\right)\delta t_{HG},$}
\end{picture}
\end{align}
where
\begin{align}\label{eq:deltatHG2}
\delta t_{HG}\,=&\,\cos(\beta - \alpha)\dth - \sin(\beta - \alpha)\dtH &\simeq\, -\dtH + \eta\,\dth + {\cal O}(\eta^2). 
\end{align}

According to the definition of the tadpole counterterms, the $\MA$ Higgs boson contribution to the tadpole counterterm for $\Mh$ is the product of the $(\Mh-\MA-\MA)$ coupling constant and the scalar one-point loop function $\fA (m)$,
\vspace{.5cm}
\begin{align}\label{eq:hA0}
\setlength{\unitlength}{1cm}
\begin{picture}(50,0)
\put(.5,.5){$\dth ^{\MA}\,=$}
\put(2,.5){$i \Big($}
\put(2.4,.1){\includegraphics[scale=.5]{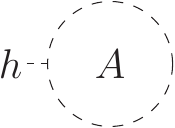}}
\put(4.1,.5){$\Big)$}
\put(.5,-.5){$=\frac{\gw}{2\,M_{W}}\fA(\MMA)\left\{2\MMA ^2 + \MMh^2 - \lFA\vac ^2 - (\MMA ^2- \lFA\vac ^2)\coefftb\eta\right\}$.}
\end{picture}
\end{align}
For $\MH$ we need to replace the $(\Mh-\MA-\MA)$ with $(\MH-\MA-\MA)$ coupling constant, and the result reads
\vspace{.9cm}
\begin{align}\label{eq:HA0}
\setlength{\unitlength}{1cm}
\begin{picture}(50,0)
\put(.5,.5){$\dtH ^{\MA}\,=$}
\put(2,.5){$i \Big($}
\put(2.4,.1){\includegraphics[scale=.5]{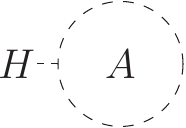}}
\put(4.1,.5){$\Big)$}
\put(.5,-.5){$=\frac{\gw}{2M_{W}}\fA(\MMA)\left\{(\MMH ^2 - \frac{\lFA\vac ^2}{2})\coefftb + (\MMH ^2 + 2\MMA ^2 - \lFA\vac ^2)\eta \right\}$.}
\end{picture}
\end{align}
By combining the previous equations we obtain for \eref{eq:deltatHG1} 
\begin{align}
\frac{i\,\gw}{2\,M_{W}}\,\delta t_{HG} \,=&\, \frac{-i\,\gw ^2}{4\,M_{W} ^2}\,\fA(\MMA)\,\times\nonumber\\
&\left\{(\MMH ^2 - \frac{\lFA\,\vac ^2}{2})\coefftb + (\MMH ^2 - \MMh ^2)\eta\right\}.
\end{align}
This shows that the $\MMA$ dependence is now only localized in $\fA (\MMA)$. On the other hand, 
after applying the quartic coupling constant of $H^\pm$-$G^\pm$-$\MA$-$\MA$, we obtain the explicit form of the inner loop of~\fref{fig:THDMA0}, with the result  
\vspace{.7cm}
\begin{align}\label{eq:THDMHGAA1}
\setlength{\unitlength}{1cm}
\begin{picture}(50,0)
\put(0,0.2){$\Big($}
\put(.5,0){\includegraphics[scale=.5]{{Fig.THDMHGAA}.pdf}}
\put(3.5,0.2){$\Big)$}
\put(4,0.2){$= \frac{i \gw ^2}{4\,M_{W} ^2} \fA(\MMA)\,\times$}
\put(4.4,-.7){$\left\{(\MMH ^2 - \frac{\lFA\,\vac ^2}{2})\coefftb + (\MMH ^2 - \MMh ^2)\,\eta\right\}$.}
\end{picture}
\end{align}
\vspace{1.5cm}
Hence, the $\MMA$ dependent parts vanish,
\vspace{-.5cm}
\begin{align}
\setlength{\unitlength}{1cm}
\begin{picture}(50,0)
\put(3,0.2){$\Big($}
\put(3.5,0){\includegraphics[scale=.5]{{Fig.THDMHGAA}.pdf}}
\put(6.5,0.2){$+$}
\put(7.2,0.2){\includegraphics[scale=.5]{{Fig.THDMCTVHG}.pdf}}
\put(9.1,0.2){$\Big)$}
\put(9.5,0.2){$= 0$.}
\end{picture}
\end{align}


\newpage

\begin{center}
{\Large\bf\boldmath Erratum: The muon magnetic moment in the $\thdm$: complete two-loop result}
\\\vspace{3em}
{Adriano Cherchiglia, Patrick Kneschke, Dominik St\"ockinger,\\ Hyejung St\"ockinger-Kim}\\[2em]
 {\sl Institut f\"ur Kern- und Teilchenphysik, TU Dresden, 01069 Dresden, Germany}
\end{center}





Here we provide corrections to our paper. The corrections are mostly typos in the formulas printed in the paper which do not affect the analytic results implemented in our codes for the numerical analyses.
In addition to the typo corrections we improve our approximation formula, eq.\ (67) and the corresponding plots in figures\ 10 and 11 as explained below.
In this context we also mention that the phenomenological discussions of the present paper are updated and superseded by the ones of our successive paper~\cite{Cherchiglia:2017uwv}. 

\begin{enumerate}
\item In the Lagrangian eq.\ (16), the abbreviation $y _f ^{H^\pm}$ for the  Yukawa coupling of the charged Higgs was not defined. 
  Here we provide a slightly rewritten version of the Lagrangian, including all necessary definitions. 
  It assumes $CP$ conservation and all appearing abbreviations to be real and reads
  \begin{align}
    \label{yukawaalign}
          {\cal{L}} _Y = &\sqrt{2} H ^+ \big({\bar{u}}[V_{\text{CKM}}\, y _d ^{\MHpm} P _{\text{R}} + y _u ^{\MHpm} V_{\text{CKM}} P _{\text{L}}] d + {\bar{\nu}} y _l ^{\MHpm} P _{\text{R}} l \big) \nonumber\\&- \sum _{f}\,h{\bar{f}} y _f ^{h} P _{\text{R}} f - \sum _{f}\,H{\bar{f}} y _f ^{H} P _{\text{R}} f +i\sum _{f}\,A{\bar{f}} y _f ^{A} P _{\text{R}} f+ h.c.. \nonumber
  \end{align}
  The Yukawa couplings in eq.\ (18) should be replaced by
  \begin{align}
    Y^{\Mh}_{f} =& \sin(\beta-\alpha)+\cos(\beta-\alpha)\zeta_{f}, \nonumber\\
    Y^{\MH}_{f} =& \cos(\beta-\alpha)-\sin(\beta-\alpha)\zeta_{f}, \nonumber\\
    Y^{\MHpm}_{d,l}=Y^{\MA}_{d,l} =& -\zeta_{d,l}, \quad Y^{\MHpm}_{u}=Y^{\MA}_{u} = \zeta_{u},  \nonumber
  \end{align}
  and those in eq.\ (19) should be replaced by
  \begin{align}
    Y^{\Mh}_{f} =& 1 + \eta \zeta_{f}, \quad Y^{\MH}_{f} = - \zeta_{f} + \eta, \quad Y^{\MHpm}_{f}=Y^{\MA}_{f} =-\Theta^{\MA} _{f}\zeta_{f},\nonumber\\
    \Theta^{\MA} _{d,l} =& 1, \qquad
    \Theta^{\MA} _{u} = -1, \qquad
    \Theta^{\MH} _{u,d,l} = 1, \nonumber
  \end{align}
  which includes $Y _f ^{\MHpm}$.
\item In eq.\ (51) $\vac$ is missing in the triple Higgs coupling constant. The correct formula is
  $$
  g_{\MH,H^\pm,H^\mp} \propto \left\{
  \left(\tb - \frac{1}{\tb}\right)\frac{\vac}{2}\left( \lFA -  2\,\frac{\MMH^2}{\vac^2} \right)\right.
  \nonumber\\
  \left.\,\, + \eta\, \vac \left( \lFA - \frac{\MMH^2}{\vac^2} - 2\frac{\MMHpm^2}{\vac^2}\right)\right\}.
  $$
\item Page 25, line 2: $M_{l}=\{(m_{e},0),(m_{\mu},0),(m_{\tau},0)\}$.
\item There was an unnecessary, extra $\TF(x_{d}^{1/2},x_{u}^{1/2},1)$
  in the second line of eq.\ (61), and its corrected version reads
  \begin{align} 
    &\mathcal{F}^{\MHpm}_{d}(M_{d})=-(x_{u}-x_{d})+\left[\frac{\bar{c}}{y}-c\left(\frac{x_{u}-x_{d}}{y}\right)\right]\TF(x_{d}^{1/2},x_{u}^{1/2},1)\nonumber\\
    &\quad\quad\quad\quad\quad\quad+c\left[\mbox{Li}_{2}\left(1-\frac{x_{d}}{x_{u}}\right)-\frac{1}{2}\ln(x_{u})\ln\left(\frac{x_{d}}{x_{u}}\right)\right]\nonumber\\
    &\quad\quad\quad\quad\quad\quad+\left(s+x_{d}\right)\ln(x_{d})+\left(s-x_{u}\right)\ln(x_{u}). \nonumber
  \end{align}
\item A factor of $-\frac{1}{2}$ was missing in eq.\ (78), and the correct expression is 
  \begin{align}
    \FFE(\vx,\vy) =& -\frac{\fZfives}{2}( 2(\vx + \vy) - (\vx - \vy)^2 -1 )\ln\left(\frac{\fSone(\vx,\vy)}{2\sqrt{\vx\vy}}\right)\nonumber\\
    &\times\left(\vx + \vy - 1 - \frac{4\vx\vy}{\fSone(\vx,\vy)}\right). \nonumber
  \end{align}
\item In the numerical evaluation of our results we have used one-loop corrected relationships between the electroweak parameters: $s_{W}$, $c_{W}$, $M_{W}$, $M_{Z}$, $v$.
  In this way, the numerical evaluation of $\amu$ contains certain terms which are formally of 3-loop order but which do not correspond to a full 3-loop calculation.
  In the following we provide results based on an evaluation which uses tree-level relationships between these electroweak parameters and which thereby corresponds to a pure 2-loop calculation in the on-shell renormalization scheme.
  
  According to the original numerical implementation, there was a slight increase with $\tb$ visible in the original figure\ 10a, and figures 10d and 11 for $\tb = 100$ were affected by the incomplete 3-loop effects mentioned above. 
  Actually, the linear large-$\tb$-enhancement vanishes in the new, strict 2-loop evaluation.
  This can be understood with the help of eq.\ (67). This equation can now be analytically evaluated by using the
  tree-level relationships for the electroweak parameters in eqs.\ (15) and (99).
  In this case, the term proportional to $M_H ^2 \zeta_l \tb$ on the far right-hand side of eq.\ (67) analytically vanishes.
  In the approximation of large $\tb$, the far right-hand side of eq.\ (67) contains only terms which decrease or approach a constant for large $\tb$.
  
  Here we provide revised versions of the plots in figures\ 10 and 11 with the modified numerical evaluation using tree-level relations between electroweak parameters: see figures~\ref{MA50} and~\ref{MA50B}.
  Here the linear increase in $\tb$ is absent and the numerical results in the large $\tb$ regime are slightly changed.
  Again, we refer to Ref.~\cite{Cherchiglia:2017uwv} for a more detailed phenomenological evaluation which also takes into account a variety of experimental constraints on the input parameters $\zeta_l$, $\tb$, and $\eta$.  
\end{enumerate}

\begin{figure}[]
  \centering
  \begin{subfigure}{0.45\textwidth}
    \centering
    \includegraphics[scale=.1]{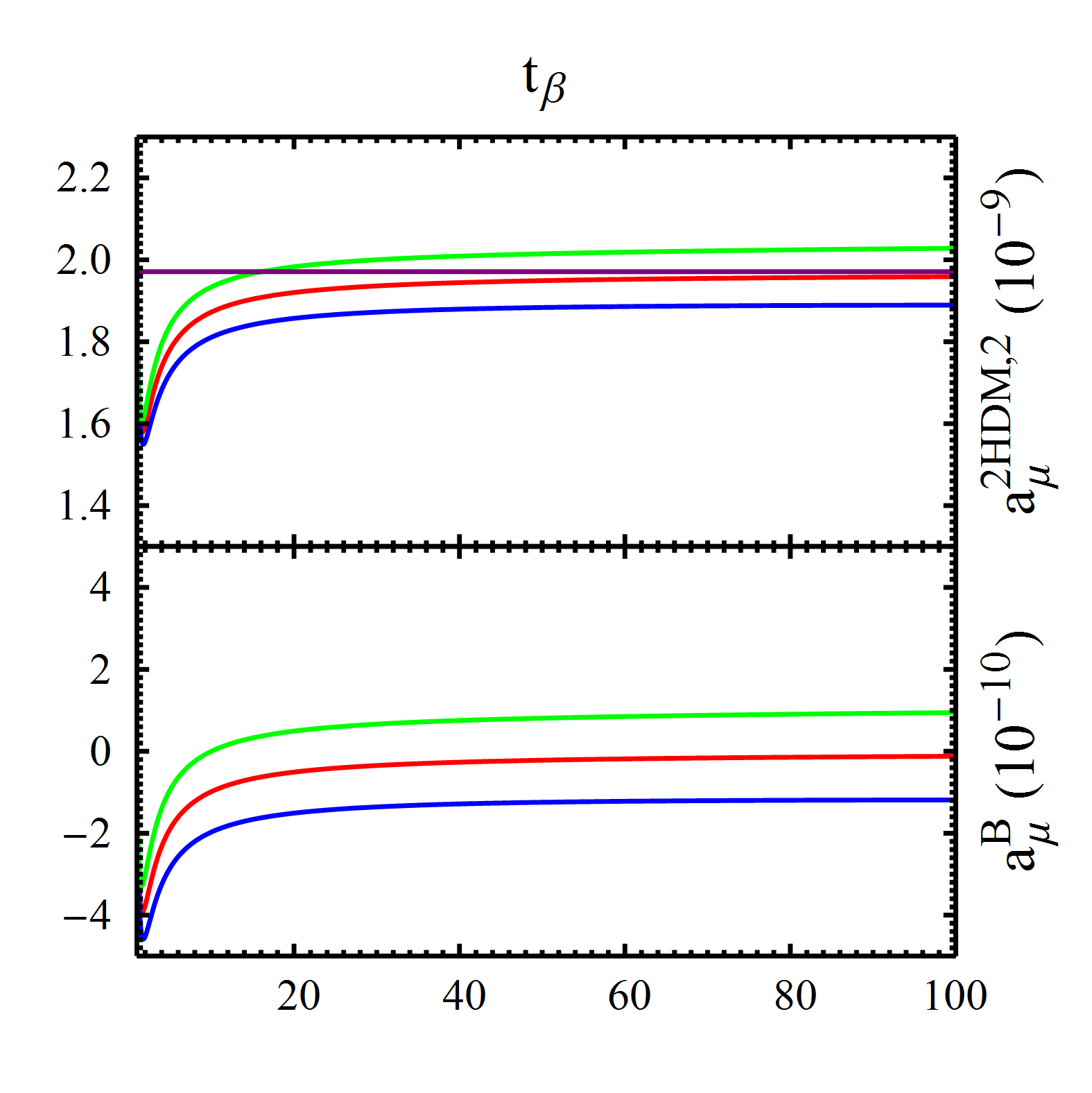}
    \vspace{-1\baselineskip}
    \subcaption{}
    \label{MA50:HighTB}
  \end{subfigure}
  \begin{subfigure}{0.45\textwidth}
    \centering
    \includegraphics[scale=.1]{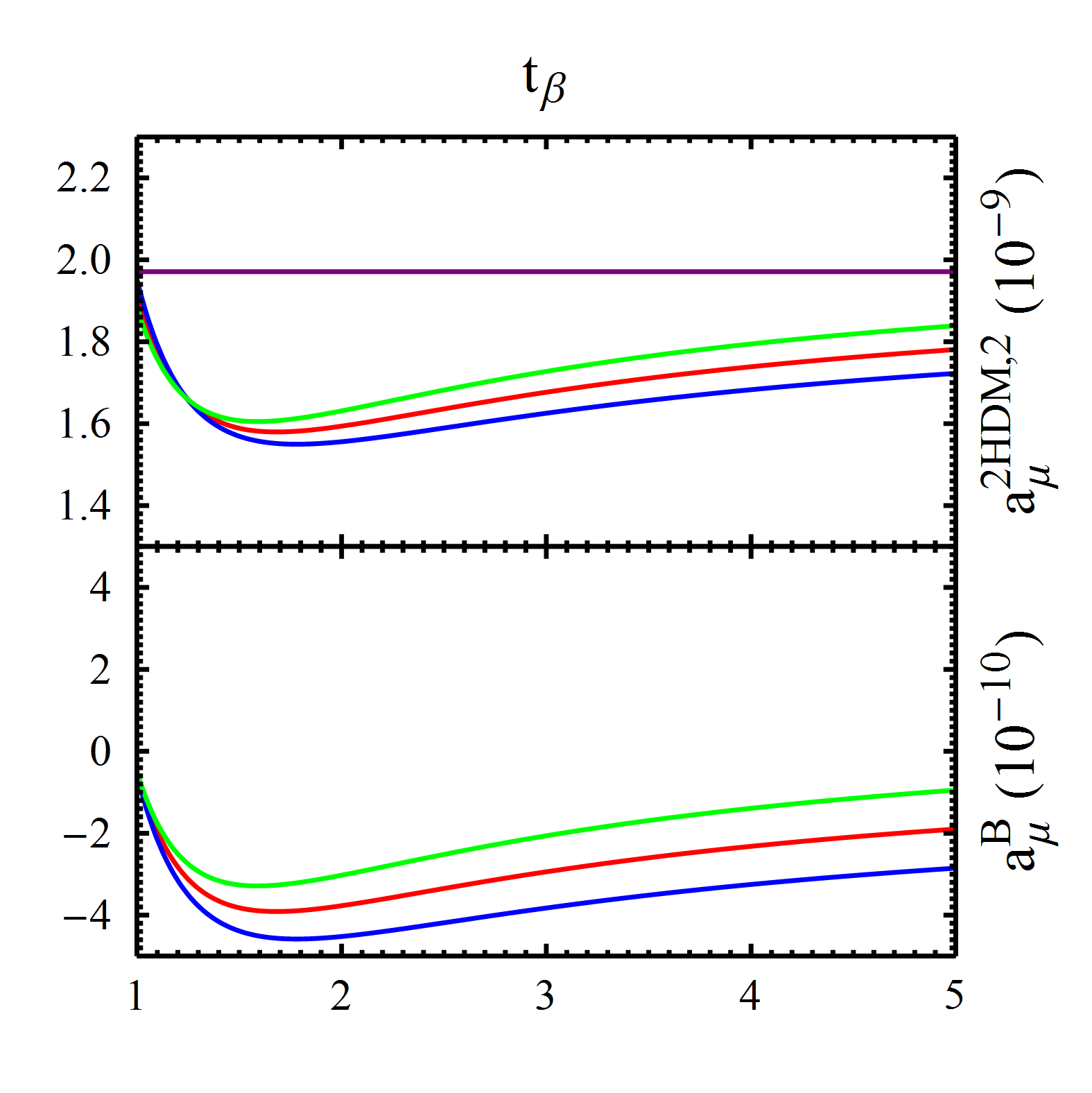}
    \vspace{-1\baselineskip}
    \subcaption{}
    \label{MA50:LowTB}
  \end{subfigure}\\\vspace{.1cm}
  \begin{subfigure}{0.45\textwidth}
    \centering
    \includegraphics[scale=.1]{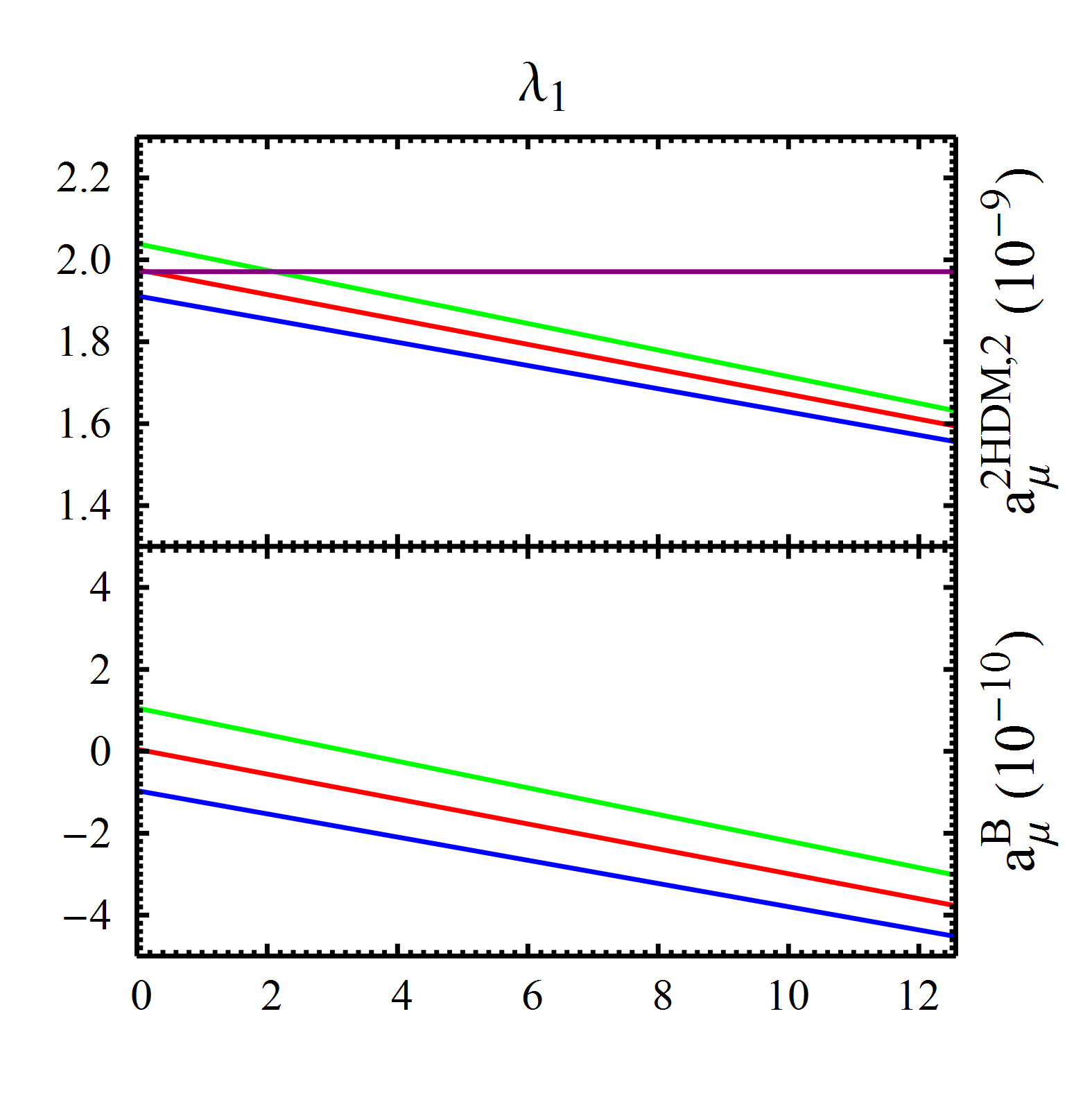}
    \vspace{-1\baselineskip}
    \subcaption{}
    \label{MA50:L1TB2}
  \end{subfigure}
  \centering
  \begin{subfigure}{0.45\textwidth}
    \centering
    \includegraphics[scale=.1]{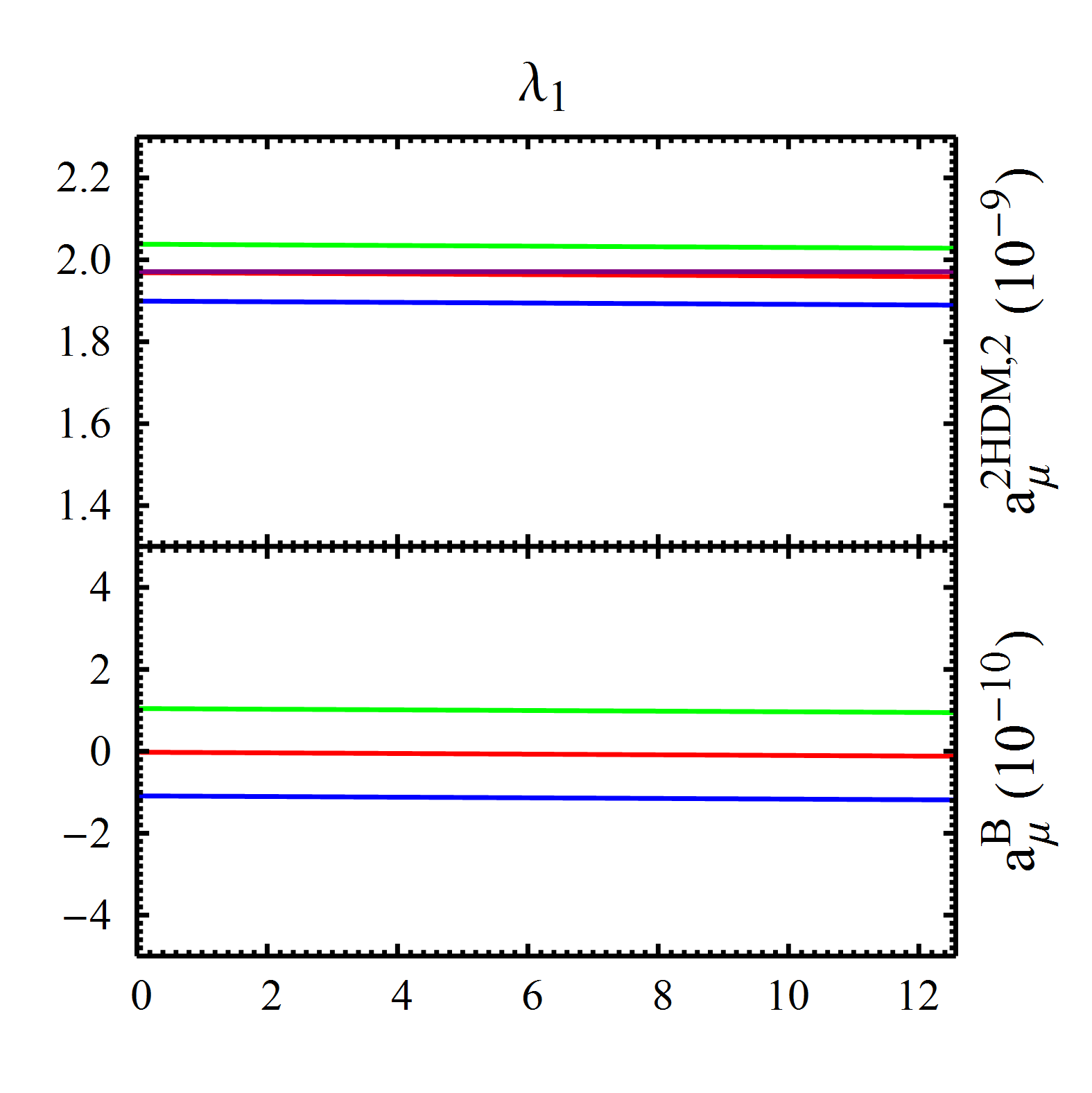}
    \vspace{-1\baselineskip}
    \subcaption{}
    \label{MA50:L1TB100}
  \end{subfigure}
  \vspace{-0.3\baselineskip}
  \caption{Revised version of figure\ 10. 
Plots showing the behavior of $\amu^{\thdm, 2}$, and $\amub$. Each red/blue/green line is for $\eta = 0/0.1/-0.1$. $\tb$ varies for (a) and (b), and $\lambda_{1}$ for (c) and (d). We consider
 the representative mass parameter point in eq.\ (4.3). $\lambda_{1}=4\pi$ for (a) and (b). We employ $\tb=2$ and $\tb=100$ for (c) and (d) respectively.
 \label{MA50}}
\end{figure}

\begin{figure}[]
  \begin{subfigure}{0.45\textwidth}
    \centering
    \includegraphics[scale=.1]{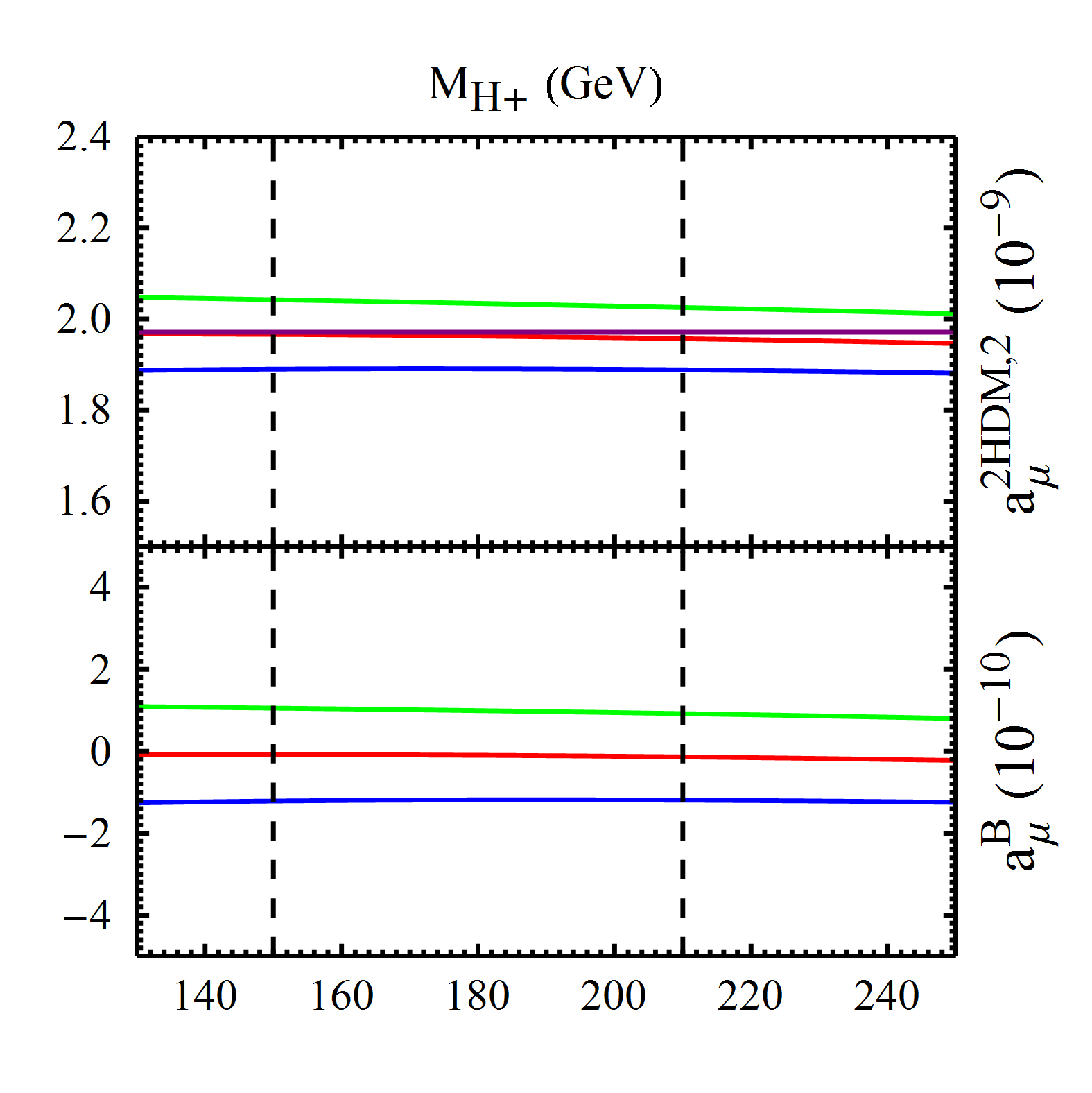}
    \vspace{-1\baselineskip}
    \subcaption{}
    \label{MA50:MHp}
  \end{subfigure}
  \begin{subfigure}{0.45\textwidth}
    \centering
    \includegraphics[scale=.1]{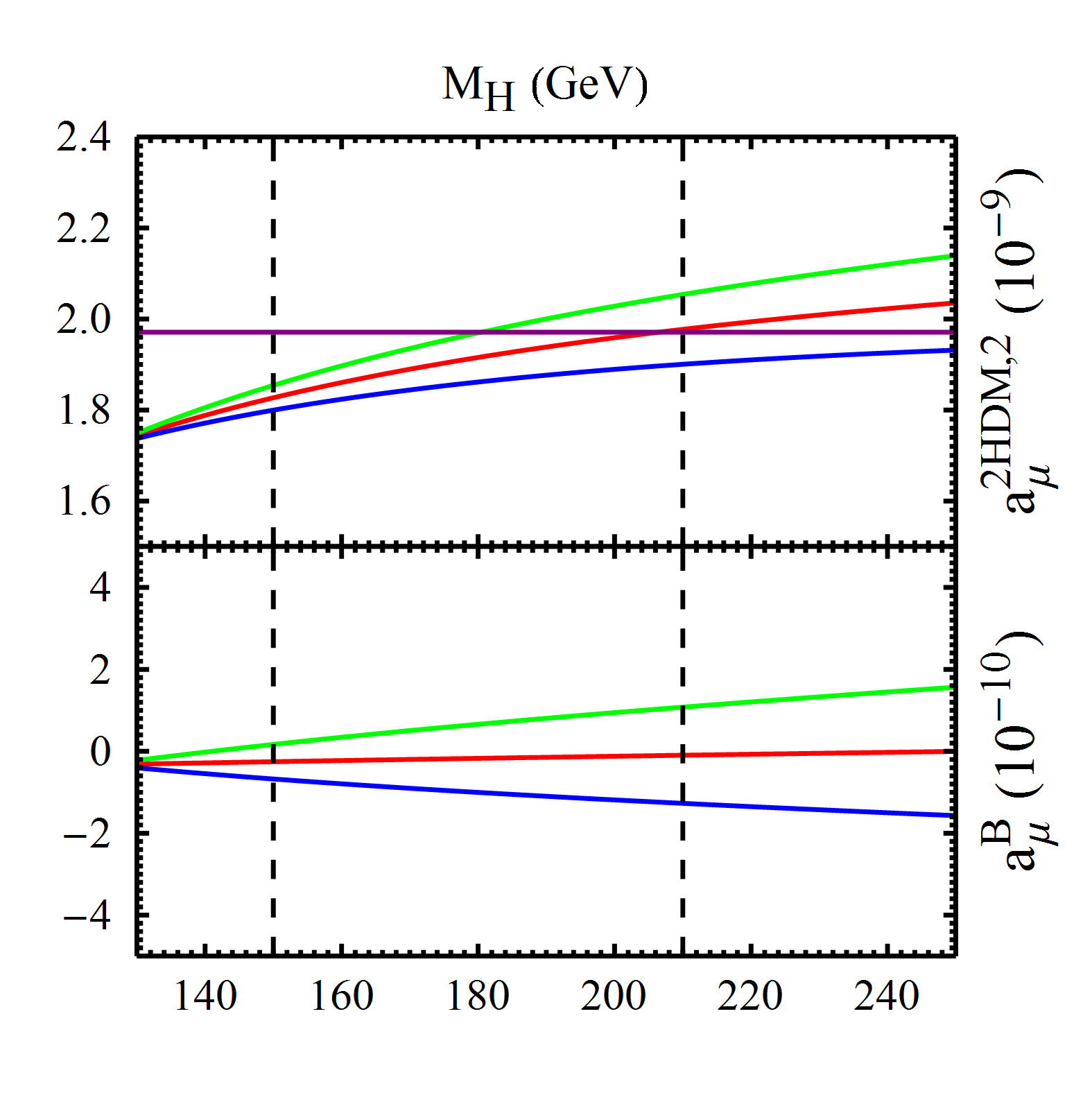}
    \vspace{-1\baselineskip}
    \subcaption{}
    \label{MA50:MHH}
  \end{subfigure}
  \vspace{-0.3\baselineskip}
  \caption{Revised version of figure\ 11. 
    Plots showing the behavior of $\amu^{\thdm, 2}$, and $\amub$. Each red/blue/green line is for $\eta = 0/0.1/-0.1$. $\MMHpm$ and $\MMH$ vary in (a) and (b) respectively. We set $\lambda_{1}=4\pi$, and $\tb=100$. The inside regions between the dashed lines are allowed by constraints. The purple line is a reference value as explained in the text.
    \label{MA50B}}
\end{figure}

\section*{Acknowledgement}
We thank  Douglas Jacob, Alexander Voigt, and Gareth Williams for pointing out the typos.


\begin{thebibliography}{99}

\bibitem{Aad:2012tfa}
  G.~Aad {\it et al.} [ATLAS Collaboration],
  Phys.\ Lett.\ B {\bf 716} (2012) 1
  [arXiv:1207.7214 [hep-ex]].

\bibitem{Chatrchyan:2012xdj}
  S.~Chatrchyan {\it et al.} [CMS Collaboration],
  Phys.\ Lett.\ B {\bf 716} (2012) 30
  [arXiv:1207.7235 [hep-ex]].

\bibitem{DuhrssenMoriond}
M.~D\"uhrssen, talk given at {\em 50th Rencontres de Moriond EW 2015};


\bibitem{CzM}
  A.~Czarnecki and W.~J.~Marciano,
  Phys.\ Rev.\ D {\bf 64} (2001) 013014
  [hep-ph/0102122].

\bibitem{WhitePaper} D.~W.~Hertzog, J.~P.~Miller, E.~de Rafael, B.~Lee
  Roberts and D.~St\"ockinger, 
arXiv:0705.4617.

\bibitem{Stockinger:1900zz}
  D.~St\"ockinger,
  in {\em Lepton Dipole Moments}, B.~L.~Roberts and W.~J.~Marciano Eds.,
  Adv.\ Ser.\ Direct.\ High Energy Phys.\  {\bf 20} (2009) 393,

\bibitem{Bennett:2006fi} G.W. Bennett, et al.,
(Muon $(g-2)$ Collaboration), Phys. Rev. D {\bf 73}, 072003 (2006).

\bibitem{Davier}
  M.~Davier, A.~Hoecker, B.~Malaescu and Z.~Zhang,
  Eur.\ Phys.\ J.\ C {\bf 71} (2011) 1515
   [Erratum-ibid.\ C {\bf 72} (2012) 1874]
  [arXiv:1010.4180 [hep-ph]].

\bibitem{HMNT}
  K.~Hagiwara, R.~Liao, A.~D.~Martin, D.~Nomura and T.~Teubner,
  J.\ Phys.\ G {\bf 38} (2011) 085003
  [arXiv:1105.3149 [hep-ph]].

\bibitem{Kinoshita2012}
  T.~Aoyama, M.~Hayakawa, T.~Kinoshita and M.~Nio,
  Phys.\ Rev.\ Lett.\  {\bf 109}, 111808 (2012)
  [arXiv:1205.5370 [hep-ph]].

\bibitem{Gnendiger:2013pva} 
  C.~Gnendiger, D.~St\"ockinger and H.~St\"ockinger-Kim,
  Phys.\ Rev.\ D {\bf 88}, 053005 (2013)
  [arXiv:1306.5546 [hep-ph]].

\bibitem{Kataev:2012kn}
  A.~L.~Kataev,
  Phys.\ Rev.\ D {\bf 86} (2012) 013010
  [arXiv:1205.6191 [hep-ph]].

\bibitem{SteinhauserQED}
  R.~Lee, P.~Marquard, A.~V.~Smirnov, V.~A.~Smirnov and M.~Steinhauser,
  JHEP {\bf 1303} (2013) 162
  [arXiv:1301.6481 [hep-ph]];
  A.~Kurz, T.~Liu, P.~Marquard and M.~Steinhauser,
  Nucl.\ Phys.\ B {\bf 879} (2014) 1
  [arXiv:1311.2471 [hep-ph]];
  A.~Kurz, T.~Liu, P.~Marquard, A.~V.~Smirnov, V.~A.~Smirnov and M.~Steinhauser,
  Phys.\ Rev.\ D {\bf 92} (2015) no.7,  073019
  [arXiv:1508.00901 [hep-ph]];
  Phys.\ Rev.\ D {\bf 93} (2016) no.5,  053017
  [arXiv:1602.02785 [hep-ph]].

\bibitem{JegerlehnerSzafron}
  F.~Jegerlehner and R.~Szafron,
  Eur.\ Phys.\ J.\ C {\bf 71} (2011) 1632
  [arXiv:1101.2872 [hep-ph]].

\bibitem{Benayoun:2012wc}
  M.~Benayoun, P.~David, L.~DelBuono and F.~Jegerlehner,
  Eur.\ Phys.\ J.\ C {\bf 73} (2013) 2453
  [arXiv:1210.7184 [hep-ph]];
  Eur.\ Phys.\ J.\ C {\bf 75} (2015) no.12,  613
  [arXiv:1507.02943 [hep-ph]];
  arXiv:1605.04474 [hep-ph].

\bibitem{Kurz:2014wya}
  A.~Kurz, T.~Liu, P.~Marquard and M.~Steinhauser,
  Phys.\ Lett.\ B {\bf 734} (2014) 144
  [arXiv:1403.6400 [hep-ph]].

\bibitem{Colangelo:2014qya}
  G.~Colangelo, M.~Hoferichter, A.~Nyffeler, M.~Passera and P.~Stoffer,
  Phys.\ Lett.\ B {\bf 735} (2014) 90
  [arXiv:1403.7512 [hep-ph]].

\bibitem{Colangelo:2014}
  G.~Colangelo, M.~Hoferichter, M.~Procura and P.~Stoffer,
  JHEP {\bf 1409} (2014) 091
  [arXiv:1402.7081 [hep-ph]];
  G.~Colangelo, M.~Hoferichter, B.~Kubis, M.~Procura and P.~Stoffer,
  Phys.\ Lett.\ B {\bf 738} (2014) 6
  [arXiv:1408.2517 [hep-ph]];
  G.~Colangelo, M.~Hoferichter, M.~Procura and P.~Stoffer,
  JHEP {\bf 1509} (2015) 074
  [arXiv:1506.01386 [hep-ph]].

\bibitem{Pauk:2014rfa}
  V.~Pauk and M.~Vanderhaeghen,
  Phys.\ Rev.\ D {\bf 90} (2014) 11,  113012
  [arXiv:1409.0819 [hep-ph]].

\bibitem{lattice}
  T.~Blum, S.~Chowdhury, M.~Hayakawa and T.~Izubuchi,
  Phys.\ Rev.\ Lett.\  {\bf 114} (2015) 1,  012001
  [arXiv:1407.2923 [hep-lat]];
  T.~Blum, N.~Christ, M.~Hayakawa, T.~Izubuchi, L.~Jin and C.~Lehner,
  Phys.\ Rev.\ D {\bf 93} (2016) no.1,  014503
  [arXiv:1510.07100 [hep-lat]].

\bibitem{Ablikim:2015orh} 
  M.~Ablikim {\it et al.} [BESIII Collaboration],
  Phys.\ Lett.\ B {\bf 753}, 629 (2016)
  [arXiv:1507.08188 [hep-ex]].

\bibitem{Chakraborty:2015ugp}
  B.~Chakraborty, C.~T.~H.~Davies, J.~Koponen, G.~P.~Lepage, M.~J.~Peardon and S.~M.~Ryan,
  Phys.\ Rev.\ D {\bf 93} (2016) no.7,  074509
  [arXiv:1512.03270 [hep-lat]].


\bibitem{JegerlehnerNyffeler}  
F.~Jegerlehner and A.~Nyffeler,
  Phys.\ Rept.\  {\bf 477}, 1 (2009)
  [arXiv:0902.3360 [hep-ph]].

\bibitem{Miller:2012opa} 
  J.~P.~Miller, E.~de Rafael, B.~L.~Roberts and D.~St\"ockinger,
  Ann.\ Rev.\ Nucl.\ Part.\ Sci.\  {\bf 62}, 237 (2012).

\bibitem{Blum:2013xva}
  T.~Blum, A.~Denig, I.~Logashenko, E.~de Rafael, B.~L.~Roberts, T.~Teubner and G.~Venanzoni,
  arXiv:1311.2198 [hep-ph].
\bibitem{Benayoun:2014tra}
  M.~Benayoun, J.~Bijnens, T.~Blum, I.~Caprini, G.~Colangelo, H.~Czyz, A.~Denig and C.~A.~Dominguez {\it et al.},
  arXiv:1407.4021 [hep-ph].

\bibitem{Melnikov:2016wdt}
  K.~Melnikov,
  EPJ Web Conf.\  {\bf 118} (2016) 01020.














\bibitem{Broggio:2014mna}
  A.~Broggio, E.~J.~Chun, M.~Passera, K.~M.~Patel and S.~K.~Vempati,
  JHEP {\bf 1411} (2014) 058
  [arXiv:1409.3199 [hep-ph]].

\bibitem{Wang:2014sda} 
  L.~Wang and X.~F.~Han,
  JHEP {\bf 1505}, 039 (2015)
  [arXiv:1412.4874 [hep-ph]].


\bibitem{Ilisie:2015tra}
  V.~Ilisie,
  JHEP {\bf 1504} (2015) 077
  [arXiv:1502.04199 [hep-ph]].

\bibitem{Abe:2015oca} 
  T.~Abe, R.~Sato and K.~Yagyu,
  JHEP {\bf 1507}, 064 (2015)
  [arXiv:1504.07059 [hep-ph]].

\bibitem{Crivellin:2015hha}
  A.~Crivellin, J.~Heeck and P.~Stoffer,
  Phys.\ Rev.\ Lett.\  {\bf 116} (2016) no.8,  081801
  [arXiv:1507.07567 [hep-ph]].

\bibitem{Chun:2015hsa}
  E.~J.~Chun, Z.~Kang, M.~Takeuchi and Y.~L.~S.~Tsai,
  JHEP {\bf 1511} (2015) 099
  [arXiv:1507.08067 [hep-ph]].

\bibitem{Han:2015yys} 
  T.~Han, S.~K.~Kang and J.~Sayre,
  JHEP {\bf 1602}, 097 (2016)
  [arXiv:1511.05162 [hep-ph]].


\bibitem{BarrZee}
  S.~M.~Barr and A.~Zee,
  Phys.\ Rev.\ Lett.\  {\bf 65} (1990) 21
  [Erratum-ibid.\  {\bf 65} (1990) 2920].

\bibitem{Chang:2000ii}
  D.~Chang, W.~F.~Chang, C.~H.~Chou and W.~Y.~Keung,
  Phys.\ Rev.\ D {\bf 63} (2001) 091301
  doi:10.1103/PhysRevD.63.091301
  [hep-ph/0009292].

\bibitem{Cheung:2001hz}
  K.~m.~Cheung, C.~H.~Chou and O.~C.~W.~Kong,
  Phys.\ Rev.\ D {\bf 64} (2001) 111301
  doi:10.1103/PhysRevD.64.111301
  [hep-ph/0103183].

\bibitem{Wu:2001vq}
  Y.~L.~Wu and Y.~F.~Zhou,
  Phys.\ Rev.\ D {\bf 64} (2001) 115018
  doi:10.1103/PhysRevD.64.115018
  [hep-ph/0104056].

\bibitem{Krawczyk:2002df}
  M.~Krawczyk,
  Acta Phys.\ Polon.\ B {\bf 33} (2002) 2621
  [hep-ph/0208076].





\bibitem{Carey:2009zzb}
  R.~M.~Carey, K.~R.~Lynch, J.~P.~Miller, B.~L.~Roberts, W.~M.~Morse, Y.~K.~Semertzides, V.~P.~Druzhinin and B.~I.~Khazin {\it et al.},
  FERMILAB-PROPOSAL-0989.
  B.~L.~Roberts,
  Chin.\ Phys.\ C {\bf 34} (2010) 741
  [arXiv:1001.2898 [hep-ex]].
%
\bibitem{Iinuma:2011zz}
  H.~Iinuma [J-PARC New g-2/EDM experiment Collaboration],
  J.\ Phys.\ Conf.\ Ser.\  {\bf 295} (2011) 012032.


\bibitem{CZKM}	
  A.~Czarnecki, B.~Krause and W.~J.~Marciano,
  Phys.\ Rev.\ Lett.\  {\bf 76} (1996) 3267
  [hep-ph/9512369].
%
  A.~Czarnecki, B.~Krause and W.~J.~Marciano,
  Phys.\ Rev.\ D {\bf 52} (1995) 2619
  [hep-ph/9506256].


\bibitem{CZMV}
  A.~Czarnecki, W.~J.~Marciano and A.~Vainshtein,
  Phys.\ Rev.\ D {\bf 67} (2003) 073006
   Erratum: [Phys.\ Rev.\ D {\bf 73} (2006) 119901]
  [hep-ph/0212229].


\bibitem{HSW04}
  S.~Heinemeyer, D.~St\"ockinger and G.~Weiglein,
  Nucl.\ Phys.\ B {\bf 699} (2004) 103.

\bibitem{HSW03}
  S.~Heinemeyer, D.~St\"ockinger and G.~Weiglein,
  Nucl.\ Phys.\ B {\bf 690} (2004) 62.
%


\bibitem{ArhribBaek}
  A.~Arhrib and S.~Baek,
  Phys.\ Rev.\ D {\bf 65} (2002) 075002
  [hep-ph/0104225].

\bibitem{ChenGeng}
  C.~H.~Chen and C.~Q.~Geng,
  Phys.\ Lett.\ B {\bf 511} (2001) 77
  [arXiv:hep-ph/0104151].


\bibitem{Cheung:2009fc}
  K.~Cheung, O.~C.~W.~Kong and J.~S.~Lee,
  JHEP {\bf 0906} (2009) 020
  [arXiv:0904.4352 [hep-ph]].	



\bibitem{vonWeitershausen:2010zr}
  P.~von Weitershausen, M.~Sch\"afer, H.~St\"ockinger-Kim and D.~St\"ockinger,
  Phys.\ Rev.\ D {\bf 81} (2010) 093004
  [arXiv:1003.5820 [hep-ph]].


\bibitem{Fargnoli:2013zda}
  H.~G.~Fargnoli, C.~Gnendiger, S.~Pa\ss{}ehr, D.~St\"ockinger and H.~St\"ockinger-Kim,
  Phys.\ Lett.\ B {\bf 726} (2013) 717
  [arXiv:1309.0980 [hep-ph]].

\bibitem{Fargnoli:2013zia}
  H.~G.~Fargnoli, C.~Gnendiger, S.~Pa\ss{}ehr, D.~St\"ockinger and H.~St\"ockinger-Kim,
  JHEP {\bf 1402} (2014) 070
  [arXiv:1311.1775 [hep-ph]].

\bibitem{DSreview} D.~St\"ockinger,
J.\ Phys.\ G {\bf 34} (2007) R45.

\bibitem{Gm2Calc}
  P.~Athron {\it et al.},
  Eur.\ Phys.\ J.\ C {\bf 76} (2016) no.2,  62
  [arXiv:1510.08071 [hep-ph]].


\bibitem{Branco:2011iw} 
  G.~C.~Branco, P.~M.~Ferreira, L.~Lavoura, M.~N.~Rebelo, M.~Sher and J.~P.~Silva,
  Phys.\ Rept.\  {\bf 516}, 1 (2012)
  [arXiv:1106.0034 [hep-ph]].

\bibitem{Gunion:2002zf} 
  J.~F.~Gunion and H.~E.~Haber,
  Phys.\ Rev.\ D {\bf 67}, 075019 (2003)
  [hep-ph/0207010].

\bibitem{Celis:2013ixa} 
  A.~Celis, V.~Ilisie and A.~Pich,
  JHEP {\bf 1312}, 095 (2013)
  [arXiv:1310.7941 [hep-ph]].

%

\bibitem{FeynArts}
  T. Hahn,
  Comput.~Phys.~Commun. {\bf 140} (2001) 418--431
  [arXiv:0012260 [hep-ph]].

\bibitem{Pich:2009sp} 
  A.~Pich and P.~Tuzon,
  Phys.\ Rev.\ D {\bf 80}, 091702 (2009)
  [arXiv:0908.1554 [hep-ph]].

\bibitem{Krause:2016oke}
  M.~Krause, R.~Lorenz, M.~Muhlleitner, R.~Santos and H.~Ziesche,
  JHEP {\bf 1609} (2016) 143
  doi:10.1007/JHEP09(2016)143
  [arXiv:1605.04853 [hep-ph]].
\bibitem{Denner:2016etu}
  A.~Denner, L.~Jenniches, J.~N.~Lang and C.~Sturm,
  JHEP {\bf 1609} (2016) 115
  doi:10.1007/JHEP09(2016)115
  [arXiv:1607.07352 [hep-ph]].


\bibitem{Ferreira:2009jb} 
  P.~M.~Ferreira and D.~R.~T.~Jones,
  JHEP {\bf 0908}, 069 (2009)
 [arXiv:0903.2856 [hep-ph]].

\bibitem{Barroso:2013awa} 
  A.~Barroso, P.~M.~Ferreira, I.~P.~Ivanov and R.~Santos,
  JHEP {\bf 1306}, 045 (2013)
  [arXiv:1303.5098 [hep-ph]].

\bibitem{Peskin:1990zt} 
  M.~E.~Peskin and T.~Takeuchi,
  Phys.\ Rev.\ Lett.\  {\bf 65}, 964 (1990).

\bibitem{Agashe:2014kda} 
  K.~A.~Olive {\it et al.} [Particle Data Group Collaboration],
  Chin.\ Phys.\ C {\bf 38}, 090001 (2014).

\bibitem{Eriksson:2009ws} 
  D.~Eriksson, J.~Rathsman and O.~Stal,
  Comput.\ Phys.\ Commun.\  {\bf 181}, 189 (2010)
  [arXiv:0902.0851 [hep-ph]].

\bibitem{Eriksson:2010zzb} 
  D.~Eriksson, J.~Rathsman and O.~Stal,
  Comput.\ Phys.\ Commun.\  {\bf 181}, 833 (2010).


%
%
%

   \bibitem{Weiglein:1993}
  G.~Weiglein, R.~Scharf and M.~B{\"o}hm,
  Nucl.\ Phys.\ B {\bf 416} (1994) 606
  [hep-ph/9310358].


\bibitem{Lautrup:1971jf}
  B.~e.~Lautrup, A.~Peterman and E.~de Rafael,
  Phys.\ Rept.\  {\bf 3} (1972) 193.

\bibitem{Leveille:1977rc}
  J.~P.~Leveille,
  Nucl.\ Phys.\ B {\bf 137} (1978) 63.

\bibitem{Dedes:2001nx}
  A.~Dedes and H.~E.~Haber,
  JHEP {\bf 0105} (2001) 006
  [hep-ph/0102297].
%
%
%
%
%

\bibitem{Bertolini:1985ia} 
  S.~Bertolini,
  Nucl.\ Phys.\ B {\bf 272}, 77 (1986).

\bibitem{LopezVal:2012zb} 
  D.~Lopez-Val and J.~Sola,
  Eur.\ Phys.\ J.\ C {\bf 73}, 2393 (2013)
  [arXiv:1211.0311 [hep-ph]].
	
\bibitem{Denner:1991kt} 
  A.~Denner,
  Fortsch.\ Phys.\  {\bf 41}, 307 (1993)
  [arXiv:0709.1075 [hep-ph]].	
	
	\bibitem{Santos:1996vt} 
  R.~Santos and A.~Barroso,
  Phys.\ Rev.\ D {\bf 56}, 5366 (1997)
  [hep-ph/9701257].

%
%

\bibitem{Caprio}	
M. A. Caprio, Comput. Phys. Commun. 171, 107 (2005)


\bibitem{Davydychev:1992mt} 
  A.~I.~Davydychev and J.~B.~Tausk,
  Nucl.\ Phys.\ B {\bf 397}, 123 (1993).

\end{thebibliography}

\begin{thebibliography}{99}

\bibitem{Cherchiglia:2017uwv}
A.~Cherchiglia, D.~St\"ockinger and H.~St\"ockinger-Kim,
\emph{Muon g-2 in the 2HDM: maximum results and detailed phenomenology},
\emph{Phys. Rev. D} {\bf{98}} (2018), 035001
[arXiv:1711.11567 [hep-ph]].

\end{thebibliography}
\end{document}